\documentclass[a4paper,11pt]{article}
\usepackage{jheppub} 

\pdfoutput=1

\usepackage{mathrsfs,amsmath,amsthm,amssymb,xcolor,epstopdf,verbatim,hyperref,enumerate,graphicx,subfigure,bm,bbm,amsfonts,subdepth,cleveref,setspace,booktabs,tabularx,placeins,multirow,mathtools,units}
\usepackage[english]{babel}
\usepackage[sort&compress]{natbib}
\usepackage[normalem]{ulem}
\usepackage{dcolumn}
\graphicspath{{Figs/}}
\usepackage{float}

\usepackage{tabularx}

\newcommand{\drawsquare}[2]{\hbox{%
		\rule{#2pt}{#1pt}\hskip-#2pt
		\rule{#1pt}{#2pt}\hskip-#1pt
		\rule[#1pt]{#1pt}{#2pt}}\rule[#1pt]{#2pt}{#2pt}\hskip-#2pt
	\rule{#2pt}{#1pt}}

\newcommand{\fund}{\raisebox{-.5pt}{\drawsquare{6.5}{0.4}}}
\newcommand{\Ysymm}{\raisebox{-.5pt}{\drawsquare{6.5}{0.4}}\hskip-0.4pt%
	\raisebox{-.5pt}{\drawsquare{6.5}{0.4}}}
\newcommand{\Yasymm}{\raisebox{-3.5pt}{\drawsquare{6.5}{0.4}}\hskip-6.9pt%
	\raisebox{3pt}{\drawsquare{6.5}{0.4}}}
\newcommand{\antifund}{\overline{\fund}}

 \renewcommand{\thetable}{\arabic{table}} 
 
\def\yzero{\smash{\hbox{$y\kern-4pt\raise1pt\hbox{${}^\circ$}$}}}

\def\s2{\frac{1}{\sqrt2}}

\def\beq{\begin{equation}}
	\def\eeq{\end{equation}}
\def\beqa{\begin{eqnarray}}
	\def\eeqa{\end{eqnarray}}

\begin{document}
	
\title{The first-order phase transitions and gravitational waves from an effective theory inspired by four-stack intersecting D6-brane models}

\author[a,b,1]{Haotian Huangfu}
\author[c,2]{Tianjun Li}
\author[a,b,3]{Qi Sun}
\author[d,4]{Rui Sun}
\author[e,f,g,5]{Wenxing Zhang}

\affiliation[a]{CAS Key Laboratory of Theoretical Physics, Institute of Theoretical Physics,\\
		Chinese Academy of Sciences, Beijing 100190, P. R. China}
\affiliation[b]{School of Physical Sciences, University of Chinese Academy of Sciences,\\
		No.19A Yuquan Road, Beijing 100049, P. R. China}
\affiliation[c]{School of Physics, Henan Normal University, Xinxiang 453007, P. R. China} 
\affiliation[d]{School of Mathematical Sciences, University of Chinese Academy of Sciences,\\
		No.19A Yuquan Road, Beijing 100049, P. R. China}
\affiliation[e]{Department of Physics, Hebei University, Baoding, 071002, P. R. China}
\affiliation[f]{Key Laboratory of High-precision Computation and Application of Quantum Field Theory of Hebei Province, Baoding, 071002, P. R. China}
\affiliation[g]{Research Center for Computational Physics of Hebei Province,\\
Baoding, 071002, P. R. China}

\emailAdd{huangfuhaotian@itp.ac.cn}
\emailAdd{tli@itp.ac.cn}
\emailAdd{sunqi@itp.ac.cn}
\emailAdd{sunrui24@ucas.ac.cn}
\emailAdd{zhangwenxing@hbu.edu.cn}

\abstract{
First-order phase transitions in the early Universe can generate stochastic gravitational-wave backgrounds in frequency bands targeted by proposed deci-Hz observatories. We study the gravitational-wave consequences of the right-handed gauge symmetry breaking $SU(2)_{R_1}\times SU(2)_{R_2}\to SU(2)_R$ in a Pati-Salam effective theory inspired by four-stack intersecting D6-brane models. The ultraviolet construction fixes the gauge structure and particle multiplicities, while the scalar masses and quartic couplings are treated as renormalized inputs below the supersymmetry-breaking scale. Using a one-loop finite-temperature potential with ring resummation, we follow a strongly supercooled transition at the few-TeV scale through nucleation, percolation, completion, and reheating. The leading-order friction diagnostic does not support a runaway wall, so sound waves dominate the gravitational-wave signal, with turbulence as a sub-leading contribution; the finite acoustic lifetime suppresses the peak. For the fast-wall reference case, the spectrum peaks at $f_{\rm peak}=0.611$~Hz with $h^2\Omega_{\rm GW,peak}=1.232\times10^{-11}$, in the deci-Hz band targeted by DECIGO and BBO. We present its dependence on the wall speed, renormalization scale, and thermal screening, together with a noise-only comparison for several proposed detector configurations.
}

\maketitle

\section{Introduction}

The first-order phase transitions in the early Universe connect particle physics at high energies with cosmological observations.
When the free energy contains a barrier, according to the first-order phase transition mechanism, the transition proceeds via nucleating bubbles of the stable phase in the false vacuum.
The expanding bubbles drive bulk motion in the plasma and can produce a stochastic gravitational-wave background through sound waves in the plasma, magnetohydrodynamic turbulence, and bubble collisions~\cite{Caprini:2015zlo, Hindmarsh:2015qta, Hindmarsh:2017gnf, Espinosa:2010hh}.
This provides a perspective on particles and interactions that differs from collider physics: gravitational waves probe the early Universe and particle physics at energy scales inaccessible to colliders.
Stochastic backgrounds can arise from inflation, cosmic strings, domain walls, and first-order phase transitions; their theoretical foundations and observational prospects across frequency bands are surveyed in~\cite{Caprini:2015zlo, Croon:2024mde, Athron:2023xlk}.
Complementary nanohertz constraints have been reported by pulsar timing arrays~\cite{Bernardo:2024bdc}, while detectability with LISA, third-generation ground-based detectors, and deci-Hz space missions has been discussed in~\cite{Caprini:2024ofd, Caprini:2025trt}.
The complete path from a particle physics model to the gravitational-wave spectrum is presented systematically in Ref.~\cite{Athron:2023xlk}.
Phase transitions during inflation can leave distinct oscillatory signatures in the gravitational-wave spectrum~\cite{An:2020fff}, and the detectability of such signals at pulsar timing arrays has been explored in Ref.~\cite{An:2023jxf}.
Theoretical studies of these signals and their detection are discussed in Refs.~\cite{Caprini:2015zlo,Caprini:2019egz}.

The construction of realistic particle physics models from intersecting D6-branes in Type II string theory has been a central topic in string phenomenology, bridging model building and phenomenological analysis~\cite{Berkooz:1996km, Blumenhagen:2000wh, Klein:2000qw, Aldazabal:2000cn, Aldazabal:2000dg, Ibanez:2001nd, Cvetic:2001tj, Blumenhagen:2002gw}.
A major goal is to realize the Minimal Supersymmetric Standard Model (MSSM) and the Standard Model (SM) from string theory~\cite{Lykken:1998ec, Cvetic:1999hb, Aldazabal:2000sa}, with chiral fermions arising from intersections of D-brane stacks.
Among these constructions, the Pati-Salam gauge symmetry $SU(4)_C\times SU(2)_L\times SU(2)_R$ is particularly attractive, as it can be broken down to the SM gauge symmetry via D6-brane splitting and the Higgs mechanism~\cite{Cvetic:2004ui}.
The search for supersymmetric Pati-Salam models from intersecting D6-branes has been carried out systematically~\cite{Li:2019nvi, Li:2021pxo, He:2021gug, Li:2025xpn}.
In particular, the complete landscape was obtained~\cite{He:2021gug}, with string-scale gauge coupling unification achieved via two-loop RGE running~\cite{He:2021kbj, Li:2022cqk}, and more recently generalized three-generation conditions led to new classes of models~\cite{Li:2025xpn}.

While the standard three-stack Pati-Salam models have been extensively studied, configurations with additional brane stacks not parallel to the orientifold planes are also allowed.
A $d$-stack of D6-branes was first introduced in the hidden sector of three-family supersymmetric Pati-Salam models~\cite{Mansha:2023kwq}.
More recently, we constructed three-family supersymmetric Pati-Salam models from non-rigid D6-branes with extra $d$-stacks of D6-branes as visible sectors~\cite{Huangfu:2025uhu}.
In this construction, the extended Pati-Salam gauge symmetry becomes $SU(4)_C\times SU(2)_L\times SU(2)_{R_1}\times SU(2)_{R_2}$ (or $SU(4)_C\times SU(2)_{L_1}\times SU(2)_R\times SU(2)_{L_2}$ in the left-handed extension), where the $d$-stack carries $SU(2)_{R_2}$ (or $SU(2)_{L_2}$).
The $SU(2)_{R_1}\times SU(2)_{R_2}$ symmetry is broken down to the diagonal $SU(2)_R$ via a bifundamental Higgs field in the $(1,1,2,2)$ representation, and subsequently the standard Pati-Salam gauge symmetry is achieved.
The extra $d$-stack reduces the number of filler branes and can even lead to models without any $USp(N)$ gauge symmetry, while providing more origins for Higgs/Higgs-like particles.
The landscape of these extended models was completed with 14 independent classes in each of the right-handed and left-handed extensions, and string/GUT-scale gauge coupling unification was realized via two-loop RGEs with additional vector-like particles from the $\mathcal{N}=2$ subsector.
These models contain a rich scalar sector, including bifundamental Higgs fields and Higgs-like particles, which play a key role in the gauge symmetry breaking patterns and may have significant cosmological implications.

These states remain active near the transition and affect the thermal potential through their multiplicities and couplings, requiring a thermal Effective Field Theory (EFT) with a particle content different from that of the minimal Pati-Salam models.
In the right-handed extension, the four-stack construction offers a special right-handed gauge sector, $SU(2)_{R_1}\times SU(2)_{R_2}$, before diagonal breaking, and the bidoublet in the transition sector carries the quantum numbers needed for
\begin{equation}
 SU(2)_{R_1}\times SU(2)_{R_2}\longrightarrow SU(2)_R.
 \label{eq:introduction-diagonal-breaking}
\end{equation}
Such a non-minimal Higgs sector can give rise to a first-order phase transition in the early Universe, and the resulting gravitational-wave signal is potentially observable by the proposed deci-Hz observatories.
The diagonal breaking also produces three heavy gauge bosons whose mass $m_A=g_R\phi$ sets the transition scale and the reference peak frequency; a global rescaling of the dimensionful parameters connects this benchmark to lower-mass points, where dilepton resonance searches constrain the neutral state $A^0$.
Gravitational waves from Pati-Salam dynamics have been studied previously~\cite{Huang:2020bbe}; here we focus on the thermal consequences of the four-stack spectrum and gauge structure.

In this work, we study the first-order phase transition associated with the right-handed gauge symmetry breaking $SU(2)_{R_1}\times SU(2)_{R_2}\to SU(2)_R$ in a four-stack intersecting D6-brane inspired Pati-Salam effective theory.
The calculation is formulated as a low-energy EFT.
The ultraviolet four-stack construction fixes the gauge group, the scalar representations, and the multiplicities of the Higgs-like and transition-sector states, and supplies the gauge-coupling relations; below the supersymmetry breaking scale, taken as $M_{\rm SUSY}=300$ TeV, the scalar masses and quartic couplings are treated as renormalized inputs of the low-energy effective theory.
The transition studied here occurs entirely in the non-supersymmetric regime.
These scalar parameters are subject to vacuum stability, perturbative control, phase transition completion, and thermal consistency requirements, which allows us to distinguish which effects originate from the string particle spectrum and which from the phenomenological parameters.
We construct the one-loop finite-temperature effective potential with the ring contribution and thermal masses fixed by the retained spectrum, and follow the transition from the critical temperature through nucleation, percolation, completion, and reheating.
Then the sound-wave and turbulence contributions to the gravitational-wave spectrum are calculated, including the finite lifetime of the acoustic source.
The wall speed is varied as a macroscopic parameter since its determination requires a separate calculation of plasma friction.
We compare the predicted signal with the sensitivity of proposed deci-Hz observatories such as DECIGO and BBO~\cite{Yagi:2011wg, Crowder:2005nr}, using power-law-integrated sensitivity curves~\cite{Thrane:2013oya, Schmitz:2020syl}.
The detailed calculation and numerical results are presented in the following sections.

The benchmark point with spectrum-informed thermal masses studied in detail provides a completed transition at temperatures of order TeV.
Percolation and completion occur at $T_p=6.397$ TeV and $T_{\rm completion}=6.375$ TeV respectively, with $T_n=6.728$ TeV, $\alpha_n=1.9013$, and $\beta/H_n=410.4$.
For the fast-wall reference convention, the resulting gravitational wave spectrum has a peak at $f_{\rm peak}=0.611$ Hz with $h^2\Omega_{\rm GW,peak}=1.232\times10^{-11}$.
The peak remains in the deci-Hz range when the wall speed is varied over the conditional scenarios considered below.
This frequency range makes the signal particularly relevant for deci-Hz observatories such as DECIGO and BBO, while the detector comparison is kept distinct from a full mission-level discovery forecast.
This is a cosmological application of the four-stack intersecting D6-brane model, and these phenomena are expected to be observed in the future.

The paper is organized as follows.
In Section~\ref{sec:construction}, we summarize the four-stack origin of the gauge and matter content studied in~\cite{Huangfu:2025uhu} and define the non-supersymmetric EFT.
In Section~\ref{sec:heavy-gauge-rescaling}, we study the heavy gauge bosons produced by diagonal breaking, the global rescaling between transitions, and the resulting dilepton constraints on the low-mass neutral state.
In Section~\ref{sec:finite-temperature-potential}, we construct the finite-temperature effective potential and its screening masses.
In Section~\ref{sec:phase-dynamics}, we analyze the phase transition dynamics from the thermal history to the completion of the transition.
In Section~\ref{sec:gw-signal}, we present the gravitational wave spectrum, wall-speed dependence, and possible detector comparison.
We summarize the physical implications and the following steps of this study in Section~\ref{sec:conclusion}.

\section{Extended Pati-Salam models with extra $d$-stack of branes}
\label{sec:construction}

The effective theory studied here is motivated by the four-stack intersecting D6-brane construction of~\cite{Huangfu:2025uhu}.
The additional D6-brane stack enlarges the visible sector of the supersymmetric Pati-Salam theory.
It also changes the gauge-coupling relation and introduces additional sources for Higgs and Higgs-like multiplets.
We use the gauge group, representations, and multiplicities of that construction to define the EFT below the supersymmetry breaking scale.
The scalar masses and quartic couplings are renormalized EFT inputs.
Their allowed values are selected by vacuum stability, perturbativity, and the requirements of a consistent phase transition.

\subsection{Intersecting-D6-brane origin}
\label{subsec:brane-origin}

We begin with Type-IIA string theory compactified on the factorizable orientifold
\begin{equation}
  T^6/(\mathbb Z_2\times\mathbb Z_2)
  =T^2_1\times T^2_2\times T^2_3/(\mathbb Z_2\times\mathbb Z_2).
\end{equation}
Writing the complex coordinates of the three two-tori as $z_i$,
the two orbifold generators can be chosen as
\begin{equation}
 \theta:(z_1,z_2,z_3)\mapsto(-z_1,-z_2,z_3),\qquad
 \omega:(z_1,z_2,z_3)\mapsto(z_1,-z_2,-z_3),
\end{equation}
while the orientifold operation $\Omega R$ acts through complex conjugation,
$R:z_i\mapsto \bar z_i$.
The O6-planes with negative Ramond--Ramond charges (RR charges) appear as a result of four types of actions $\Omega R$, $\Omega R\theta$, $\Omega R\omega$, and $\Omega R\theta\omega$.
In order to cancel the RR charges, D6-brane stacks wrapping factorizable three-cycles are introduced, satisfying RR tadpole cancellation conditions and ${\cal N}=1$ supersymmetry conditions~\cite{Huangfu:2025uhu}.

For a stack $x$, a one-cycle on the $i$-th two-torus is specified by wrapping numbers $(n_x^i,l_x^i)$.
With $\beta_i=0$ ($1$) for a rectangular (tilted) two-torus, its homology class is
\begin{equation}
 [\Pi_x]=\prod_{i=1}^3\left(n_x^i[a_i]+2^{-\beta_i}l_x^i[b_i]\right),
 \qquad
 [\Pi_{x'}]=\prod_{i=1}^3\left(n_x^i[a_i]-2^{-\beta_i}l_x^i[b_i]\right),
 \label{eq:wrapped-cycles}
\end{equation}
where $x'$ denotes the orientifold image.
Chiral multiplicities are fixed by topological intersection numbers.
With $k=\sum_i\beta_i$, the intersection numbers are given by
\begin{equation}
 I_{xy}=2^{-k}\prod_{i=1}^3
 \left(n_x^i l_y^i-n_y^i l_x^i\right),\qquad
 I_{xy'}=-2^{-k}\prod_{i=1}^3
 \left(n_x^i l_y^i+n_y^i l_x^i\right).
 \label{eq:intersection-numbers}
\end{equation}
The generic massless particle spectrum for the intersecting D6-branes at generic angles is obtained in terms of intersection numbers~\eqref{eq:intersection-numbers}, as shown in Table~\ref{tab:representation}.

\begin{table}[h] 
	\renewcommand{\arraystretch}{1.25}
	\begin{center}
		\begin{tabular}{|c|c|}
			\hline {\bf Sector} & \phantom{more space inside this box}{\bf
				Representation}
			\phantom{more space inside this box} \\
			\hline\hline
			$aa$   & $U(N_a/2)$ vector multiplet  \\
			& 3 adjoint chiral multiplets  \\
			\hline
			$ab+ba$   & $I_{ab}$ $(\fund_a,\antifund_b)$ fermions   \\
			\hline
			$ab'+b'a$ & $I_{ab'}$ $(\fund_a,\fund_b)$ fermions \\
			\hline $aa'+a'a$ &$\frac 12 (I_{aa'} - \frac 12 I_{a,O6})\;\;
			\Ysymm\;\;$ fermions \\
			& $\frac 12 (I_{aa'} + \frac 12 I_{a,O6}) \;\;
			\Yasymm\;\;$ fermions \\
			\hline
		\end{tabular}
	\end{center}
	\caption{ 
	The general massless particle spectrum arising from intersecting D6-branes at generic angles. The second column specifies the representations of the gauge group $\mathrm{U}(N_a/2)$ under the $\mathbb{Z}_2\times \mathbb{Z}_2$ orbifold projections.  
}
	\label{tab:representation}
\end{table}

\subsection{Construction and spectrum of four-stack Pati-Salam models}
\label{subsec:four-stack-spectrum}

Before diagonal breaking, the visible brane stacks $a,b,c$ and $d$ carry
\begin{equation}
 U(4)_C\times U(2)_L\times U(2)_{R_1}\times U(2)_{R_2},
\end{equation}
whose non-Abelian part is
\begin{equation}
 G_{\rm vis}=SU(4)_C\times SU(2)_L\times SU(2)_{R_1}\times SU(2)_{R_2}.
 \label{eq:Gvis}
\end{equation}
The additional $d$ stack is the essential difference from the familiar three-stack Pati--Salam construction~\cite{He:2021gug}.
A bifundamental scalar in the $(\mathbf{1},\mathbf{1},\mathbf{2},\mathbf{2})$ representation can acquire a vacuum expectation value and realizes
\begin{equation}
 SU(2)_{R_1}\times SU(2)_{R_2}\longrightarrow SU(2)_R.
 \label{eq:diagonal-breaking}
\end{equation}
A first-order phase transition can indeed occur accordingly, as discussed in Sections~\ref{sec:finite-temperature-potential} and~\ref{sec:phase-dynamics}.
Subsequent brane splitting and Higgs breaking yield the ordinary
Pati--Salam and Standard-Model gauge groups.
In particular,
\begin{equation}
 SU(4)_C\times SU(2)_L\times SU(2)_R
 \longrightarrow SU(3)_C\times SU(2)_L\times U(1)_Y.
 \label{eq:hypercharge-embedding}
\end{equation}
The hypercharge is given by
\begin{equation}
 Q_Y=Q_{I_{3R}}+\frac{Q_B-Q_L}{2}.
\end{equation}

The extra $d$ stack also modifies the three-family fermion condition.
The three left-handed families arise from the $ab$ and $ab'$ sectors, whereas the right-handed families may be shared between the $ac$, $ac'$, $ad$, and $ad'$ sectors.
In the conventions of~\cite{Huangfu:2025uhu}, the condition is
\begin{equation}
 I_{ab}+I_{ab'}=3,\qquad
 I_{ac}+I_{ac'}+I_{ad}+I_{ad'}=-3.
 \label{eq:family-condition}
\end{equation}
For the representative spectrum used here, the three right-handed families of fermions are distributed between one $ac$ and two $ad$ intersections.
The same construction produces bifundamental Higgs from the $cd$ and $cd'$ sectors and Higgs/Higgs-like fields from the $bc$, $bc'$, $bd$, and $bd'$ sectors.
Consequently, the presence of the $d$ stack is directly relevant for the thermal scalar sector, rather than merely for high-scale gauge unification.

\begin{table}[htb]
\footnotesize
\renewcommand{\arraystretch}{1.0}
\caption{The chiral spectrum in the open string sector of Model 1~\cite{Huangfu:2025uhu}}
\label{tab:spectrum-model-1}
\begin{center}
\begin{tabular}{|c||c||c|c|c|c||c|c|c|}\hline
\scriptsize{Model 1} &\scriptsize{$SU(4)\times SU(2)_L\times SU(2)_{R_1}$}
& \scriptsize{$Q_4$} & \scriptsize{$Q_{2L}$} & \scriptsize{$Q_{2R_1}$} & \scriptsize{$Q_{2R_2}$} & $Q_{em}$ & $B-L$ & Field \\
 &\scriptsize{$\times SU(2)_{R_2}\times USp(2)^2$}
&   &   &   &   &   &   &   \\
\hline\hline
$ab$ & $3 \times (4,\overline{2},1,1,1,1)$ & $1$ & -$1$ & 0 & 0 &  $-\frac 13, \frac 23,-1, 0$ & $\frac 13,-1$ & $Q_L, L_L$\\
$ac$ & $1 \times (\overline{4},1,2,1,1,1)$ & -$1$ & 0 & $1$ & 0 &  $\frac 13, -\frac 23, 1, 0$ & $-\frac 13, 1$ & $Q_R, L_R$\\
$ad$ & $2 \times (\overline{4},1,1,2,1,1)$ & -$1$ & 0 & 0 & $1$  & $\frac 13, -\frac 23, 1, 0$ & $-\frac 13, 1$ & $Q_R, L_R$\\
$bc'$ & $2 \times(1,\overline{2},\overline{2},1,1,1)$ & 0 & -$1$ & -$1$ & 0  & $0,\;\mp 1$ & 0 & $H'$\\
$bd$ & $2 \times(1,2,1,\overline{2},1,1)$ & 0 & $1$ & 0 & -$1$  & $0,\;\pm 1$ & 0 & $H$\\
$bd'$ & $1 \times(1,2,1,2,1,1)$ & 0 & $1$ & 0 & $1$   & $0,\;\pm 1$ & 0 & $H'$\\
$cd$ & $2 \times(1,1,\overline{2},2,1,1)$ & 0 & 0 & -$1$ & $1$   & $0,\;\mp1$ & 0 & $\Phi$\\
$cd'$ & $1 \times(1,1,2,2,1,1)$ & 0 & 0 & $1$ & $1$   & $0,\;\pm1$ & 0 & $\Phi$\\
$b1$ & $3\times (1,\overline{2},1,1,2,1)$ & 0 & -1 & 0 & 0 & $\pm\frac 12$ & 0 & \\
$c4$ & $1\times (1,1,2,1,1,\overline{2})$ & 0 & 0 & 1 & 0 & $\pm\frac 12$ & 0 & \\
$d1$ & $3\times (1,1,1,2,\overline{2},1)$ & 0 & 0 & 0 & 1 & $\pm\frac 12$ & 0 & \\
$d4$ & $1\times (1,1,1,\overline{2},1,2)$ & 0 & 0 & 0 & -1 & $\mp\frac 12$ & 0 & \\
$b_{\Ysymm}$ & $2\times(1,3,1,1,1,1)$ & 0 & 2 & 0 & 0  & $0,\;\pm 1$ & 0 & \\
$b_{\overline{\Yasymm}}$ & $2\times(1,\overline{1},1,1,1,1)$ & 0 & -2 & 0 & 0   & 0 & 0 & \\
$d_{\overline{\Ysymm}}$ & $2\times(1,1,1,\overline{3},1,1)$ & 0 & 0 & 0 & -2   & $0,\;\mp 1$ & 0 & \\
$d_{\Yasymm}$ & $2\times(1,1,1,1,1,1)$ & 0 & 0 & 0 & 2   & 0 & 0 & \\
	\hline\hline
$bc$ & $4\times (1,2,\overline{2},1,1,1)$ & 0 & 1 & -1 & 0   &  $0,\;\pm 1$ &0 &${H}_u^i, {H}_d^i$\\
 & $4\times (1,\overline{2},2,1,1,1)$ & 0 & -1 & 1 & 0   &  & &  \\
\hline
\end{tabular}
\end{center}
\end{table}

Following the bookkeeping of the spectrum of Model 1 in~\cite{Huangfu:2025uhu} (see Table~\ref{tab:spectrum-model-1}), our transition sector contains three
$\Phi\sim(\mathbf{1},\mathbf{1},\mathbf{2},\mathbf{2})$ bidoublets, one from $cd'$ and two from $cd$.
One radial direction is selected as the order parameter and the remaining components are retained as spectator degrees of freedom.
The Higgs/Higgs-like sector is represented by thirteen bidoublets after the diagonal breaking in Eq.~\eqref{eq:diagonal-breaking}.  This number follows directly from the open-string entries of the representative spectrum:
\begin{equation}
 N_H=2_{bc'}+\left(1_{bd'}+2_{bd}\right)
 +2\times4_{bc,\,N=2}=2+3+8=13.
 \label{eq:higgs-like-counting}
\end{equation}
The last term consists of four $N=2$ $bc$ hypermultiplets, each providing a vector-like pair customarily denoted by $H_u$ and $H_d$.
After diagonal breaking Eq.~\eqref{eq:diagonal-breaking}, ten of the thirteen bidoublets originate from the $bc$ and $bc'$ sectors, whereas the remaining three originate from the $bd$ and $bd'$ sectors.
It is important to stress that this is a phenomenological low-energy organization of the string-motivated spectrum:
the mass ordering, possible pairing, and decoupling of composite states are not fixed by the topological intersection numbers alone.
We therefore state explicitly which degrees of freedom are thermally active in Table~\ref{tab:thermal-spectrum} and impose positivity and EFT-validity conditions on their masses.

\subsection{Gauge coupling relations with $d$-stack D6-branes}
\label{subsec:matching-eft}

The gauge couplings inherit a geometric boundary condition through the holomorphic gauge kinetic functions.
For $x=a,b,c$ and $d$,
\begin{equation}
 f_x=\frac{1}{4\kappa_x}\left[
 n_x^1n_x^2n_x^3 S-
 \sum_{i\neq j\neq k}2^{-\beta_j-\beta_k}
 n_x^il_x^jl_x^k U^i\right],
 \qquad g_x^{-2}=\left|\mathrm{Re}\,f_x\right|,
 \label{eq:gauge-kinetic-function}
\end{equation}
where $S$ and $U^i$ are the dilaton and complex-structure moduli and $\kappa_x=1$ for an $SU(N)$ factor.
Diagonal breaking implies
\begin{equation}
 g_R^{-2}=g_{R_1}^{-2}+g_{R_2}^{-2},\qquad
 f_R=f_c+f_d,
 \label{eq:diagonal-coupling}
\end{equation}
and the hypercharge kinetic function is
\begin{equation}
 f_Y=\frac{3}{5}\left(f_a+f_R\right).
 \label{eq:hypercharge-kinetic}
\end{equation}
The representative four-stack spectrum selected in this work obeys the Model 1 boundary relation of~\cite{Huangfu:2025uhu},
\begin{equation}
 g_a^2=g_b^2=2g_R^2=
 \frac{8}{5}\left(\frac{5}{3}g_Y^2\right).
 \label{eq:model-one-boundary-relation}
\end{equation}
This relation is a string-scale ($M_{\rm string}\simeq 5\times 10^{17}$ TeV) input.
The couplings entering the finite-temperature potential are obtained by renormalization-group evolution and are not identified with their boundary values.

\begin{figure}[h!]
    \centering
    \includegraphics[]{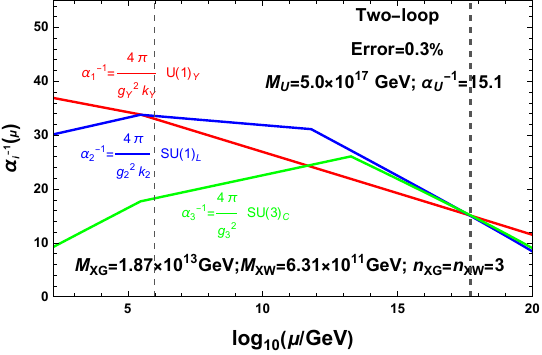}
    \caption{The evolution of two-loop gauge couplings in Model 1~\cite{Huangfu:2025uhu} with vector-like 3($X_G$+$X_W$).
    The masses are set as $M_{X_G}=1.87\times10^{13}$ GeV and $M_{X_W}=6.31\times10^{11}$ GeV.
    For this model, $k_2=1$ and $k_Y=2$.
    Gauge coupling unification is achieved at $5\times 10^{17}$ GeV.
    The dashed line on the left indicates the SUSY-breaking scale, while the dashed line on the right indicates the unification scale.}
    \label{fig:rge}
\end{figure}

In contrast to the illustrative low supersymmetry-breaking scale used in the RGE examples of~\cite{Huangfu:2025uhu}, we choose
\begin{equation}
 M_{\rm SUSY}=300~\mathrm{TeV}.
 \label{eq:msusy}
\end{equation}
This setting places the whole process of transition studied below inside a non-supersymmetric domain, with a central reference point $T_c\simeq 13.9~\mathrm{TeV}$ and $T_n\simeq 6.7~\mathrm{TeV}$.
We match the supersymmetric completion to the non-supersymmetric EFT at $M_{\rm SUSY}$ and use the non-supersymmetric running below this threshold.
Specifically, the two-loop RGEs evolve as
\begin{equation}
    \frac{{\rm d} g_i}{{\rm d} \ln{\mu}} =\frac{b_i}{(4\pi)^2}g_i^3+\sum_j\frac{g_i^3}{(4\pi)^4}\left[\sum_{j=1}^3 B_{ij}g_j^2-\sum_{\alpha=u,d,e}d_i^{\alpha}Tr(h^{\alpha\dagger}h^{\alpha})\right]~, ~
    \label{eq:RGE}
\end{equation}
in which $g_i(i=1,2,3)$ and $h^\alpha(\alpha=u,d,e)$ stand for the SM gauge couplings and the Yukawa couplings.
The coefficients $B_{ij}$ and $b_i$ for the beta functions can be found in~\cite{Machacek:1983tz,Machacek:1983fi,Machacek:1984zw,Cvetic:1998uw,Gogoladze:2010in} for the SM, and in~\cite{Barger:1992ac,Barger:1993gh,Martin:1993zk} for supersymmetric models.
The $Z$-boson mass, the top-quark pole mass, the Higgs vacuum expectation value (VEV), the strong coupling constant, the fine-structure constant, and the mixing-angle relation are also used as inputs in Eq.~\eqref{eq:RGE}; see~\cite{Huangfu:2025uhu,Li:2022cqk} for details.
The RGE analysis includes the vector-like threshold content required to realize the string-scale coupling relation.
More specifically, the RGE setup used in this work contains three $X_G$ multiplets at $M_{X_G}=1.87\times10^{13}~\mathrm{GeV}$
and three $X_W$ multiplets at $M_{X_W}=6.31\times10^{11}~\mathrm{GeV}$, as shown in Fig.~\ref{fig:rge}.
The quantum numbers of these additional particles are $X_G=(\mathbf{8},\mathbf{1},\mathbf{0})$ and $X_W=(\mathbf{1},\mathbf{3},\mathbf{0})$, and their contributions to the one-loop and two-loop beta functions are
\begin{equation}
    \Delta b^{(X_G)}= \left(0,0,3\right)~,~~
    \Delta b^{(X_W)}=\left(0,2,0\right)~,~
\end{equation}
\begin{equation}
    \Delta B^{(X_G)}=
    \begin{pmatrix}
        0& 0& 0 \\
             0& 0& 0\\
        0& 0& 54
    \end{pmatrix}~,~
    \Delta B^{(X_W)}=
    \begin{pmatrix}
        0& 0& 0\\
        0& 24& 0\\
        0& 0& 0
    \end{pmatrix}.
\end{equation}
These thresholds are needed only for the ultraviolet RGE evolution and the resulting string-scale coupling relation
\footnote{Note that $M_{X_G},~M_{X_W}\gg M_{\rm SUSY}\gg T_c$, therefore they are neither active fields in the non-supersymmetric EFT nor thermally populated degrees of freedom in the phase-transition calculation. And thus, they do not enter into the following section.}.

The running supplies the scale-dependent inputs
\begin{equation}
 g_{R_1}(Q),\quad g_{R_2}(Q),\quad y_t(Q),\quad y_b(Q),\quad y_\tau(Q)
 \label{eq:rge-inputs}
\end{equation}
to the effective potential.
Representative values are listed in Table~\ref{tab:rge-couplings}.
Here we use the central renormalization scale $Q=65~\mathrm{TeV}$ and assess the combined RGE and Coleman--Weinberg scale dependence over $30~\mathrm{TeV}\leq Q\leq150~\mathrm{TeV}$.

\begin{table}[h!]
  \centering
  \caption{Representative running couplings below the supersymmetry-breaking threshold $M_{\rm SUSY}=300~\mathrm{TeV}$.  The two right-handed factors are matched symmetrically, $g_{R_1}=g_{R_2}$, and the diagonal coupling is $g_R=(g_{R_1}^{-2}+g_{R_2}^{-2})^{-1/2}=g_{R_1}/\sqrt{2}$.
  The highlighted row is the central Coleman--Weinberg scale adopted for the reference transition;
  the $300~\mathrm{TeV}$ row denotes the non-supersymmetric limit immediately below the matching threshold.}
  \label{tab:rge-couplings}
  \begin{tabular}{c c c c c c}
    \hline
    $Q$ [TeV] & $g_{R_1}=g_{R_2}$ & $g_R$ & $y_t$ & $y_b$ & $y_\tau$ \\
    \hline
     10  & 0.7265 & 0.5137 & 0.7882 & 0.01916 & 0.01041 \\
     30  & 0.7126 & 0.5039 & 0.7591 & 0.01825 & 0.01044 \\
     50  & 0.7068 & 0.4998 & 0.7467 & 0.01786 & 0.01045 \\
    \textbf{65} & \textbf{0.7039} & \textbf{0.4977} & \textbf{0.7405} & \textbf{0.01766} & \textbf{0.01046} \\
    100  & 0.6993 & 0.4945 & 0.7307 & 0.01736 & 0.01047 \\
    150  & 0.6952 & 0.4916 & 0.7219 & 0.01709 & 0.01047 \\
    200  & 0.6924 & 0.4896 & 0.7157 & 0.01690 & 0.01048 \\
    300  & 0.6885 & 0.4869 & 0.7074 & 0.01664 & 0.01048 \\
    \hline
  \end{tabular}
\end{table}

\section{Heavy gauge bosons, global rescaling, and low-mass constraints}
\label{sec:heavy-gauge-rescaling}

In this section we establish the heavy-gauge-boson mass scale and the global
rescaling that connects the benchmark transition at $m_A\simeq 42.76$~TeV to
lower-mass points.  The rescaling is exact at the level of the one-loop
effective potential when the gauge couplings, Yukawa couplings, and the
spectrum are held fixed; the residual $g_R$ running is quantified separately.
The last subsection constrains the low-mass neutral state with dilepton
resonance searches.

\subsection{Heavy gauge bosons after diagonal breaking}
\label{subsec:heavy-gauge-mass}

The breaking
\begin{equation}
 SU(2)_{R_1}\times SU(2)_{R_2}\longrightarrow SU(2)_R
 \label{eq:breaking-again}
\end{equation}
has $6-3=3$ broken generators, so three gauge bosons acquire masses.
The corresponding three Goldstone modes are eaten by these gauge bosons, each
of which then has three polarizations, giving nine degrees of freedom in the
one-loop potential.  Along the radial direction $\phi$ of the active bidoublet
$X\sim(\mathbf{1},\mathbf{1},\mathbf{2},\mathbf{2})$, the field-dependent
mass is
\begin{equation}
 m_A^2(\phi)=\frac14\left[g_{R_1}^2(Q)+g_{R_2}^2(Q)\right]\phi^2 .
 \label{eq:mA-phi}
\end{equation}
At the central renormalization scale $Q=65$~TeV the two right-handed
couplings are matched symmetrically (see Table~\ref{tab:rge-couplings}),
\begin{equation}
 g_{R_1}(Q)=g_{R_2}(Q)=0.703883,
 \qquad
 g_R(Q)=\left(g_{R_1}^{-2}+g_{R_2}^{-2}\right)^{-1/2}=0.497719 .
 \label{eq:gR-values}
\end{equation}
The critical temperature and the order parameter at $T_c$ are
\begin{equation}
 T_c=13.8791~\text{TeV},
 \qquad
 \frac{\phi_c}{T_c}=6.1902,
 \qquad
 \phi_c=85.914~\text{TeV}.
 \label{eq:Tc-phic}
\end{equation}
Substituting into Eq.~\eqref{eq:mA-phi} gives
\begin{equation}
 m_A(\phi_c)
 =\frac12\sqrt{g_{R_1}^2+g_{R_2}^2}\,\phi_c
 =g_R\,\phi_c
 \simeq 42.76~\text{TeV}.
 \label{eq:mA-value}
\end{equation}
The three broken generators give three degenerate states,
\begin{equation}
 m_{A_1}=m_{A_2}=m_{A_3}\simeq 42.8~\text{TeV}
 \qquad\text{at }T_c.
 \label{eq:three-degenerate}
\end{equation}
Varying the renormalization scale over $30~\text{TeV}\le Q\le 150~\text{TeV}$
changes $m_A(\phi_c)$ by at most about $1.3\%$ relative to the central value,
with a total spread of about $2.6\%$:
$m_A=43.3$, $42.76$, and $42.2$~TeV at $Q=30$, $65$, and $150$~TeV,
respectively.  The mass at zero temperature is
$m_A(v_\phi)=g_R v_\phi$ with
$v_\phi=\mu_\phi/\sqrt{\lambda_\phi}$; for the benchmark parameters of
Table~\ref{tab:reference-eft-inputs} in Section~\ref{subsec:scalar-eft} this
gives $v_\phi\simeq 89.48$~TeV and $m_A(v_\phi)\simeq 44.6$~TeV, consistent
with Eq.~\eqref{eq:mA-value} up to the $O(\lambda_\phi)$ shift of the one-loop
vacuum expectation value.  Here and below, \(m_A\equiv m_A(\phi_c)\) denotes
the thermal mass at the critical temperature; the zero-temperature mass
\(M_{A_0}\equiv m_A(v_\phi)\) is used only in Section~\ref{sec:A0-dilepton},
where \(M_{A_0}/m_A(\phi_c)=v_\phi/\phi_c\simeq1.042\).

\subsection{Global rescaling}
\label{subsec:global-rescaling}

The one-loop effective potential used in the finite-temperature analysis of
Section~\ref{sec:finite-temperature-potential} is a function of
$(\phi,T;Q,\mu_\phi,\lambda_\phi,m_{0\Phi},m_{0H},a_\Phi,a_H,g_R,y_f)$,
where $a_\Phi$ and $a_H$ are the portal couplings of
Eq.~\eqref{eq:spectator-masses}.
Rescaling the dimensionful parameters by a common factor $s$,
\begin{equation}
 \phi'=s\phi,\qquad
 T'=sT,\qquad
 Q'=sQ,\qquad
 \mu_\phi'=s\mu_\phi,\qquad
 m_{0i}'=s\,m_{0i},
 \label{eq:rescaling-fields}
\end{equation}
while keeping $\lambda_\phi$, $a_\Phi$, $a_H$, $g_R$, and the Yukawa
couplings fixed, makes each term in $V_{\rm eff}$ scale as $s^4$:
\begin{equation}
 V_{\rm eff}(s\phi;sT;Q',\mu_\phi',m_{0i}')
 =s^4\,V_{\rm eff}(\phi;T;Q,\mu_\phi,m_{0i}).
 \label{eq:rescaling-potential}
\end{equation}
Consequently
\begin{equation}
 T_c'=s\,T_c,\qquad
 \phi_c'=s\,\phi_c,\qquad
 \frac{\phi_c'}{T_c'}=\frac{\phi_c}{T_c},
 \label{eq:rescaling-Tc}
\end{equation}
and the dimensionless transition parameters are invariant,
\begin{equation}
 \alpha'=\alpha,\qquad
 \frac{\beta}{H}'=\frac{\beta}{H},\qquad
 h^2\Omega_{\rm GW}'=h^2\Omega_{\rm GW},
 \label{eq:rescaling-dimensionless}
\end{equation}
while the peak frequency scales linearly,
\begin{equation}
 f_{\rm peak}'=s\,f_{\rm peak}.
 \label{eq:rescaling-fpeak}
\end{equation}
The heavy gauge boson mass follows the same rule,
\begin{equation}
 m_A'=s\,m_A,
 \qquad
 s=\frac{m_A'}{42.76~\text{TeV}},
 \label{eq:rescaling-mA}
\end{equation}
with $m_A\equiv m_A(\phi_c)$ defined in
Section~\ref{subsec:heavy-gauge-mass}.  The zero-temperature mass
\(M_{A_0}\equiv m_A(v_\phi)\) is used only in Section~\ref{sec:A0-dilepton},
where \(M_{A_0}/m_A(\phi_c)=v_\phi/\phi_c\simeq1.042\).

For the reference transition the benchmark parameters are
\begin{equation}
 \begin{aligned}
 \lambda_\phi&=0.0291649, \quad
 \mu_\phi^{\rm ref}=15.2809~\text{TeV}, \quad
 v_\phi^{\rm ref}=89.4787~\text{TeV},\\
 Q^{\rm ref}&=65~\text{TeV}, \quad
 m_{0\Phi}^{\rm ref}=8.29077~\text{TeV}, \quad
 m_{0H}^{\rm ref}=2.42098~\text{TeV}.
 \end{aligned}
 \label{eq:benchmark-params}
\end{equation}
For a target mass $m_A'$ the rescaled parameters are therefore
\begin{equation}
 \begin{aligned}
 \mu_\phi'&=15.2809\,s~\text{TeV},\\
 Q'&=65\,s~\text{TeV},\\
 m_{0\Phi}'&=8.29077\,s~\text{TeV},\\
 m_{0H}'&=2.42098\,s~\text{TeV},\\
 \lambda_\phi'&=0.0291649,
 \end{aligned}
 \qquad
 s=\frac{m_A'}{42.76~\text{TeV}}.
 \label{eq:rescaled-params}
\end{equation}
The gauge couplings, Yukawa couplings, multiplicities, and portal couplings
$a_\Phi=0.181016$, $a_H=0.196814$ are held fixed at their reference values in
the rescaling; the residual $g_R$ running is quantified in
Section~\ref{subsec:gR-running}.  As a representative example, $m_A'=10$~TeV
corresponds to $s=0.2339$ and
\begin{equation}
 \mu_\phi'=3.574~\text{TeV},\quad
 Q'=15.20~\text{TeV},\quad
 m_{0\Phi}'=1.939~\text{TeV},\quad
 m_{0H}'=0.5663~\text{TeV}.
 \label{eq:example-10TeV}
\end{equation}

\subsection{Numerical verification of the rescaling}
\label{subsec:rescaling-verification}

To verify Eq.~\eqref{eq:rescaling-potential} numerically, we solve the
finite-temperature effective potential for
$m_A'=10,9,\dots,1$~TeV using the rescaled parameters
Eq.~\eqref{eq:rescaled-params}, with $g_{R_1}=g_{R_2}$ held at $0.703883$.
The results are collected in Table~\ref{tab:rescaling-check}.  In every case
the broken minimum remains the global minimum at $T=0$ (column ``zero-$T$''),
and the ratio $T_c/m_A$ is constant to four significant digits,
\begin{equation}
 \frac{T_c}{m_A}=0.3458
 \qquad\text{for }1~\text{TeV}\le m_A\le 10~\text{TeV}.
 \label{eq:Tc-over-mA-constant}
\end{equation}
This is the direct numerical confirmation of
Eq.~\eqref{eq:rescaling-potential}.

\begin{table}[h]
 \centering
 \caption{
 Numerical check of the global rescaling with the updated benchmark
 parameters of Eq.~\eqref{eq:benchmark-params}.
 The parameters are obtained from Eq.~\eqref{eq:rescaled-params} with
 $g_{R_1}=g_{R_2}$ fixed at $0.703883$.
 The column ``zero-$T$'' indicates whether the broken minimum
 is the global minimum of the zero-temperature potential.
 The ratio $T_c/m_A$ is computed from the unrounded values of $T_c$ and is
 constant, as required by Eq.~\eqref{eq:rescaling-potential}.
 The column $f_{\rm peak}^{\rm single-field}$ is the single-field truncation
 value $0.611\,T_c/13.879$~Hz; it is shown only to establish
 Eq.~\eqref{eq:Tc-over-mA-constant} and is not the physical prediction.
 The physical predictions use the full multi-field result
 $T_c=13.879\,s$~TeV and $f_{\rm peak}=0.611\,s$~Hz.
 The absolute value $0.3458$ differs from the full multi-field result
 $13.879/42.76=0.3246$ by $6.5\%$; this residual offset originates from the
 truncation of the single-field potential and does not affect the rescaling.
 }
 \label{tab:rescaling-check}
 \begin{tabular}{cccccc}
  \toprule
  $m_A'$ [TeV] & $Q'$ [TeV] & $T_c$ [TeV] & $T_c/m_A$ & zero-$T$ & $f_{\rm peak}^{\rm single-field}$ [Hz] \\
  \midrule
  10 & 15.20 & 3.458 & 0.3458 & \checkmark & 0.1522 \\
   9 & 13.68 & 3.112 & 0.3458 & \checkmark & 0.1370 \\
   8 & 12.16 & 2.767 & 0.3458 & \checkmark & 0.1218 \\
   7 & 10.64 & 2.421 & 0.3458 & \checkmark & 0.1066 \\
   6 &  9.12 & 2.075 & 0.3458 & \checkmark & 0.0913 \\
   5 &  7.60 & 1.729 & 0.3458 & \checkmark & 0.0761 \\
   4 &  6.08 & 1.383 & 0.3458 & \checkmark & 0.0609 \\
   3 &  4.56 & 1.037 & 0.3458 & \checkmark & 0.0457 \\
   2 &  3.04 & 0.692 & 0.3458 & \checkmark & 0.0304 \\
   1 &  1.52 & 0.346 & 0.3458 & \checkmark & 0.0152 \\
  \bottomrule
 \end{tabular}
\end{table}

The column $f_{\rm peak}^{\rm single-field}$ is obtained from
$f_{\rm peak}=0.611\,(T_c/13.879)$~Hz, which reduces to
$f_{\rm peak}=0.611\,s$~Hz under Eq.~\eqref{eq:rescaling-fpeak}.  We emphasize
that the physical predictions for $T_c$ and $f_{\rm peak}$ use the full
multi-field result $T_c=13.879\,s$~TeV and $f_{\rm peak}=0.611\,s$~Hz; the
numerical check in Table~\ref{tab:rescaling-check} is used only to establish
Eq.~\eqref{eq:Tc-over-mA-constant}.

\subsection{Gauge-coupling running and the low-mass limit}
\label{subsec:gR-running}

The rescaling Eq.~\eqref{eq:rescaling-fields} keeps $g_{R_1}=g_{R_2}$ fixed,
and hence $g_R$ at its reference value.  In practice $g_R$ runs.  Using the
two-loop running of Section~\ref{sec:construction}, the deviation
\begin{equation}
 \delta g_R(Q')\equiv\frac{g_R(Q')}{g_R(65~\text{TeV})}-1
 \label{eq:gR-deviation}
\end{equation}
is $2.7\%$ at $Q'=15.2$~TeV ($m_A'=10$~TeV), $3.0\%$ at $Q'=12.16$~TeV
($m_A'=8$~TeV), $3.7\%$ at $Q'=7.6$~TeV ($m_A'=5$~TeV), $4.6\%$ at
$Q'=4.56$~TeV ($m_A'=3$~TeV), and $6.6\%$ at $Q'=1.52$~TeV
($m_A'=1$~TeV).  We classify the rescaled points by the size of
$\delta g_R$:
\begin{itemize}
 \item $\delta g_R<3\%$: \emph{fully safe}.  The rescaled transition is
       physically equivalent to the benchmark up to a $3\%$ uncertainty in
       $m_A$.  This covers $m_A'>8$~TeV.
 \item $3\%\le\delta g_R\le 5\%$: \emph{acceptable}.  The $g_R$ running is
       comparable to the one-loop truncation error of the potential.
       This covers $3~\text{TeV}\le m_A'\le 8$~TeV.
 \item $5\%<\delta g_R\le 7\%$: \emph{marginal}.  The rescaled point should
       be used only after either correcting $s$ by $g_R(65)/g_R(Q')$ or
       rerunning the RGE matching below $M_{\rm SUSY}$.  This covers
       $1~\text{TeV}\le m_A'<3$~TeV.
 \item $\delta g_R>7\%$: \emph{not recommended}.  The direct rescaling is no
       longer reliable and the transition should be recomputed from the
       RGE-matched parameters; this category is retained for completeness,
       and no scanned point falls in it.
\end{itemize}
If an exact alignment of $m_A'$ is required at the lower end, the rescaling
factor should be corrected to
\begin{equation}
 s_{\rm exact}=s\,\frac{g_R(65~\text{TeV})}{g_R(Q')},
 \label{eq:s-exact}
\end{equation}
where $g_R(Q')$ is evaluated iteratively, or to first order with $Q'=65\,s$
from the uncorrected $s$.  This keeps the physical $m_A'$ equal to its
nominal value at the cost of a small deviation of $T_c/m_A$ from
Eq.~\eqref{eq:Tc-over-mA-constant}.

In summary, the global rescaling Eq.~\eqref{eq:rescaling-fields} provides a
controlled way to connect the benchmark transition at $m_A\simeq 42.76$~TeV
to lower-mass points.  The rescaling is exact at fixed $g_{R_1}=g_{R_2}$ and
is verified numerically in Table~\ref{tab:rescaling-check}, with
$T_c/m_A=0.3458$ constant across the entire mass range.  The absolute value
differs from the full multi-field calculation by $6.5\%$; this residual
offset originates from the truncation of the single-field potential---the
reduced thermal spectrum and the omission of multi-field mixing terms---and
is not removed by varying single-field parameters.  For the physical
gravitational-wave predictions we therefore use the full result
$T_c=13.879\,s$~TeV and $f_{\rm peak}=0.611\,s$~Hz.  The gauge-coupling
running limits the direct applicability of the rescaling to
$m_A'\gtrsim 3$~TeV for quantitative predictions, with $m_A'>8$~TeV being
fully safe; below $3$~TeV the rescaling should be combined with
Eq.~\eqref{eq:s-exact} or replaced by a dedicated RGE-matched calculation.

\subsection{Dilepton constraints on the neutral heavy gauge boson}
\label{sec:A0-dilepton}

This subsection constrains the low-mass rescaled points of Section~\ref{subsec:global-rescaling}; it does not affect the benchmark at \(m_A \simeq 42.76\)~TeV. 
The reduction of the symmetry-breaking scale discussed above brings the heavy
vectors into a mass range accessible to direct searches. We therefore examine
the constraints on the neutral state $A^0\equiv A^3$ from dilepton resonance
searches. We express the collider constraints in terms of the zero-temperature
mass $M_{A_0}\equiv m_A(v_\phi)$. For the reference benchmark and its fixed-coupling
rescalings, this is related to the thermal mass parameter by
$M_{A_0}/m_A(\phi_c)=v_\phi/\phi_c\simeq1.042$. The following interpretation assumes alignment between the
brane-sector and fermion mass bases, negligible additional neutral-vector
mixing, and no non-fermionic decay channels. These assumptions specify a
low-energy collider benchmark rather than a complete matching of the subsequent
Pati--Salam breaking sector.

The relevant couplings follow from the same diagonal breaking that generates
the vector mass. Writing $g_i\equiv g_{R_i}$ and
$G=\sqrt{g_1^2+g_2^2}$, the heavy combination and its coupling to the two
right-handed currents are
\begin{equation}
 A_\mu^a=\frac{g_1W_{1\mu}^a-g_2W_{2\mu}^a}{G},
 \qquad
 \mathcal L_{\rm int}\supset
 A_\mu^a\left(\frac{g_1^2}{G}J_1^{a\mu}
             -\frac{g_2^2}{G}J_2^{a\mu}\right).
 \label{eq:A0-currents}
\end{equation}
At the symmetric reference point this gives
\begin{equation}
 \mathcal L_{A^0}=A^0_\mu\sum_f\bar f\gamma^\mu
       (g_{Af}^{L}P_L+g_{Af}^{R}P_R)f,
 \qquad
 g_{Af}^{L}=0,\qquad g_{Af}^{R}=\eta_f g_R T^3_{R,f},
 \label{eq:A0-chiral}
\end{equation}
where $P_{L,R}=(1\mp\gamma_5)/2$ and $\eta_f=+1$ ($-1$) for a right-handed
multiplet originating from the first (second) sector. With
$g_1=g_2=0.703883$ and $g_R=g_1/\sqrt2=0.497720$, the magnitude of every
charged-fermion right-handed coupling is $|g_{Af}^{R}|=0.248860$.
We keep these reference couplings fixed throughout the mass scan, consistently
with the fixed-coupling rescaling. The relative sector signs do not suppress
the pure resonant rate: distinct quark initial states contribute incoherently.

The total width is calculated from these chiral couplings. Neglecting the
final-state fermion mass, the partial width is
\begin{equation}
 \Gamma(A^0\to f\bar f)=\frac{N_c^f M_{A_0}}{24\pi}
 \left(|g_{Af}^{L}|^2+|g_{Af}^{R}|^2\right).
 \label{eq:A0-partial-width}
\end{equation}
The numerical calculation retains $m_t=172.5\,\mathrm{GeV}$ and neglects the
other SM fermion masses. We consider two limiting spectra: $m_{N_i}>M_{A_0}/2$,
for which $A^0\to N_iN_i$ is kinematically forbidden, and $m_{N_i}\ll M_{A_0}/2$,
for which the three right-handed-neutrino decay modes are accessible
($i=1,2,3$). In the large-$M_{A_0}$ limit, respectively,
\begin{equation}
 \left(\frac{\Gamma_A}{M_{A_0}},\,
       \mathrm{BR}(A^0\to\ell^+\ell^-)\right)
 \simeq
 \begin{cases}
 (0.01725,\,1/21), & m_{N_i}>M_{A_0}/2,\\
 (0.01971,\,1/24), & m_{N_i}\ll M_{A_0}/2,
 \end{cases}
 \label{eq:A0-width-benchmarks}
\end{equation}
for either $\ell=e$ or $\mu$. Thus, opening the neutrino channels moderately
reduces the dilepton branching fraction while leaving a narrow resonance.

We use the model-independent Run~3 limits of
ref.~\cite{ATLAS:2026Dilepton}, obtained at $\sqrt{s}=13.6\,\mathrm{TeV}$
with $165\,\mathrm{fb}^{-1}$, rather than importing its model-dependent
Run~2+3 mass bounds. The particle-level fiducial region requires both leptons
to have $p_T>30\,\mathrm{GeV}$ and $|\eta|<2.5$, with
$m_{\ell\ell}>M_{A_0}-2\Gamma_A$. The combined $ee$ and $\mu\mu$ limit is expressed
per single-flavour branching fraction; no additional factor of two is applied
to the prediction. The published observed and expected curves are extracted
from the vector version of figure~3(c) of that reference. At each mass we
interpolate $\log(\sigma_{\rm fid}\mathrm{BR})_{95}$ linearly in $\Gamma_A/M_{A_0}$
between the $1.2\%$ and $3\%$ width templates.

The Born rates use the central
\texttt{NNPDF23\_lo\_as\_0130\_qed} parton distribution set~\cite{NNPDF:2013QED}
with $\mu_F=M_{A_0}$ and separate $u,d,s,c,b$ luminosities.
We normalize the signal prediction according to
\begin{equation}
 \sigma_{A^0}(M_{A_0})\simeq
 \sigma_{\rm SSM}^{\rm ATLAS}(M_{A_0})
 \frac{\sigma_{A^0}^{\rm Born}(M_{A_0})}
      {\sigma_{\rm SSM}^{\rm Born}(M_{A_0})},
 \qquad
 \sigma_V\equiv\sigma_{\rm fid}(pp\to V)\,
                   \mathrm{BR}(V\to\ell^+\ell^-).
 \label{eq:A0-reweighting}
\end{equation}
Here $V$ denotes either $A^0$ or the Sequential Standard Model (SSM)
reference resonance, a hypothetical
heavy vector with the same fermionic couplings as the SM $Z$.
We define $\sigma_V^{\rm Born}$ as the tree-level hadronic cross section for
$pp\to V\to\ell^+\ell^-$ within the fiducial region specified above.
We first evaluate the partonic amplitude for
$q\bar q\to V\to\ell^+\ell^-$ using the chiral couplings of each model,
retaining the finite resonance width through the Breit--Wigner denominator
$\hat s-M_{A_0}^2+i M_{A_0}\Gamma_V$, where $\hat s$ denotes the square of the
partonic centre-of-mass energy. The squared amplitude is then convoluted with the quark
and antiquark parton distributions and integrated over the parton momentum
fractions and two-body phase space, subject to the lepton transverse-momentum,
pseudorapidity and invariant-mass cuts. This calculation includes the decay
into a single lepton flavour and the fiducial acceptance. We therefore
apply no additional branching-fraction factor. No higher-order QCD corrections or detector reconstruction efficiencies
are included. In eq.~\eqref{eq:A0-reweighting}, the two Born cross sections are computed
at the same resonance mass $M_{A_0}$ and collider energy $\sqrt{s}$. For each
resonance $V$, its total width $\Gamma_V$ enters both the propagator and
the selection $m_{\ell\ell}>M_{A_0}-2\Gamma_V$.

The factor $\sigma_{\rm SSM}^{\rm ATLAS}(M_{A_0})$ denotes the ATLAS theoretical
prediction, normalized at NNLO, for the fiducial signal cross section of
$pp\to Z'_{\rm SSM}\to\ell^+\ell^-$ at $\sqrt{s}=13.6\,\mathrm{TeV}$,
where $Z'_{\rm SSM}$ is the SSM reference resonance with mass $M_{A_0}$ and
$\ell=e$ or $\mu$. The reference rate includes the branching fraction to a single lepton
flavour and the fiducial acceptance. At each mass $M_{A_0}$, we extract an
effective correction factor by dividing the ATLAS SSM prediction
$\sigma_{\rm SSM}^{\rm ATLAS}(M_{A_0})$ by the SSM Born rate
$\sigma_{\rm SSM}^{\rm Born}(M_{A_0})$ evaluated at the same mass and with the
same fiducial definition. We assume that the relative higher-order
corrections to the $A^0$ and SSM fiducial rates are approximately equal
at fixed $M_{A_0}$. Multiplying $\sigma_{A^0}^{\rm Born}(M_{A_0})$ by this factor
then gives the estimate in eq.~\eqref{eq:A0-reweighting}. To extend the prediction below the visible range of the published SSM
curve, we use the correction factor at $M_{A_0}=1.56\,\mathrm{TeV}$ for all lower
masses. This extrapolation does not affect the mass limits near
$5.2\,\mathrm{TeV}$. To compare with the ATLAS limits for non-interfering
resonance signals~\cite{ATLAS:2026Dilepton}, we include the squared resonance
amplitude without its interference with the SM Drell--Yan amplitude.

\begin{figure}[h]
 \centering
 \includegraphics[width=0.98\textwidth]{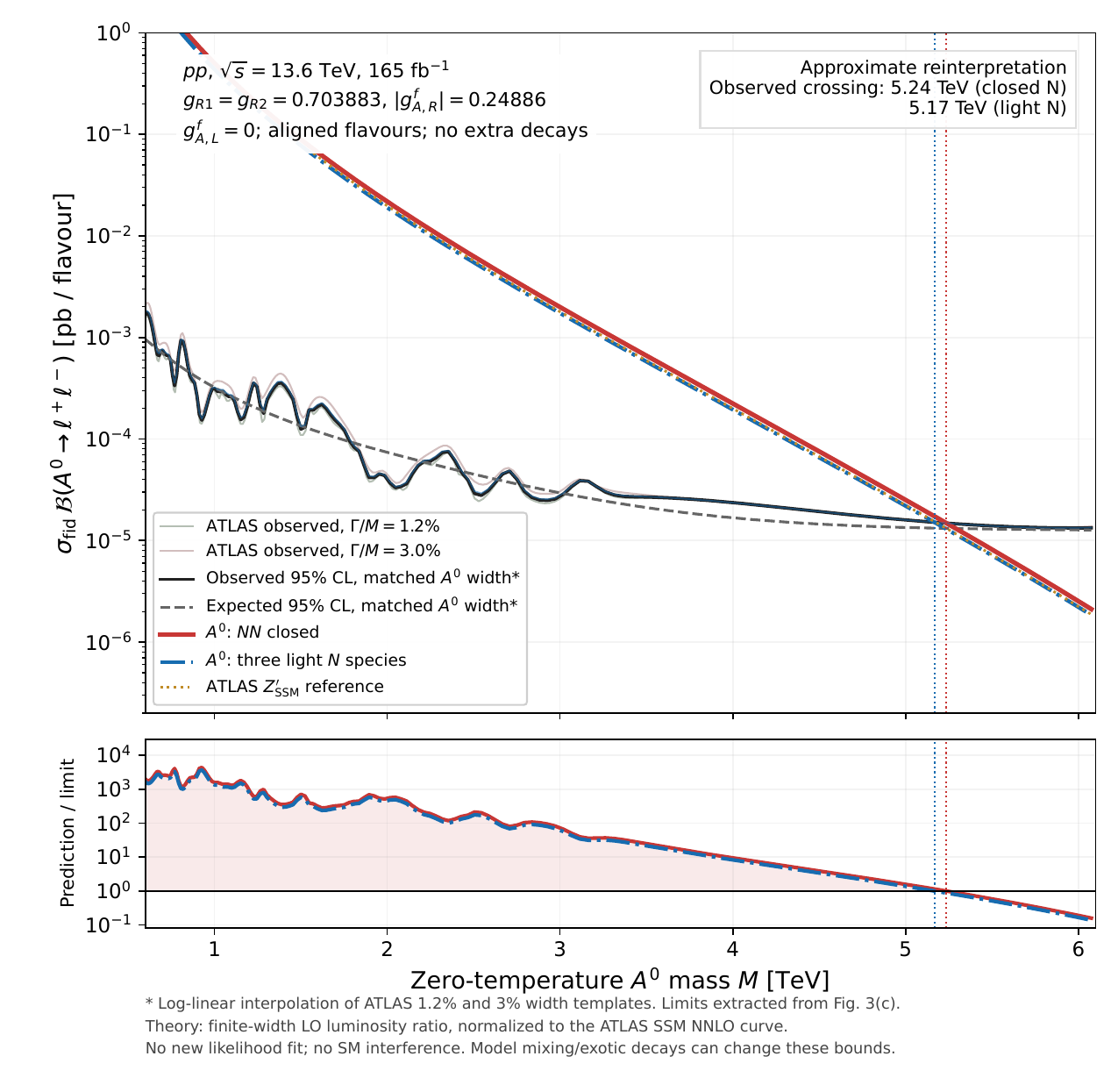}
 \caption{ATLAS Run~3 constraints~\cite{ATLAS:2026Dilepton} on
$pp\to A^0\to\ell^+\ell^-$, with $\ell=e$ or $\mu$, at
$\sqrt{s}=13.6\,\mathrm{TeV}$ and $165\,\mathrm{fb}^{-1}$.
Red and blue curves show
$\sigma_{\rm fid}(pp\to A^0)\,\mathrm{BR}(A^0\to\ell^+\ell^-)$
for $m_{N_i}>M_{A_0}/2$ and $m_{N_i}\ll M_{A_0}/2$ ($i=1,2,3$), respectively.
Black solid and grey dashed curves give the observed and expected 95\% CL
upper limits for the former spectrum; the thin blue curve gives the observed
upper limit for the latter. Vertical lines indicate the observed mass limits.
The lower panel shows the predicted rate divided by its observed upper limit;
values above unity are excluded.}
 \label{fig:A0-dilepton}
\end{figure}

Figure~\ref{fig:A0-dilepton} compares the predicted signal rates with the
ATLAS 95\% CL upper limits interpolated to the relative width
$\Gamma_A/M_{A_0}$ predicted for each neutrino mass spectrum. The observed
95\% CL lower limits on $M_{A_0}$ are
\begin{equation}
 M_{A_0,\rm lim}^{\rm obs}\simeq
 \begin{cases}
 5.24\,\mathrm{TeV}, & m_{N_i}>M_{A_0}/2,\\
 5.17\,\mathrm{TeV}, & m_{N_i}\ll M_{A_0}/2,
 \end{cases}
 \label{eq:A0-mass-limits}
\end{equation}
For the couplings and decay assumptions specified above, the observed
95\% CL lower limit on the $A^0$ mass is approximately $5.2\,\mathrm{TeV}$
for either neutrino mass spectrum.


\section{Finite-temperature effective potential}
\label{sec:finite-temperature-potential}

In this section we construct the finite-temperature effective potential of the low-energy EFT.
We include the tree-level potential, the Coleman-Weinberg correction, finite counterterms, thermal corrections, and the ring contribution, following the approach of~\cite{Huang:2020bbe}.
The particle representations and multiplicities follow the four-stack D6-brane construction of~\cite{Huangfu:2025uhu}; the scalar parameters and renormalization conditions are specified below.
Unlike the previous studies, here we study the EFT that contains the particle content and the renormalization prescription arising from four stacks of intersecting D6-branes~\cite{Huangfu:2025uhu}.

We work in Landau gauge, which simplifies the gauge and Goldstone contributions.
The transition parameters obtained from this one-loop potential retain a dependence on the gauge choice.
Our numerical tests examine the renormalization scale and thermal screening coefficients; the size of the gauge dependence remains to be calculated.

\subsection{Non-supersymmetric EFT}
\label{subsec:scalar-eft}

The low-energy scalar sector consists of the active bidoublet $X\sim(\mathbf{1},\mathbf{1},\mathbf{2},\mathbf{2})$, the remaining $\Phi$-like bidoublets $X_a$, and the Higgs-like bidoublets $H_A$.
The gauge-invariant building blocks for the active field are
\begin{equation}
 I_X=\mathrm{Tr}(X^\dagger X),\qquad
 J_X=\mathrm{Tr}(X^\dagger X X^\dagger X).
\end{equation}
At the renormalizable level, the scalar EFT can be written schematically as
\begin{align}
 V_0={}&-m_X^2I_X+\lambda_1 I_X^2+\lambda_2 J_X \notag\\
 &+\sum_a\left[m_a^2I_{X_a}+\lambda_{Xa}I_XI_{X_a}
 +\widetilde\lambda_{Xa}\,
 \mathrm{Tr}(X^\dagger X_a)\mathrm{Tr}(X_a^\dagger X)\right] \notag\\
 &+\sum_{A\in H_{R_1},H_{R_2}}
\left[
M_A^2 I_{H_A}+\lambda_{HA} I_X I_{H_A}
\right]
+V_{\rm allowed}^{(X,H)}\\
 &+V_{\rm self}+V_{\rm mix}.
 \label{eq:gauge-invariant-scalar-eft}
\end{align}
The term $V_{\rm allowed}^{(X,H)}$ denotes additional quartic contractions whose existence depends on the gauge representation of each Higgs-like multiplet.
We retain only the universal portal $I_X I_{H_A}$ in the flavor-degenerate reduced EFT.
Here $V_{\rm self}$ and $V_{\rm mix}$ denote spectator self-interactions and all additional gauge-allowed interactions, respectively.
The brane construction fixes the available representations and multiplicities, but does not uniquely fix the renormalized coefficients in Eq.~\eqref{eq:gauge-invariant-scalar-eft}.
We therefore assume that the phase transition follows one common direction in scalar field space, while the remaining scalars in each multiplet class are assigned the same mass and coupling parameters.

With the kinetic term $\mathrm{Tr}[(D_\mu X)^\dagger D^\mu X]$, we choose $X=(\phi/2)\mathbf{1}_2$ and restrict the bounce to this radial direction.
This gives the canonical kinetic term $(\partial_\mu\phi)^2/2$, with $I_X=\phi^2/2$ and $J_X=\phi^4/8$.
The radial parameters are therefore $\mu_\phi^2=m_X^2$ and $\lambda_\phi=\lambda_1+\lambda_2/2$.
Along this trajectory the tree potential reduces to
\begin{equation}
 V_{\rm tree}(\phi)=-\frac12\mu_\phi^2\phi^2
 +\frac14\lambda_\phi\phi^4,
 \qquad
 v_\phi^{\rm tree}=\frac{\mu_\phi}{\sqrt{\lambda_\phi}},
 \label{eq:radial-tree-potential}
\end{equation}
while the two classes of spectators are parameterized by
\begin{equation}
 m_{\Phi,\rm spec}^2(\phi)=m_{0\Phi}^2+a_\Phi\phi^2,
 \qquad
 m_{H,\rm spec}^2(\phi)=m_{0H}^2+a_H\phi^2.
 \label{eq:spectator-masses}
\end{equation}
Thus $(m_{0\Phi},m_{0H},a_\Phi,a_H)$ are the mass and portal-coupling~\cite{Patt:2006fw,Lebedev:2021xey} parameters scanned in the numerical analysis.

The reference parameters are listed in Table~\ref{tab:reference-eft-inputs}.
We take $v_\phi$ and $\lambda_\phi$ as independent inputs and set $\mu_\phi=v_\phi\sqrt{\lambda_\phi}$.
The numerical potential retains the radial quartic and the two universal portals in Eq.~\eqref{eq:spectator-masses}, with diagonal spectator mass matrices.
Additional spectator self-quartics, cross-quartics, and interactions that mix or split these modes are set to zero in this reduced calculation.
Only $t,b,\tau$ are retained in the field-dependent fermion masses.
The gauge and Yukawa couplings are held fixed at the chosen $Q$ during the temperature evolution; the variation of $g_R$ and $y_t$ over $T_n\le T\le T_c$ is below the percent level and does not affect the transition parameters at the quoted precision.

\begin{table}[htbp]
 \centering
 \caption{Inputs for the reference transition at $Q=65~\mathrm{TeV}$.
 Mass parameters are given in TeV; all couplings are dimensionless.
 The value of $\mu_\phi$ is derived from $v_\phi$ and $\lambda_\phi$.
 The spectator screening estimate uses $g_L=0.6190$.}
 \label{tab:reference-eft-inputs}
 \small
 \begin{tabular}{lrlr}
  \toprule
  Parameter & Value & Parameter & Value \\
  \midrule
  $v_\phi$ & $89.4787088728$ & $\lambda_\phi$ & $0.0291648691042$ \\
  $m_{0\Phi}$ & $8.29077423797$ & $a_\Phi$ & $0.181015861427$ \\
  $m_{0H}$ & $2.42098281612$ & $a_H$ & $0.196813958612$ \\
  $\mu_\phi$ & $15.2809278313$ & $Q$ & $65$ \\
  $g_{R_1}=g_{R_2}$ & $0.703882988424$ & $y_t$ & $0.740533791090$ \\
  $y_b$ & $0.0176645857198$ & $y_\tau$ & $0.0104600074446$ \\
  $g_L$ & $0.6190$ & $M_{\rm SUSY}$ & $300$ \\
  \bottomrule
 \end{tabular}
\end{table}

Here in the extended Pati-Salam model with extra $d$-stack of branes, the three broken generators of the breaking in Eq.~\eqref{eq:diagonal-breaking} yield three Goldstone modes. The field-dependent heavy-vector mass is given in Eq.~\eqref{eq:mA-phi},
\begin{equation}
m_A^2(\phi)=\frac14\left[g_{R_1}^2(Q)+g_{R_2}^2(Q)\right]\phi^2,
 \label{eq:heavy-vector-mass}
\end{equation}

We reduce the scalar sector to one effective transition field by restricting the tunneling path to the radial direction associated with $SU(2)_{R_1}\times SU(2)_{R_2}\to SU(2)_R$, while the other scalar degrees of freedom are treated as spectators.
Their field-dependent masses modify the effective potential, while the background values are fixed to zero along the tunneling path.
By this means, only a few parameters are considered, and it is clear to find out how the particles in the spectrum affect the effective potential.

With the diagonal spectator, the one-field reduction is a directly testable local-stability condition.
That is, the orthogonal curvatures are
\begin{equation}
 m_{\Phi,\rm spec}^2(\phi)+\Pi_{\Phi}(T)>0,
 \qquad
 m_{H,\rm spec}^2(\phi)+\Pi_H(T)>0
 \label{eq:transverse-stability-condition}
\end{equation}
throughout the field range followed by the bounce solution.
This condition can be verified numerically for the reference point; the smallest effective spectator masses along the tunneling path are listed in Table~\ref{tab:transverse-stability}.
The spectator fields are preserved in the effective potential, since their field-dependent masses, their loop contributions, and thermal contributions are unchanged.
Therefore, here the bounce is solved only for the radial transition field.

The one-loop zero-temperature corrections, thermal functions, counterterms, and ring contribution that complete the effective potential are introduced in the next section.

\subsubsection{Fields in the one-loop potential}
\label{subsec:field-dependent-spectrum}

The background field is the canonically normalized radial component $\phi$ of the active bidoublet $X$.
The false vacuum is at $\phi=0$, whereas the broken phase has $\phi>0$.
The field-dependent squared masses retained in the effective potential are
\begin{align}
 m_h^2(\phi) &=-\mu_\phi^2+3\lambda_\phi\phi^2,
 &m_G^2(\phi)&=-\mu_\phi^2+\lambda_\phi\phi^2, \notag\\
 m_A^2(\phi) &=\frac14\left[g_{R_1}^2(Q)+g_{R_2}^2(Q)\right]\phi^2,
 &m_{\Phi,\mathrm{spec}}^2(\phi)&=m_{0\Phi}^2+a_\Phi\phi^2, \notag\\
 m_{H,\mathrm{spec}}^2(\phi)&=m_{0H}^2+a_H\phi^2,
 &m_f^2(\phi)&=\frac12y_f^2(Q)\phi^2,
 \qquad f=t,b,\tau .
 \label{eq:field-dependent-masses}
\end{align}
Here $h$ is the radial fluctuation, $G$ denotes the three Goldstone modes, and $A$ denotes the three gauge bosons associated with the broken generators in Eq.~\eqref{eq:diagonal-breaking}.
The multiplicities used in the one-loop potential are collected in Table~\ref{tab:thermal-spectrum}.
In particular, the three $\Phi$ bidoublets yield one radial mode, three Goldstones, and twenty spectator real degrees of freedom.  The effective Higgs-like sector contributes $13\times(2\times 2\times 2)=104$ real spectator degrees of freedom.
The three broken gauge bosons contribute nine degrees of freedom to the zero-temperature and thermal one-loop potentials, but only their three longitudinal zero modes enter the ring term.\footnote{At nonzero $\phi$ each broken gauge boson has two transverse and one longitudinal polarizations, giving nine degrees of freedom in $V_{\rm CW}$ and $V_T$. Only the longitudinal zero mode receives a Debye mass at leading order, so the gauge contribution to $V_{\rm ring}$ contains one mode per broken generator.}

\begin{table}[h!]
  \centering
  \caption{Fields retained in the one-loop potential at $Q=65$~TeV.
  The last column gives the degrees of freedom in $V_{\rm CW}$ and $V_T$;
  for the gauge contribution the number ``3" in parentheses is the longitudinal zero-mode multiplicity used in $V_{\rm ring}$.}
  \label{tab:thermal-spectrum}
  \begin{tabular}{c c c}
    \hline
    species & $m_i^2(\phi)$ & $n_i$ \\
    \hline
    radial scalar $h$ & $-\mu_\phi^2+3\lambda_\phi\phi^2$ & $1$ \\
    Goldstones $G$ & $-\mu_\phi^2+\lambda_\phi\phi^2$ & $3$ \\
    $\Phi$ spectators & $m_{0\Phi}^2+a_\Phi\phi^2$ & $20$ \\
    Higgs-like spectators & $m_{0H}^2+a_H\phi^2$ & $104$ \\
    broken vectors $A$ & $[g_{R_1}^2(Q)+g_{R_2}^2(Q)]\phi^2/4$ & $9\;(3)$ \\
    top quark $t$ & $y_t^2(Q)\phi^2/2$ & $12$ \\
    bottom quark $b$ & $y_b^2(Q)\phi^2/2$ & $12$ \\
    tau lepton $\tau$ & $y_\tau^2(Q)\phi^2/2$ & $4$ \\
    \hline
  \end{tabular}
\end{table}

\subsubsection{Zero-temperature potential}
\label{subsec:zero-temperature-potential}

Along the radial trajectory, the tree-level potential is
\begin{equation}
 V_{\rm tree}(\phi)=-\frac12\mu_\phi^2\phi^2
 +\frac14\lambda_\phi\phi^4~,~
 \label{eq:tree-potential-section-three}
\end{equation}
as explained in Section~\ref{subsec:scalar-eft}.
Unlike the classically scale-invariant example of~\cite{Huang:2020bbe}, we retain the renormalized quadratic parameter $\mu_\phi^2$.

\subsubsection*{Coleman-Weinberg potential} 
The Coleman-Weinberg contribution in the $\overline{\rm MS}$ scheme is
\begin{equation}
 V_{\rm CW}(\phi;Q)=\frac{1}{64\pi^2}
 \left[
 \sum_{b}n_bm_b^4(\phi)
 \left(\ln\frac{m_b^2(\phi)}{Q^2}-c_b\right)
 -\sum_fn_fm_f^4(\phi)
 \left(\ln\frac{m_f^2(\phi)}{Q^2}-c_f\right)
 \right],
 \label{eq:cw-potential}
\end{equation}
where $c_b=3/2$ for scalars, $c_b=5/6$ for gauge bosons, and $c_f=3/2$ for fermions.
A negative field-dependent curvature makes the fixed-order logarithm and bosonic thermal function complex.
In the numerical implementation we use the positive regulator $m_i^2\to\max(m_i^2,\epsilon)$ inside these loop functions, with $\epsilon$ much smaller than any physical mass scale.
This keeps the numerical potential real near the origin.

\subsubsection*{Renormalization Conditions} 

To preserve the chosen vacuum expectation value and radial curvature at one-loop order, we include finite counterterms
\begin{equation}
 V_{\rm CT}(\phi)=-\frac12\delta\mu_\phi^2\phi^2
 +\frac14\delta\lambda_\phi\phi^4,
 \label{eq:counterterm-potential}
\end{equation}
fixed by the on-shell-like renormalization conditions
\begin{equation}
 \left.\frac{\partial(V_{\rm CW}+V_{\rm CT})}{\partial\phi}
 \right|_{\phi=v_\phi}=0,
 \qquad
 \left.\frac{\partial^2(V_{\rm CW}+V_{\rm CT})}{\partial\phi^2}
 \right|_{\phi=v_\phi}=0,
 \label{eq:counterterm-conditions}
\end{equation}
where $v_\phi=\mu_\phi/\sqrt{\lambda_\phi}$ at tree level.
Thus the one-loop corrections do not shift the chosen zero-temperature radial vacuum or its tree-level curvature.
The central scale is $Q=65~\mathrm{TeV}$;
the range $30~\mathrm{TeV}\leq Q\leq150~\mathrm{TeV}$ is used for the scale-variation test.

\subsubsection{Finite-temperature potential}
\label{subsec:thermal-contribution}

The thermal functions entering the finite-temperature correction are defined by
\begin{align}
 J_B(x)&=\int_0^\infty {\rm d}z\,z^2
 \ln\left[1-\exp\left(-\sqrt{z^2+x}\right)\right], \notag\\
 J_F(x)&=\int_0^\infty {\rm d}z\,z^2
 \ln\left[1+\exp\left(-\sqrt{z^2+x}\right)\right].
 \label{eq:thermal-functions}
\end{align}
With this convention $J_F(x)>0$.
The thermal free energy is therefore
\begin{equation}
 V_T(\phi,T)=\frac{T^4}{2\pi^2}
 \left[
 \sum_b n_bJ_B\!\left(\frac{m_b^2(\phi)}{T^2}\right)
 -\sum_f n_fJ_F\!\left(\frac{m_f^2(\phi)}{T^2}\right)
 \right].
 \label{eq:thermal-potential}
\end{equation}
The minus sign in the fermionic contribution follows from Eq.~\eqref{eq:thermal-functions}.
For efficient evaluation, we interpolate a table built from the Bessel-function series, using the small-mass expansion for $m^2/T^2<0.02$.
The table and interpolation settings are specified in Appendix~\ref{app:numerical-implementation}.

\subsubsection{Ring contribution to the effective potential}
\label{subsec:ring-resummation}

Infrared-sensitive bosonic zero modes are resummed in the Arnold-Espinosa prescription~\cite{Arnold:1992rz,Carrington:1991hz},
\begin{equation}
 V_{\rm ring}(\phi,T)=-\frac{T}{12\pi}
 \sum_{i}n_i
 \left\{
 \left[m_i^2(\phi)+\Pi_i(T)\right]^{3/2}
 -\left[m_i^2(\phi)\right]^{3/2}
 \right\},
 \label{eq:ring-potential}
\end{equation}
where $i\in\{h,G,\Phi_{\rm spec},H_{\rm spec},A_L\}$.
For the scalar zero modes, the screening masses used in the present EFT are
\begin{align}
 \Pi_h(T)&=\Pi_G(T)=c_\phi^{\rm ring}T^2, \notag\\
 c_\phi^{\rm ring}&=c_\phi^{\rm base}
 +\frac{n_{\Phi}a_\Phi+n_Ha_H}{12}, \notag\\
 \Pi_{\Phi,\rm spec}(T)&=c_{\Phi}T^2,\qquad
 \Pi_{H,\rm spec}(T)=c_HT^2,\qquad
 \Pi_{A,L}(T)=c_AT^2,
 \label{eq:screening-masses}
\end{align}
with $n_{\Phi}=20$ and $n_H=104$.
The same portal couplings that make the spectator masses depend on $\phi$ also generate thermal self-energies for the active scalar.
Their contribution, $(n_\Phi a_\Phi+n_Ha_H)T^2/12$, is therefore included in $\Pi_\phi$.
Omitting it would retain the spectator contribution to the thermal free energy while discarding the corresponding screening of the $\phi$ zero mode.

\subsubsection*{Spectrum-informed screening coefficients}
\label{subsubsec:spectrum-screening}

Rather than adopting arbitrary screening coefficients, we determine the terms fixed by the retained representation content in the high-temperature limit.
For $X\sim(\mathbf{2},\mathbf{2})$ under $SU(2)_{R_1}\times SU(2)_{R_2}$, the secondary Casimir invariant $C_2(\mathbf{2})=3/4$, because the generator $T^a=\frac{\sigma^a}{2}$ and $\sum_{a=1}^3 T^a T^a =\frac{1}{4}\sum_{a=1}^3 \sigma^a\sigma^a=\frac{3}{4}{\mathbf 1}$.
Since the gauge thermal mass of $X$ is
\begin{equation}
    \Pi_X^{gauge}=\frac{T^2}{4}[g_{R_1}^2 C_2(\mathbf{2})+g_{R_2}^2 C_2(\mathbf{2})],
\end{equation}
we have
\begin{equation}
 c_\phi^{\rm gauge}=\frac{3}{16}\left(g_{R_1}^2+g_{R_2}^2\right),
 \label{eq:active-scalar-screening}
\end{equation}
With the convention $m_f^2(\phi)=y_f^2\phi^2/2$, a Dirac fermion with color multiplicity $N_c$ gives the leading high-temperature scalar self-energy
\begin{equation}
 \Pi_{\phi,f}(T)=\frac{N_c y_f^2}{12}T^2.
\end{equation}
Keeping the phenomenologically relevant third-generation Yukawa couplings, we therefore obtain
\begin{equation}
 c_\phi^{\rm Yukawa}
 =\frac14 y_t^2+\frac14 y_b^2+\frac1{12}y_\tau^2 .
\end{equation}
The factors $1/4$ and $1/12$ follow from $N_c=3$ for quarks and $N_c=1$ for the charged lepton, respectively.
Describing the active radial mode and the three Goldstone modes by an $O(4)$-symmetric quartic potential, we obtain $c_\phi^{\rm self}=\lambda_\phi/2$.
This is because the high-temperature self-interaction thermal mass for an $O(N)$ theory is~\cite{Dolan:1973qd,Quiros:1999jp}
\begin{equation}
    \Pi_{\rm self}=\frac{N+2}{12}\lambda_{\phi} T^2.
\end{equation}
Details of the calculation are shown in Appendix~\ref{app:thermal-tadpole}.
The full $O(8)$ treatment of the complex $X$ doublet is not applied. In the one-field reduction, one radial mode and three Goldstones are retained, while the remaining four real components are counted in the $\Phi$ spectator sector.
At $Q=65~\mathrm{TeV}$ this yields
\begin{equation}
 c_\phi^{\rm base}=c_\phi^{\rm gauge}+c_\phi^{\rm Yukawa}
 +c_\phi^{\rm self}=0.3376,
 \qquad
 c_\phi^{\rm ring}=2.3450
 \label{eq:active-screening-numerical}
\end{equation}
for the central reference point, while $c_\phi^{\rm ring}$ includes the portal contribution in Eq.~\eqref{eq:screening-masses}.

The known gauge-plus-active-portal contributions to $\Phi$ and Higgs-like spectator screening coefficients are (at $Q=65$ TeV)
\begin{equation}
 c_{\Phi}=\frac{3}{16}(g_{R_1}^2+g_{R_2}^2)
 +\frac{8a_\Phi}{12}=0.3065,
 \qquad
 c_H=\frac{3}{16}\left[g_L^2+\frac{g_{R_1}^2+g_{R_2}^2}{2}\right]
 +\frac{8a_H}{12}=0.2959,
 \label{eq:spectator-screening-numerical}
\end{equation}
where\footnote{Note that below the string scale, the relation $g_b^2=g_R^2$ no longer exists.} $g_L=g_b\simeq0.619$ is the $SU(2)_L$ gauge coupling at the central scale $Q=65$ TeV, with $g_b$ denoting the coupling of the $b$-stack.
The factor 8 is the number of real components of the active bidoublet.
Each component gives a tadpole correction of $\frac{aT^2}{12}$ at the high temperature.
The remaining spectator self-quartics and cross-quartics are not fixed by the reduced parameterization and are not included in Eq.~\eqref{eq:spectator-screening-numerical}.

For the longitudinal gauge screening, the leading high-temperature Debye coefficient of a non-Abelian factor is
\begin{equation}
 \frac{\Pi_{L,a}}{g_a^2T^2}=\frac{C_A}{3}
 +\frac{1}{3}\sum_{\text{complex }S}T_a(R_S)
 +\frac{1}{6}\sum_{\text{Weyl }F}T_a(R_F).
 \label{eq:nonabelian-debye}
\end{equation}
For later convenience, we define the dimensionless coefficients $K_{R1}$ and $K_{R2}$ through
\begin{equation}
 \Pi_{L,Ri}(T)\equiv g_{Ri}^2 K_{Ri}T^2,
 \qquad i=1,2.
 \label{eq:define-right-debye-coefficients}
\end{equation}
Thus, $K_{Ri}$ is the total leading-order longitudinal Debye coefficient of the $SU(2)_{R_i}$ plasma after factoring out $g_{Ri}^2T^2$.
It is fixed by the light particle content and its $SU(2)_{R_i}$ representations through Eq.~\eqref{eq:nonabelian-debye}; it is not an independent Lagrangian parameter.
For the retained spectrum, three $\Phi$ bidoublets contribute Dynkin index 3 to each right-handed factor.
The thirteen Higgs-like bidoublets split as ten fields charged under $SU(2)_{R_1}$ and three under $SU(2)_{R_2}$, while the three right-handed families split as one and two, respectively.
Hence
\begin{equation}
 K_{R_1}=\frac23+\frac{3+10}{3}+\frac{2}{6}=\frac{16}{3},
 \qquad
 K_{R_2}=\frac23+\frac{3+3}{3}+\frac{4}{6}=\frac{10}{3}.
 \label{eq:right-debye-coefficients}
\end{equation}
The secondary Casimir for the adjoint representation of $SU(N)$ is $C_A=N$.
In each expression of Eq.~\eqref{eq:right-debye-coefficients}, the first term is the $SU(2)$ gauge contribution $C_A/3=2/3$, the second term is the contribution of the complex scalar bidoublets, and the final term is the contribution of the Weyl fermions.
For a bidoublet, the Dynkin index with respect to either relevant $SU(2)$ factor is $2T({\bf 2})=1$, where the factor 2 is the dimension of the doublet under the other $SU(2)$ factor.
A right-handed Pati-Salam family has index $4T({\bf 2})=2$, with the factor 4 supplied by its $SU(4)$ multiplicity.
Projecting the diagonal Debye matrix onto the heavy gauge-boson direction in Eq.~\eqref{eq:diagonal-breaking} gives
\begin{equation}
 c_A=\frac{g_{R_1}^4K_{R1}+g_{R_2}^4K_{R2}}
 {g_{R_1}^2+g_{R_2}^2}=2.1470.
 \label{eq:heavy-vector-screening}
\end{equation}
The resulting reference coefficients are summarized in Table~\ref{tab:thermal-parameters}.

\begin{table}[h]
  \centering
  \caption{Spectrum-informed high-temperature screening inputs in the reduced one-field EFT at $Q=65~\mathrm{TeV}$.
  The portal contribution to $c_\phi^{\rm ring}$ is not an independent parameter:
  it is determined by $a_\Phi$, $a_H$, and the fixed spectator multiplicities.
  The spectator coefficients omit unknown scalar self- and cross-quartics, while the gauge coefficient is the high-temperature representation-counting estimate.}
  \label{tab:thermal-parameters}
  \begin{tabular}{c c l}
    \hline
    quantity & reference value & role \\
    \hline
    $c_\phi^{\rm base}$ & $0.3376$ & active-radial and Goldstone baseline screening \\
    $c_{\Phi}$ & $0.3065$ & $\Phi$ spectator gauge-plus-portal screening \\
    $c_H$ & $0.2959$ & Higgs-like spectator gauge-plus-portal screening \\
    $c_A$ & $2.1470$ & longitudinal broken-vector high-$T$ screening \\
    $n_{\Phi}$ & $20$ & real $\Phi$ spectator degrees of freedom \\
    $n_H$ & $104$ & real Higgs-like spectator degrees of freedom \\
    \hline
  \end{tabular}
\end{table}

Eqs.~\eqref{eq:active-scalar-screening}--\eqref{eq:heavy-vector-screening}
are high-temperature and spectrum-informed estimates.
Finite-mass threshold suppression, especially for the $\Phi$ spectators, and unlisted scalar self/cross-quartics remain sources of uncertainty.
A UV-resolved treatment would specify the total scalar potential, construct the full scalar and longitudinal-gauge Debye matrices, and use their eigenvalues in Eq.~\eqref{eq:ring-potential}, as in the matrix treatment of~\cite{Huang:2020bbe}.  We quantify the residual screening uncertainty by explicit variations around the spectrum-informed benchmark.

\subsection{Total potential and thermal equation of state}
\label{subsec:total-potential}

The potential used for the phase-transition calculation is
\begin{equation}
 V_{\rm eff}(\phi,T)=V_{\rm tree}(\phi)+V_{\rm CW}(\phi;Q)
 +V_{\rm CT}(\phi)+V_T(\phi,T)+V_{\rm ring}(\phi,T).
 \label{eq:total-effective-potential}
\end{equation}
The radiation and entropy degrees of freedom are evaluated in the false phase,
\begin{equation}
 g_\rho(T)=\frac{30\rho(T)}{\pi^2T^4},\qquad
 g_s(T)=\frac{45[\rho(T)+p(T)]}{2\pi^2T^4}.
 \label{eq:gstar-definitions}
\end{equation}
The distinct functions $g_\rho$ and $g_s$ are used, respectively, in the Hubble rate and in the redshift of the gravitational-wave spectrum.
These definitions follow by writing the radiation energy density as $\rho(T)=(\pi^2/30)g_\rho(T)T^4$ and using the zero-chemical-potential thermodynamic identity $s(T)=[\rho(T)+p(T)]/T$ together with $s(T)=(2\pi^2/45)g_s(T)T^3$; see, e.g.~\cite{Laine:2016hma,KolbTurner1990}.
They remain valid away from the strictly relativistic limit and therefore allow $g_\rho(T)\ne g_s(T)$ when particles cross mass thresholds.
Here $\rho$ and $p$ denote the thermal plasma contributions, with any temperature-independent vacuum-energy offset subtracted.
Their threshold treatment is particularly relevant when the spectator masses are comparable to the transition temperature.

For a species with multiplicity $n_i$ and false-phase mass $m_i(T)$, we use the equilibrium integrals
\begin{align}
 \rho_i(T)&=\frac{n_iT^4}{2\pi^2}\int_0^\infty {\rm d}q\,
 \frac{q^2 E_i}{e^{E_i}\mp1}, &
 p_i(T)&=\frac{n_iT^4}{6\pi^2}\int_0^\infty {\rm d}q\,
 \frac{q^4}{E_i(e^{E_i}\mp1)},
 \label{eq:thermal-bath-integrals}\\
 E_i&=\sqrt{q^2+[m_i(T)/T]^2}, & q&=|\mathbf p|/T.\notag
\end{align}
The minus sign applies to bosons and the plus sign to fermions.
We add these contributions to a relativistic background with the Standard Model normalization $g_{\rho,\rm bg}=g_{s,\rm bg}=106.75$.
Table~\ref{tab:false-phase-bath} specifies the additions used in the calculation, including all 104 Higgs-like spectator modes.
The $t,b,\tau$ thermal degrees of freedom are already included in the background.
Superpartners and the heavy $X_G,X_W$ states are thermally decoupled at these temperatures; no hidden-sector radiation is added.

\begin{table}[htbp]
 \centering
 \caption{Thermal bath used for the reference calculation. The last two columns give the contributions at $T_n=6.72766952~\mathrm{TeV}$.
 The background entry is its effective relativistic count; $n_i$ for the other entries counts bosonic degrees of freedom.}
 \label{tab:false-phase-bath}
 \small
 \begin{tabular}{lclrr}
  \toprule
  Component & $n_i$ & False-phase mass & $g_\rho(T_n)$ & $g_s(T_n)$ \\
  \midrule
  Relativistic background & --- & $0$ & $106.750000$ & $106.750000$ \\
  Extra gauge modes & $6$ & $0$ & $6.000000$ & $6.000000$ \\
  Active scalar modes & $4$ & $m_{\rm act}(T)$ & $3.748196$ & $3.676154$ \\
  $\Phi$ spectators & $20$ & $m_{0\Phi}$ & $16.826884$ & $16.165292$ \\
  Higgs-like spectators & $104$ & $m_{0H}$ & $102.340102$ & $101.721969$ \\
  \midrule
  Total & & & $235.665182$ & $234.313415$ \\
  \bottomrule
 \end{tabular}
\end{table}

For the four active scalar modes we use the common curvature mass
$m_{\rm act}(T)=\sqrt{\max[V_{\rm eff}''(0,T),0]}$, which is $4.95346~\mathrm{TeV}$ at $T_n$.
The spectator masses in these integrals are $m_{0\Phi}$ and $m_{0H}$; their ring masses enter Eq.~\eqref{eq:ring-potential} separately.
The screening scan in Section~\ref{subsec:numerical-checks} varies the ring coefficients while preserving this bare-mass prescription for the spectators in the equation of state.
The six extra gauge degrees of freedom count two transverse polarizations for each of the three $SU(2)_R$ gauge bosons that remain massless at $\phi=0$; the three broken gauge bosons are instead included through their field-dependent masses in $V_{\rm CW}$ and $V_T$.
The phase-transition observables obtained from Eq.~\eqref{eq:total-effective-potential} are discussed in the next section.

\section{Phase transition dynamics}
\label{sec:phase-dynamics}

Having constructed the finite-temperature effective potential, we now determine the cosmological history of the transition.
We follow the thermal minima, calculate the critical temperature, and evaluate the thermal bounce action.
The nucleation rate and the false-vacuum survival probability then determine percolation and completion.
Finally, we extract the released energy and the inverse duration that enter the gravitational-wave calculation.

The calculation is carried out in the single-field, diagonally aligned EFT described in the previous section.
The spectator masses remain positive along the radial tunneling path, as shown in Table~\ref{tab:transverse-stability}; this supports the single-field description within the reduced EFT.
We construct the one-loop thermal potential, including ring corrections and the thermal-mass contributions from the particles listed below.
We also retain the temperature dependence of the energy and entropy degrees of freedom, $g_\rho(T)$ and $g_s(T)$, defined in \eqref{eq:gstar-definitions}.
The $g_\rho(T)$ dependence is included in the Hubble rate, while $g_s(T)$ is retained in the later gravitational wave redshift calculation.
In the percolation integral below, we use the standard $aT={\rm const.}$ approximation over the short interval between nucleation and completion.
This is relevant because several spectator masses are comparable to the transition temperatures.

\subsection{Thermal history and phase structure}
\label{subsec:thermal-history}

At a fixed temperature, let $\phi_f(T)$ and $\phi_t(T)$ denote the false-phase and true-phase local minima of the effective potential, respectively.
We define the free-energy density released by conversion of the false phase into the true phase as
\begin{equation}
 \Delta\mathcal{F}(T)\equiv
 V_{\rm eff}\bigl(\phi_f(T),T\bigr)
 -V_{\rm eff}\bigl(\phi_t(T),T\bigr).
 \label{eq:free-energy-difference}
\end{equation}
Below the critical temperature, this convention gives $\Delta\mathcal{F}(T)>0$.
Since the equilibrium pressure is $p=-\mathcal{F}$, the corresponding pressure difference that drives bubble growth is
\begin{equation}
 \Delta p(T)\equiv p_t(T)-p_f(T)=\Delta\mathcal{F}(T).
 \label{eq:driving-pressure}
\end{equation}
Both quantities are invariant under a temperature-independent additive shift of the effective potential.

The critical temperature is determined by the coexistence conditions
\begin{align}
 \left.\frac{\partial V_{\rm eff}}{\partial\phi}\right|_{\phi_f,T_c}
 &=\left.\frac{\partial V_{\rm eff}}{\partial\phi}\right|_{\phi_t,T_c}=0,
 \qquad
 \left.\frac{\partial^2 V_{\rm eff}}{\partial\phi^2}\right|_{\phi_f,T_c}>0,
 \qquad
 \left.\frac{\partial^2 V_{\rm eff}}{\partial\phi^2}\right|_{\phi_t,T_c}>0,
 \notag\\
 V_{\rm eff}(\phi_f,T_c)&=V_{\rm eff}(\phi_t,T_c).
 \label{eq:critical-temperature-conditions}
\end{align}
For the central reference point \(Q=65\) TeV, the two minima are separated by a barrier, so the conversion proceeds through thermal bubble nucleation rather than a continuous roll of the homogeneous field.
Using the critical temperature and order parameter of
Eq.~\eqref{eq:Tc-phic}, we find that the high-temperature minimum is at
$\phi_f=0$ and the broken minimum is at
\begin{equation}
 \phi_c\equiv\phi_t(T_c),
 \label{eq:benchmark-critical-data}
\end{equation}
with $T_c=13.8791~\mathrm{TeV}$ and $\phi_c/T_c=6.1902$ taken from
Eq.~\eqref{eq:Tc-phic}.

The large value of $\phi_c/T_c$ shows that the order parameter changes sharply at $T_c$ in the chosen gauge.
We determine whether the transition completes by following bubble nucleation and growth below.
Table~\ref{tab:reference-phase-values} gives the minima and free-energy difference at the key temperatures.
The minima are found by a field scan followed by bounded minimization, and $T_c$ is obtained from a root of $\Delta\mathcal F(T)$; the numerical settings are given in Appendix~\ref{app:numerical-implementation}.

\begin{table}[htbp]
 \centering
 \caption{Phase values for the reference transition. The false-phase minimum is $\phi_f=0$ in every row, and $\phi_c=\phi_t(T_c)$.
 The dynamical temperatures are defined in Sections~\ref{subsec:thermal-tunnelling} and~\ref{subsec:percolation}.}
 \label{tab:reference-phase-values}
 \begin{tabular}{lrrr}
  \toprule
  Stage & $T$ [TeV] & $\phi_t(T)$ [TeV] & $\Delta\mathcal F(T)$ [$\mathrm{TeV}^4$] \\
  \midrule
  Critical & $13.879134$ & $85.913984$ & $0$ \\
  Nucleation & $6.727670$ & $89.918527$ & $2.4274432\times10^5$ \\
  Percolation & $6.396665$ & $89.877021$ & $2.4540579\times10^5$ \\
  Completion & $6.374565$ & $89.874174$ & $2.4556603\times10^5$ \\
  \bottomrule
 \end{tabular}
\end{table}

\begin{figure}[h]
 \centering
 \includegraphics[width=0.82\textwidth]{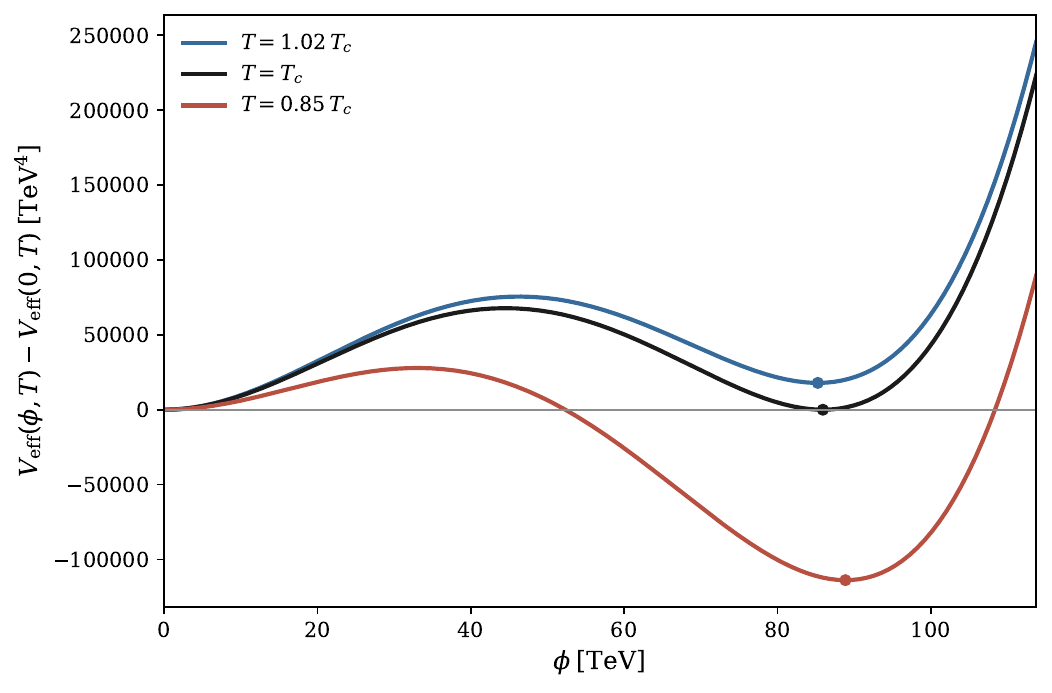}
 \caption{Finite-temperature effective potential relative to the symmetric phase, $V_{\rm eff}(\phi,T)-V_{\rm eff}(0,T)$, at three temperatures around  the critical temperature.
 At $T_c$, the symmetric and broken minima are  degenerate.
 Below $T_c$, the broken minimum has lower free-energy density,
 while the barrier between the two minima persists and makes the transition proceed through bubble nucleation.}
 \label{fig:thermal-potential-slices}
\end{figure}

Below $T_c$, the free-energy difference between the two phases increases as the universe cools, making conversion to the broken phase increasingly favorable.
Nevertheless, the system remains temporarily trapped in the false phase because producing a critical bubble costs gradient energy and must overcome the intervening barrier.
The physically relevant temperatures are consequently the nucleation temperature $T_n$, the percolation temperature $T_p$, and the completion temperature $T_{\rm completion}$, rather than $T_c$ alone.
They are obtained from the thermal bounce and false-vacuum survival probability in the following subsections.

\begin{figure}[h]
 \centering
 \includegraphics[width=0.82\textwidth]{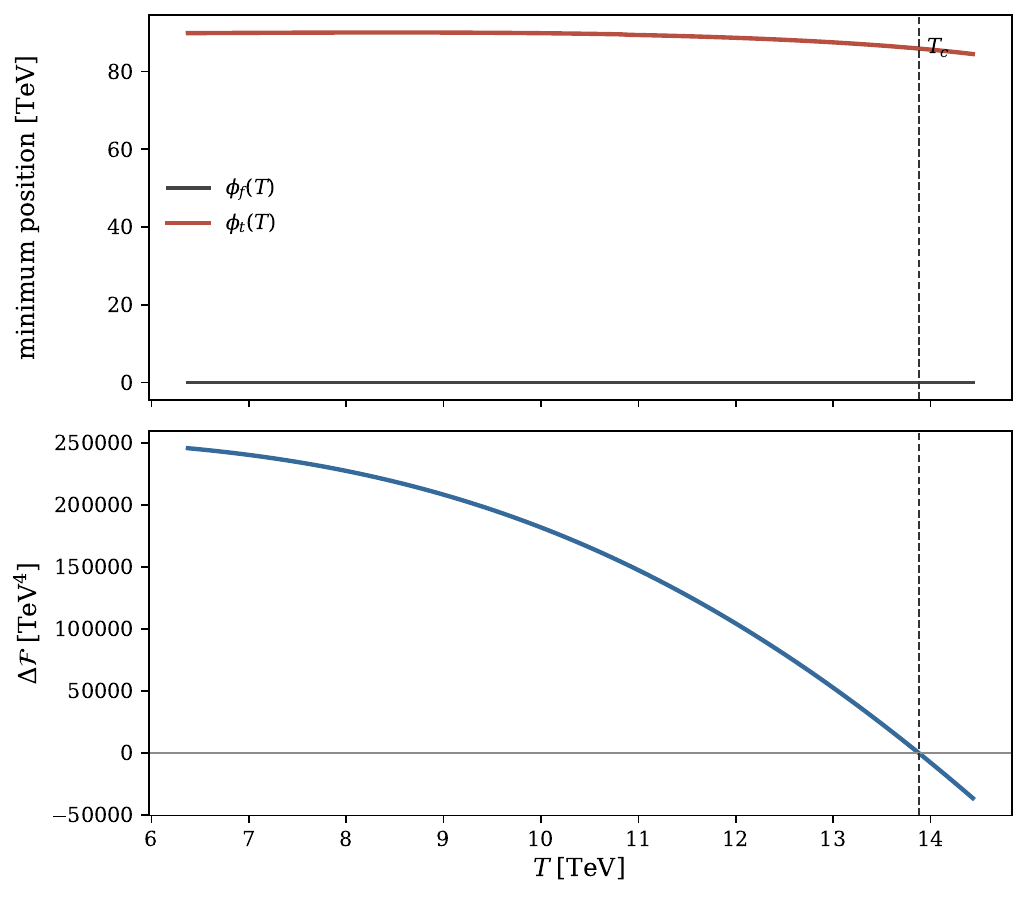}
 \caption{Temperature evolution of the false-phase and true-phase minima (upper panel) and the free-energy density difference defined in \eqref{eq:free-energy-difference} (lower panel).
 The vertical line marks $T_c$.
 The sign change of $\Delta\mathcal{F}$ identifies the interchange of the thermodynamically preferred phase; $\Delta\mathcal{F}>0$ below $T_c$.
 The dynamical temperatures at which bubbles nucleate, percolate, and complete the transition are determined in the following subsections.}
 \label{fig:thermal-history}
\end{figure}

\subsection{Bubble nucleation}
\label{subsec:thermal-tunnelling}

Below $T_c$, the false phase is metastable rather than immediately unstable.
The field must fluctuate across the barrier and form a bubble large enough for the gain of free energy in its interior to overcome the gradient energy of its wall.
At the temperatures of interest this process is described by the finite-temperature, $O(3)$-symmetric bounce~\cite{Linde:1981zj,Quiros:1999jp}.
Its action is
\begin{equation}
 S_3(T)=4\pi\int_0^\infty {\rm d}r\,r^2
 \left[\frac12\left(\frac{{\rm d}\phi}{{\rm d}r}\right)^2
 +V_{\rm eff}(\phi,T)-V_{\rm eff}(\phi_f,T)\right],
 \label{eq:thermal-bounce-action}
\end{equation}
where the subtraction ensures that the action is measured relative to the homogeneous false phase.
The bounce profile obeys
\begin{equation}
 \frac{{\rm d}^2\phi}{{\rm d}r^2}+\frac{2}{r}\frac{{\rm d}\phi}{{\rm d}r}
 =\frac{\partial V_{\rm eff}(\phi,T)}{\partial\phi},
 \qquad
 \left.\frac{{\rm d}\phi}{{\rm d}r}\right|_{r=0}=0,
 \qquad
 \phi(r\rightarrow\infty)=\phi_f(T).
 \label{eq:thermal-bounce-eom}
\end{equation}
The profile describes a bubble with the field on the broken-phase side of the barrier at its center and approaching the false vacuum at large radius.
It is the saddle-point configuration with the smallest action and therefore gives the dominant thermal tunneling rate.
We solve the radial equation with the overshoot/undershoot method in \texttt{CosmoTransitions}, using its $O(3)$ single-field solver.
Field rescaling, profile tolerances, and the temperature search are specified in Appendix~\ref{app:numerical-implementation}.

We use the standard semiclassical form
\begin{equation}
 \Gamma(T)=T^4
 \left(\frac{S_3(T)}{2\pi T}\right)^{3/2}
 \exp\left[-\frac{S_3(T)}{T}\right].
 \label{eq:nucleation-rate}
\end{equation}
As the universe cools, the decrease in $S_3/T$ makes the nucleation rate rise rapidly through the exponential factor.
We evaluate $\Gamma(T)$ using the calculated action and prefactor at each temperature.

\begin{figure}[h]
 \centering
 \includegraphics[width=0.82\textwidth]{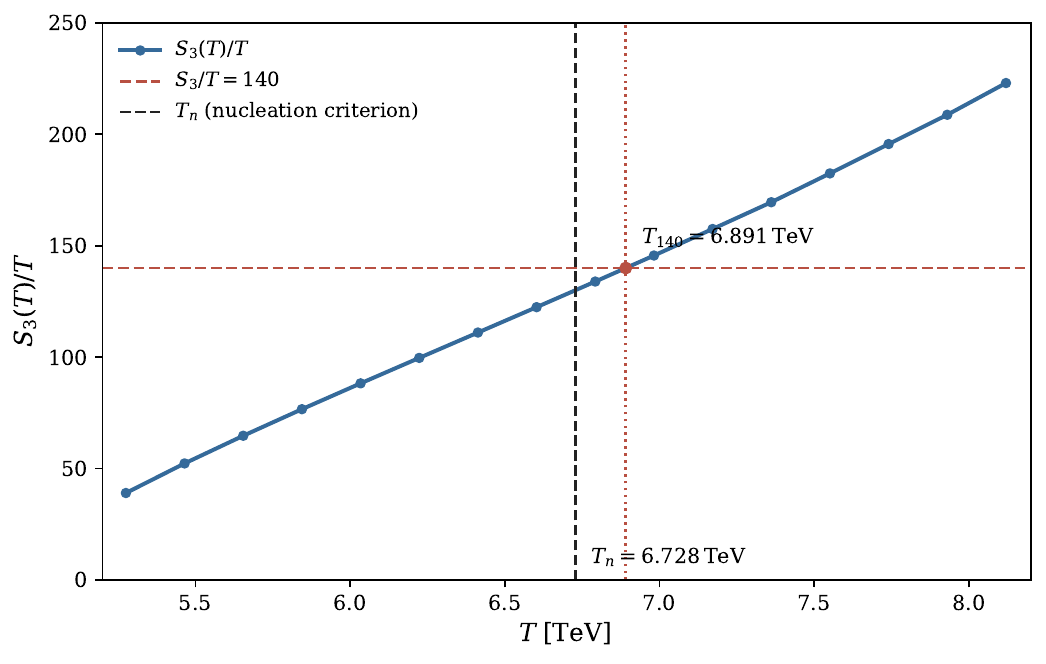}
 \caption{Thermal bounce action as a function of temperature.  The horizontal line marks $S_3/T=140$, a commonly adopted rule-of-thumb value for the onset of nucleation.
 The red line marks the nucleation temperature obtained instead from $\Gamma/H^4=1$.
 The small difference between the two crossings illustrates why a fixed value of $S_3/T$ is used only as a rule of thumb.}
 \label{fig:nucleation-action}
\end{figure}

\begin{figure}[h]
 \centering
 \includegraphics[width=0.82\textwidth]{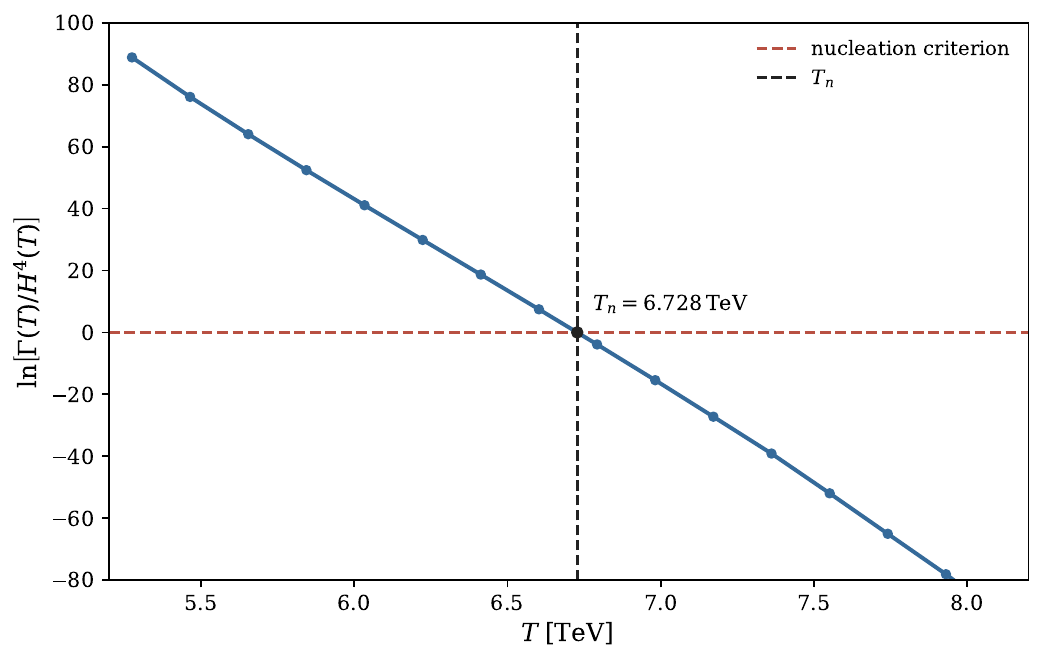}
 \caption{Temperature dependence of the dimensionless nucleation rate.
 We plot $\ln(\Gamma/H^4)$ because $\Gamma/H^4$ changes exponentially during cooling.
 The horizontal line at zero is equivalent to $\Gamma/H^4=1$; its intersection with the curve defines $T_n$.}
 \label{fig:nucleation-rate}
\end{figure}

For the numerical results presented here, the nucleation temperature is defined by the standard local criterion
\begin{equation}
 \left.\frac{\Gamma(T)}{H^4(T)}\right|_{T=T_n}=1,
 \label{eq:nucleation-condition}
\end{equation}
where $H$ is the Hubble expansion rate.
This criterion estimates when the nucleation probability in a Hubble four-volume becomes order unity.
We then follow the growing bubbles to determine how much of space has converted to the broken phase.
For the parameter point studied below, this condition gives
\begin{equation}
 T_n=6.728~\mathrm{TeV}.
 \label{eq:benchmark-nucleation-temperature}
\end{equation}
At this temperature, $S_3(T_n)/T_n=130.0$.  Figure~\ref{fig:nucleation-action} compares the resulting action with a commonly adopted rule-of-thumb value of 140.
The crossing at $S_3/T=140$ occurs at a slightly higher temperature because the condition $\Gamma/H^4=1$ also includes the Hubble rate and the nucleation prefactor.
Figure~\ref{fig:nucleation-rate} shows the same crossing directly in terms of $\Gamma/H^4$.
The next subsection follows the subsequent expansion and overlap of these bubbles.

\subsection{Percolation, completion, and reheating}
\label{subsec:percolation}

The condition $\Gamma/H^4=1$ shows that bubbles start to appear, but it does not tell us how much of the universe has changed phase.
A bubble must grow, and many bubbles must overlap, before the false phase disappears.
We track this process through the false-vacuum survival probability~\cite{Guth:1982pn},
\begin{equation}
 P_f(T)=e^{-I(T)},
 \label{eq:false-vacuum-probability}
\end{equation}
where $I(T)$ is the accumulated volume fraction of bubbles, including their growth after nucleation.
The approximation $aT={\rm const.}$ used below follows from adiabatic expansion:
entropy conservation, $s(T)a^3={\rm const.}$, with $s(T)=(2\pi^2/45)g_s(T)T^3$, gives the more general relation
$aTg_s^{1/3}={\rm const.}$~\cite{KolbTurner1990}.
We neglect the small variation of $g_s$ across the short nucleation-to-completion interval and use $aT={\rm const.}$ in the numerical percolation integral.
For the present point, $g_s$ changes from $234.31$ at $T_n$ to $233.77$ at completion, a relative change of $0.23\%$.
We then evaluate
\begin{equation}
 I(T)=\frac{4\pi}{3}\int_T^{T_c}{\rm d}T'\,
 \frac{\Gamma(T')}{T'^4H(T')}
 \left[v_w\int_T^{T'}\frac{{\rm d}\widetilde T}{H(\widetilde T)}\right]^3.
 \label{eq:percolation-integral}
\end{equation}
The inner integral gives the distance traveled by a bubble wall nucleated at $T'$ before the universe cools to $T$, while the outer integral sums the volumes of all such bubbles.
The overlap of bubbles is already included by the exponential in \eqref{eq:false-vacuum-probability}.
For the central result, both temperature integrals are evaluated by the trapezoidal rule on a 96-point grid of bounce solutions.
We interpolate the resulting $I(T)$ to locate the percolation and completion thresholds, as detailed in Appendix~\ref{app:numerical-implementation}.

We define the percolation temperature by
\begin{equation}
 I(T_p)=0.34,
 \qquad P_f(T_p)\simeq0.71.
 \label{eq:percolation-condition}
\end{equation}
The value $0.34$ is the continuum percolation threshold for equal-size, randomly distributed overlapping spheres in three dimensions \cite{Ellis:2018mja,Megevand:2016lpr,Rintoul:1997tze}.
In $I$, overlapping regions are counted once for each bubble that covers them; the converted volume fraction is $1-e^{-I}$.
With the Poisson survival probability in \eqref{eq:false-vacuum-probability}, it corresponds to $P_f=e^{-0.34}=0.712$ and hence to a converted fraction $1-P_f=0.288$.
It is a conventional point at which bubbles have become common enough to form a connected broken-phase network.
It is still not sufficient by itself:
if the false-vacuum volume grows faster than bubbles can remove it, the transition does not finish.
We therefore also require
\begin{equation}
 \frac{{\rm d}}{{\rm d}t}\left[a^3(t)P_f(t)\right]<0.
 \label{eq:completion-volume-condition}
\end{equation}
With the temperature variable used in \eqref{eq:percolation-integral}, this condition becomes
\begin{equation}
 3+T\frac{{\rm d}I}{{\rm d}T}<0.
 \label{eq:completion-temperature-condition}
\end{equation}
For definiteness, we quote $T_{\rm completion}$ at $I=4.6$, for which $P_f=e^{-4.6}\simeq0.01$, after verifying \eqref{eq:completion-temperature-condition}.

The Hubble rate in \eqref{eq:percolation-integral} includes the released energy,
\begin{equation}
H(T)=H_{\rm rad}(T)\sqrt{1+\frac{\Delta\rho(T)}{\rho_R(T)}},
\qquad
H_{\rm rad}(T)=1.66\sqrt{g_\rho(T)}\frac{T^2}{M_{\rm Pl}},
\label{eq:percolation-hubble-rate}
\end{equation}
where
\begin{equation}
 \Delta\rho(T)=\Delta\mathcal{F}(T)
 -T\frac{d\Delta\mathcal{F}(T)}{dT},
 \qquad
 \rho_R(T)=\frac{\pi^2}{30}g_\rho(T)T^4.
 \label{eq:released-energy-density}
\end{equation}
The quantity $\Delta\rho$ is the energy-density difference between the false and true phases.
Indeed, for a homogeneous phase with free-energy density $\mathcal{F}$, the thermodynamic energy density is $\rho=\mathcal{F}-T\,d\mathcal{F}/dT$.
With the sign convention in \eqref{eq:free-energy-difference},
\begin{equation}
 \Delta\rho(T)=\rho_f(T)-\rho_t(T)>0
 \label{eq:released-energy-meaning}
\end{equation}
is the energy made available when a volume element converts from the false phase to the true phase.
It includes both the free-energy difference and the entropy difference between the phases.
The radiation density $\rho_R(T)$ is instead the energy density of the thermal plasma in the false phase, evaluated with the threshold-resolved $g_\rho(T)$.
It excludes the temperature-independent vacuum-energy offset and is the background against which the released energy is normalized.
The energy in $\Delta\rho$ need not become heat instantaneously: it can be partitioned among plasma bulk motion, thermal excitations, and scalar-field gradients.
The reheating estimate below assumes that this energy eventually thermalizes.
This correction is important here because the transition is sufficiently supercooled that the released energy is comparable to, or larger than, the radiation energy density.
Here $M_{\rm Pl}$ is the non-reduced Planck mass.

For the point considered in this work, the percolation calculation gives
\begin{equation}
 T_p=6.3967~\mathrm{TeV},
 \qquad
 T_{\rm completion}=6.3746~\mathrm{TeV}.
 \label{eq:percolation-results}
\end{equation}
At $T_p$ we find $3+T_p(dI/dT)=-3.73\times10^2$, so the physical false-vacuum volume is already decreasing.
Thus the transition both percolates and completes, rather than remaining trapped in the symmetric phase.
In this completion check we take $v_w=1$ in \eqref{eq:percolation-integral}.
This is a fast-wall reference choice, not a prediction for the microscopic wall velocity; the wall-speed dependence of the gravitational wave signal is discussed below.

Once the transition completes, part of the released energy reheats the plasma.
We estimate the reheating temperature from energy conservation,
\begin{equation}
 \rho_R(T_{\rm reh})=\rho_R(T_p)+\Delta\rho(T_p),
 \label{eq:reheating-condition}
\end{equation}
which gives
\begin{equation}
 T_{\rm reh}=8.5729~\mathrm{TeV},
 \label{eq:reheating-result}
\end{equation}
using the same thermal-bath prescription as in Section~\ref{subsec:total-potential}.
This temperature lies above $T_p$, as expected, but remains below $T_c$; the transition is therefore not thermally restored after reheating.

\subsection{Transition strength and duration}
\label{subsec:transition-parameters}

The calculation of gravitation waves requires the energy released by the transition and the time over which bubbles form.
We use the released-energy convention for the strength parameter,
\begin{equation}
 \alpha(T)\equiv\frac{\Delta\rho(T)}{\rho_R(T)},
 \label{eq:alpha-definition}
\end{equation}
where $\Delta\rho$ and $\rho_R$ are given in \eqref{eq:released-energy-density}.
Thus, $\alpha$ measures the energy released by the transition relative to the radiation energy density at the same temperature.
This is the convention used in our numerical gravitational wave estimates.
Other conventions based on the trace anomaly are sometimes used in the literature; they should not be mixed with \eqref{eq:alpha-definition} without translating the corresponding gravitational-wave efficiency factors.

For the present point, we find
\begin{equation}
 \alpha_n\equiv\alpha(T_n)=1.9013,
 \qquad
 \alpha_p\equiv\alpha(T_p)=2.2543.
 \label{eq:strength-results}
\end{equation}
The released energy is therefore already larger than the radiation energy density at nucleation and grows further before percolation.
The value $\alpha_n>1$ reflects the large energy release after supercooling.
Whether the bubble wall keeps accelerating depends on the balance between the driving pressure and plasma friction, discussed below.

The inverse duration parameter is defined by the rate at which the tunneling exponent changes,
\begin{equation}
 \beta\equiv-\left.\frac{{\rm d}(S_3/T)}{{\rm d}t}\right|_{T=T_*}.
 \label{eq:beta-definition}
\end{equation}
Using the same $aT={\rm const.}$ approximation as in the percolation calculation, $dT/dt=-HT$, this becomes
\begin{equation}
 \frac{\beta}{H_*}
 =\left.T\frac{{\rm d}(S_3/T)}{{\rm d}T}\right|_{T=T_*}.
 \label{eq:beta-over-H}
\end{equation}
We take $T_*=T_n$ and evaluate the action derivative by a central finite difference, solving the bounce at both neighboring temperatures.
The derivative of $\Delta\mathcal F$ in $\alpha$ is evaluated in the same way, with the minima recomputed at each temperature.
The step sizes and numerical samples are given in Appendix~\ref{app:numerical-implementation}.
This gives
\begin{equation}
 \frac{\beta}{H_n}=410.4.
 \label{eq:duration-result}
\end{equation}
Equivalently, the characteristic transition time is $\beta^{-1}\simeq2.4\times10^{-3}H_n^{-1}$, so the transition is short on a Hubble time scale even though it is strongly supercooled.

We use $(T_n,\alpha_n,\beta/H_n)$ to calculate the reference gravitational-wave spectra.
The value of $\alpha_p$ shows how the released energy grows between nucleation at $6.728~\mathrm{TeV}$ and percolation at $6.397~\mathrm{TeV}$.
The large value of $\beta/H_n$ is consistent with this short temperature interval.
We study the effect of the wall speed separately in Section~\ref{subsec:gw-spectrum}.

\subsection{Bubble-wall velocity and runaway diagnostic}
\label{subsec:wall-dynamics}

The bubble-wall velocity $v_w$ is not fixed by the effective potential alone.
As the wall passes through the plasma, particle masses change and the plasma exerts friction on the wall.
Determining $v_w$ requires a microscopic transport calculation together with hydrodynamic matching across the wall.
Such a calculation is beyond the present EFT analysis.
We therefore keep $v_w$ as a conditional input in the gravitational-wave calculation, rather than presenting a value of $v_w$ as a prediction.

We first compare the driving pressure with a leading-order estimate of the friction to test whether the wall can keep accelerating~\cite{Bodeker:2009qy}.
In the high-temperature approximation, the corresponding friction pressure is estimated as
\begin{equation}
 P_{\rm LO}(T)=\frac{T^2}{24}\sum_{b}n_b\,\Delta m_b^2
 +\frac{T^2}{48}\sum_{f}n_f\,\Delta m_f^2,
 \qquad
 \Delta m_i^2\equiv
 \max\!\left[m_i^2(\phi_t)-m_i^2(\phi_f),0\right].
 \label{eq:leading-order-friction}
\end{equation}
The bosonic sum includes the radial mode, three Goldstones, 20 $\Phi$ spectators, 104 Higgs-like spectators, and six transverse gauge modes.
The fermionic sum includes $t,b,\tau$ with multiplicities $12,12,4$, respectively.
The gauge contribution counts two transverse polarizations per vector in this friction estimate.
The factor of two difference between bosons and fermions follows from their thermal statistics.
The force balance involves the pressure difference $\Delta\mathcal F(T)$ that drives the wall.
A leading-order runaway is possible only if the driving pressure exceeds this friction estimate,
\begin{equation}
 \Delta\mathcal F(T)>P_{\rm LO}(T).
 \label{eq:runaway-condition}
\end{equation}

At the nucleation temperature, we obtain
\begin{equation}
 P_{\rm LO}=4.177\times10^5~\mathrm{TeV}^4,
 \qquad
 \Delta\mathcal F=2.427\times10^5~\mathrm{TeV}^4,
 \qquad
 \frac{P_{\rm LO}}{\Delta\mathcal F}=1.721.
 \label{eq:runaway-diagnostic-result}
\end{equation}
The leading-order friction is therefore larger than the driving pressure, so the present point does not support a runaway wall at this level of analysis.
Determining the terminal speed and the associated fluid flow requires a transport calculation, including the departure of the plasma from thermal equilibrium and higher-order friction effects.

The choice $v_w=1$ used above represents the limiting case of a wall moving close to the speed of light.
In the gravitational-wave section we show explicitly how the signal changes with $v_w$, and keep that uncertainty separate from the potential, bounce, and percolation calculations.

\subsection{Numerical checks and scope of the result}
\label{subsec:numerical-checks}

Several checks support the transition history obtained above. 
First, we locate both minima of the thermal potential and solve the bounce at temperatures on either side of the nucleation crossing.
Second, the transition is followed beyond nucleation:
the false-vacuum probability reaches the percolation threshold,
the physical false-vacuum volume is decreasing, and the false-phase fraction falls to one percent.
Third, Table~\ref{tab:transverse-stability} shows that
the spectator directions remain locally stable along the radial path at the temperatures relevant for tunneling and completion.
Within the reduced EFT, these checks give a consistent single-field transition history from $T_c$ to
$T_{\rm reh}$.

\begin{table}[h]
  \centering
  \caption{Transverse-stability check for the spectrum-informed benchmark.
  The entries are the smallest spectator masses along the radial interval from the false vacuum to the broken minimum at each temperature.}
  \label{tab:transverse-stability}
  \begin{tabular}{c c c c}
    \hline
    temperature & $T$ [TeV] & $\min m_{\Phi,\rm spec}^{\rm eff}$ [TeV] & $\min m_{H,\rm spec}^{\rm eff}$ [TeV] \\
    \hline
    $0.999T_c$ & $13.865$ & $11.298$ & $7.922$ \\
    $T_n$ & $6.728$ & $9.089$ & $4.388$ \\
    $T_p$ & $6.397$ & $9.015$ & $4.239$ \\
    $T_{\rm completion}$ & $6.375$ & $9.011$ & $4.229$ \\
    \hline
  \end{tabular}
\end{table}

We have also tested the sensitivity of this history to two numerical inputs that are unavoidable in a one-loop calculation.
Varying the renormalization scale over
\begin{equation}
 30~\mathrm{TeV}\leq Q\leq150~\mathrm{TeV}
 \label{eq:scale-variation-range}
\end{equation}
leaves all sampled points nucleating, percolating, and completing.
Over this range, the sound wave peak evaluated at $v_w=1$ varies from
\begin{equation}
 9.8\times10^{-12}\ \lesssim\ h^2\Omega_{\rm sw,peak}\
 \lesssim\ 1.6\times10^{-11}.
 \label{eq:scale-variation-result}
\end{equation}
We vary each of $(c_\phi^{\rm base},c_\Phi,c_H,c_A)$ individually by $\pm30\%$, and also vary all four together, giving eight individual and two joint variations.
The portal contribution to $c_\phi^{\rm ring}$ is held fixed.
All ten variations yield a completed transition; their ranges are listed separately in Table~\ref{tab:index678-uncertainty}.
The combined peak range is
\begin{equation}
 1.0\times10^{-11}\ \lesssim\ h^2\Omega_{\rm sw,peak}\
 \lesssim\ 1.5\times10^{-11}.
 \label{eq:screening-variation-result}
\end{equation}
Across these tests, the transition continues to complete and the predicted peak remains of order $10^{-11}$.

These results describe the one-loop EFT in Landau gauge with the diagonal spectator ansatz.
The positive transverse curvatures support local stability of the radial path within this ansatz.
Testing other tunneling paths requires a scalar potential that includes the corresponding mixing interactions.
The scale and screening tests measure the response to the specified variations, while the wall speed is varied separately in the gravitational-wave calculation.

\section{Gravitational wave signal}
\label{sec:gw-signal}

The transition described in Section~\ref{sec:phase-dynamics} releases more energy than is present in the radiation background and completes at a temperature of a few TeV.
We now use the calculated transition parameters to obtain the stochastic gravitational-wave spectrum today.
The relevant inputs include the transition temperature, the released-energy fraction $\alpha$, the inverse duration $\beta/H$, the wall velocity $v_w$, and the energy and entropy degrees of freedom that control the cosmological redshift.

We calculate the reference spectra with the nucleation inputs $(T_n,\alpha_n,\beta/H_n)$ from Section~\ref{subsec:transition-parameters}.
The percolation values show how the released energy increases as the bubbles grow and overlap.
The leading-order wall-friction test favors energy transfer to the plasma.
We therefore calculate the signal from sound waves and turbulence, with sound waves providing the dominant contribution.

The wall velocity has a substantial effect on the signal.
We show the spectrum for several values of $v_w$ and use $v_w=1$ for the ultrarelativistic limit.

\subsection{Source parameters and cosmological redshift}
\label{subsec:gw-inputs}

For the reference spectra, we take the source parameters at the nucleation temperature,
\begin{equation}
 T_*=T_n=6.728~\mathrm{TeV},
 \qquad
 \alpha_*=\alpha_n=1.9013,
 \qquad
 \frac{\beta}{H_*}=410.4 ,
 \label{eq:gw-reference-inputs}
\end{equation}
as convention in the numerical spectrum analysis.
This provides a well-defined reference point at which the tunneling rate turns out to be  efficient.
The actual fluid source develops as the bubbles grow and overlap, so we also keep in mind the nearby percolation values
\begin{equation}
 T_p=6.397~\mathrm{TeV},
 \qquad \alpha_p=2.2543.
 \label{eq:gw-percolation-inputs}
\end{equation}
The difference between $T_n$ and $T_p$ is about 5\%.
We use the nucleation values as common inputs for the spectral fits; Eq.~\eqref{eq:gw-percolation-inputs} shows their subsequent evolution as the fluid source develops.

The symbol $T_*$ denotes the temperature used to evaluate the source parameters and the cosmological redshift.
Here we set $T_*=T_n$.
The redshift depends on two different effective degrees of freedom, whose values at $T_*$ are
\begin{equation}
 g_{\rho,*}=235.67,
 \qquad
 g_{s,*}=234.31.
 \label{eq:gw-gstar-inputs}
\end{equation}
For a specified temperature, $g_\rho$ fixes the radiation energy density,
\begin{equation}
 \rho_R(T)=\frac{\pi^2}{30}g_\rho(T)T^4,
 \qquad
 H_{\rm rad}^2(T)=\frac{8\pi G}{3}\rho_R(T).
 \label{eq:gw-radiation-hubble}
\end{equation}
It therefore fixes the radiation contribution to the Hubble rate and enters the redshift normalization of the gravitational wave energy density.
The full Hubble rate used in Section~\ref{sec:phase-dynamics} also includes the released energy.
The entropy degrees of freedom $g_s$ do not fix $a$ or $T$ separately;
instead, they fix their relation during  the adiabatic expansion.
Entropy conservation gives
\begin{equation}
 \frac{a_*}{a_0}=
 \frac{T_0}{T_*}
 \left(\frac{g_{s,0}}{g_{s,*}}\right)^{1/3}.
 \label{eq:gw-redshift}
\end{equation}
We calculate $g_\rho$ and $g_s$ separately because spectators with masses comparable to the temperature contribute differently to the energy and entropy densities.

For example, the present-day sound wave peak frequency used below is
\begin{equation}
 f_{\rm sw,0}^{\rm peak}=1.9\times10^{-5}~\mathrm{Hz}\,
 \frac{1}{v_w}\frac{\beta}{H_*}
 \left(\frac{T_*}{100~\mathrm{GeV}}\right)
 \left(\frac{g_{\rho,*}}{100}\right)^{1/2}
 \left(\frac{100}{g_{s,*}}\right)^{1/3}.
 \label{eq:sound-wave-peak-frequency}
\end{equation}
For $v_w=1$, this gives
\begin{equation}
 f_{\rm sw,0}^{\rm peak}=0.606~\mathrm{Hz}.
 \label{eq:reference-sound-peak-frequency}
\end{equation}
The inverse dependence on $v_w$ means that a slower wall shifts the spectrum to a higher frequency, as well as changing its amplitude.

\subsection{Energy transfer and gravitational-wave sources}
\label{subsec:gw-sources}

The expanding bubbles transfer energy to the surrounding plasma.
The resulting fluid motion and magnetohydrodynamic turbulence generate gravitational waves.
We therefore write the present signal as
\begin{equation}
 \Omega_{\rm GW}(f)=\Omega_{\rm sw}(f)+\Omega_{\rm turb}(f).
 \label{eq:gw-source-sum}
\end{equation}
Scalar-field collisions can produce a sizeable signal when the walls retain a large fraction of the released energy until they collide.
Our leading-order friction estimate instead favors energy transfer to the plasma.
We therefore retain the two plasma sources in Eq.~\eqref{eq:gw-source-sum} and neglect the expected smaller scalar-field contribution.

The fraction of released energy converted into bulk fluid motion is described by the sound wave efficiency factor $\kappa_{\rm sw}$.
For the fast, non-runaway reference treatment, we use the standard fit~\cite{Espinosa:2010hh,Huang:2020bbe}
\begin{equation}
 \kappa_{\rm sw}(\alpha)=
 \frac{\alpha}{0.73+0.083\sqrt{\alpha}+\alpha}.
 \label{eq:kappa-sound}
\end{equation}
This is a hydrodynamic fit, not a separate prediction of the scalar
potential.
It tells us what fraction of the available energy is put into coherent fluid motion rather than being immediately thermalized.
At the reference point, we have
\begin{equation}
 \kappa_{\rm sw}=0.692.
 \label{eq:kappa-sound-result}
\end{equation}
We estimate the energy in turbulence by taking
\begin{equation}
 \kappa_{\rm turb}=\epsilon_{\rm turb}\kappa_{\rm sw},
 \qquad \epsilon_{\rm turb}=0.05.
 \label{eq:kappa-turbulence}
\end{equation}
This choice takes the conservative lower end of the commonly used $\epsilon_{\rm turb}\simeq0.05$--$0.10$ range~\cite{Caprini:2019egz}.
In ~\cite{Caprini:2015zlo}, $\epsilon_{\rm turb}=0.05$ is used directly in its eLISA phase-transition forecasts.
The conversion from coherent flow to turbulence depends on the plasma dynamics and magnetic fields, and is not fixed by the present EFT.
We therefore retain the turbulence contribution as a sub-leading component rather than using it to drive the main conclusion.

The corresponding MHD turbulence spectrum today is~\cite{Caprini:2015zlo,Caprini:2019egz}
\begin{equation}
 h^2\Omega_{\rm turb}(f)=3.35\times10^{-4}
 \frac{v_w}{\beta/H_*}
 \left(\frac{\kappa_{\rm turb}\alpha_*}{1+\alpha_*}\right)^{3/2}
 \left(\frac{g_{\rho,*}}{100}\right)
 \left(\frac{100}{g_{s,*}}\right)^{4/3}
 S_{\rm turb}(f),
 \label{eq:turbulence-spectrum}
\end{equation}
where
\begin{equation}
 S_{\rm turb}(f)=
 \frac{x^3}{(1+x)^{11/3}\left(1+8\pi f/h_{*,0}\right)},
 \qquad
 x\equiv\frac{f}{f_{\rm turb,0}^{\rm peak}}.
 \label{eq:turbulence-shape}
\end{equation}
This fit uses the characteristic turbulence frequency and the redshifted Hubble frequency,
\begin{align}
 f_{\rm turb,0}^{\rm peak}={}&2.7\times10^{-5}\,\mathrm{Hz}\,
 \frac{\beta/H_*}{v_w}
 \left(\frac{T_*}{100\,\mathrm{GeV}}\right)
 \left(\frac{g_{\rho,*}}{100}\right)^{1/2}
 \left(\frac{100}{g_{s,*}}\right)^{1/3},
 \label{eq:turbulence-peak-frequency}\\
 h_{*,0}={}&1.65\times10^{-5}\,\mathrm{Hz}\,
 \left(\frac{T_*}{100\,\mathrm{GeV}}\right)
 \left(\frac{g_{\rho,*}}{100}\right)^{1/2}
 \left(\frac{100}{g_{s,*}}\right)^{1/3}.
 \label{eq:redshifted-hubble-frequency}
\end{align}
Here $h_{*,0}$ is a frequency, not the reduced Hubble constant.
The last factor in \eqref{eq:turbulence-shape} suppresses the high-frequency tail once the finite source duration is taken into account.
Distinguishing $g_{\rho,*}$ from $g_{s,*}$ gives the energy and entropy redshifts used throughout this work;
the usual $(100/g_*)^{1/3}$ dependence is recovered when they are set equal.

For the fast-wall reference point, the above prescription gives $\kappa_{\rm turb}=0.0346$, $f_{\rm turb,0}^{\rm peak}=0.862\,\mathrm{Hz}$, and $h_{*,0}=1.28\times10^{-3}\,\mathrm{Hz}$.
The numerical maximum of the turbulence curve is $h^2\Omega_{\rm turb}=1.00\times10^{-14}$ at $f\simeq1.03\,\mathrm{Hz}$.
At the sound-wave peak it changes the total signal by only $0.07\%$.
Sound waves therefore determine the peak of the total spectrum, while turbulence contributes mainly to its high-frequency tail.

The acoustic source lasts until the fluid motion develops shocks and loses coherence, which can happen within a Hubble time.
We account for this by multiplying the sound wave spectrum by the finite-lifetime factor
\cite{Ellis:2018mja},
\begin{equation}
 \Upsilon_{\rm sw}=\min\!\left(1,\frac{H_*R_*}{\bar U_f}\right),
 \qquad
 \bar U_f^2=\frac34\frac{\kappa_{\rm sw}\alpha_*}{1+\alpha_*}.
 \label{eq:sound-wave-lifetime}
\end{equation}
Here $\bar U_f$ is the enthalpy-weighted root-mean-square fluid
four-velocity generated by the transition.
More generally, it can be defined by
\begin{equation}
 \bar U_f^2\equiv \frac{1}{\bar w\,\mathcal V}
 \int_{\mathcal V}d^3x\,w(\mathbf{x})\,\gamma^2(\mathbf{x})v^2(\mathbf{x}),
 \qquad w\equiv e+p,
 \label{eq:rms-fluid-four-velocity}
\end{equation}
where $w$ is the local plasma enthalpy density, $v$ is the local three-velocity, and $\gamma=(1-v^2)^{-1/2}$.
The second relation in~\eqref{eq:sound-wave-lifetime} is the standard hydrodynamic estimate of this quantity in terms of the fraction of released energy transferred to bulk fluid motion.
It is this fluid velocity that determines how quickly the acoustic flow steepens into shocks.
Here $R_*$ is the characteristic bubble separation when bubbles begin to overlap; it is not the critical radius of the thermal bounce.
For a short transition with an approximately exponential nucleation rate and nearly constant $v_w$, the average bubble number density $n_b$ gives
\begin{equation}
 R_*\equiv n_b^{-1/3}\simeq(8\pi)^{1/3}\frac{v_w}{\beta},
 \qquad
 H_*R_*\simeq(8\pi)^{1/3}\frac{v_w}{\beta/H_*}.
 \label{eq:mean-bubble-separation}
\end{equation}
The numerical factor $(8\pi)^{1/3}\simeq2.93$ is an order-one consequence of the exponential nucleation estimate;
other definitions of a characteristic bubble size differ by comparable order-one factors.
The ratio $R_*/\bar U_f$ estimates the shock-formation time, which limits the duration of the coherent acoustic source.
For $v_w=1$, the reference point gives
\begin{equation}
 \bar U_f=0.583,
 \qquad
 \Upsilon_{\rm sw}=1.22\times10^{-2}.
 \label{eq:sound-wave-lifetime-result}
\end{equation}
The small value of $\Upsilon_{\rm sw}$ means that the sound wave source is short-lived compared with a Hubble time.
This suppression is included for obtaining a conservative estimate of the signal strength.

\subsection{Spectrum and wall-speed dependence}
\label{subsec:gw-spectrum}

With the source parameters fixed, the sound wave peak amplitude can be evaluated as~\cite{Caprini:2015zlo}
\begin{equation}
 h^2\Omega_{\rm sw,peak}=2.65\times10^{-6}
 \frac{1}{\beta/H_*}
 \left(\frac{\kappa_{\rm sw}\alpha_*}{1+\alpha_*}\right)^2
 \left(\frac{g_{\rho,*}}{100}\right)
 \left(\frac{100}{g_{s,*}}\right)^{4/3}
 v_w\,\Upsilon_{\rm sw}.
 \label{eq:sound-wave-peak-amplitude}
\end{equation}
The corresponding spectral shape is
\begin{equation}
 h^2\Omega_{\rm sw}(f)=h^2\Omega_{\rm sw,peak}
 \left(\frac{f}{f_{\rm sw,0}}\right)^3
 \left[\frac{7}{4+3(f/f_{\rm sw,0})^2}\right]^{7/2}.
 \label{eq:sound-wave-spectrum}
\end{equation}
The turbulence contribution is evaluated utilizing~\eqref{eq:turbulence-spectrum}--\eqref{eq:redshifted-hubble-frequency}.
With the energy fraction in Eq.~\eqref{eq:kappa-turbulence}, this component makes a small correction near the sound wave peak.

\begin{figure}[h]
 \centering
 \includegraphics[width=0.82\textwidth]{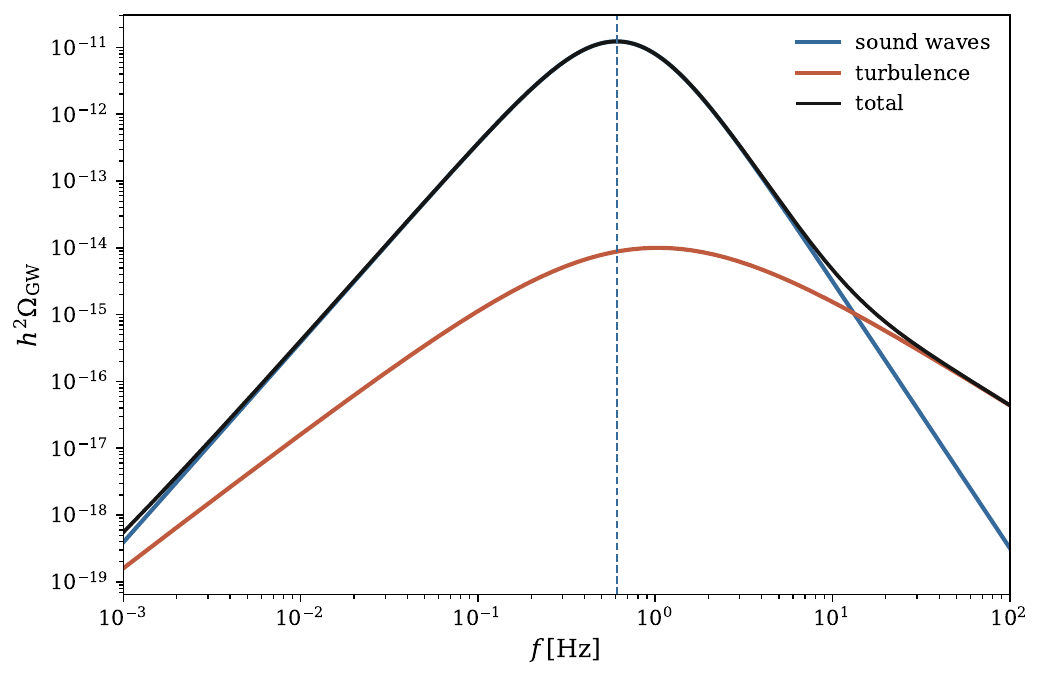}
 \caption{Gravitational-wave spectrum in the $v_w\to1$ limit.
 The sound wave  component includes the finite-lifetime suppression in \eqref{eq:sound-wave-lifetime}.
 Turbulence is sub-leading near the peak but gives a broader high-frequency tail.
 The black line represents the sum of the two contributions.}
 \label{fig:reference-gw-spectrum}
\end{figure}

For the fast-wall reference choice $v_w=1$, the sound wave component peaks at
\begin{equation}
 f_{\rm sw,0}^{\rm peak}=0.606~\mathrm{Hz},
 \qquad
 h^2\Omega_{\rm sw,peak}=1.232\times10^{-11}.
 \label{eq:reference-gw-peak}
\end{equation}
Incorporating the turbulence component, the total spectrum peaks at
\begin{equation}
 f_{\rm peak}=0.611~\mathrm{Hz},
 \qquad
 h^2\Omega_{\rm GW,peak}=1.232\times10^{-11}.
 \label{eq:reference-total-gw-peak}
\end{equation}
Figure~\ref{fig:reference-gw-spectrum} shows all the gravitational wave components.
The sound wave curve includes the finite acoustic lifetime.
This reduces its peak by about two orders of magnitude relative to the estimate for a source lasting a full Hubble time.

\begin{figure}[h]
 \centering
 \includegraphics[width=0.94\textwidth]{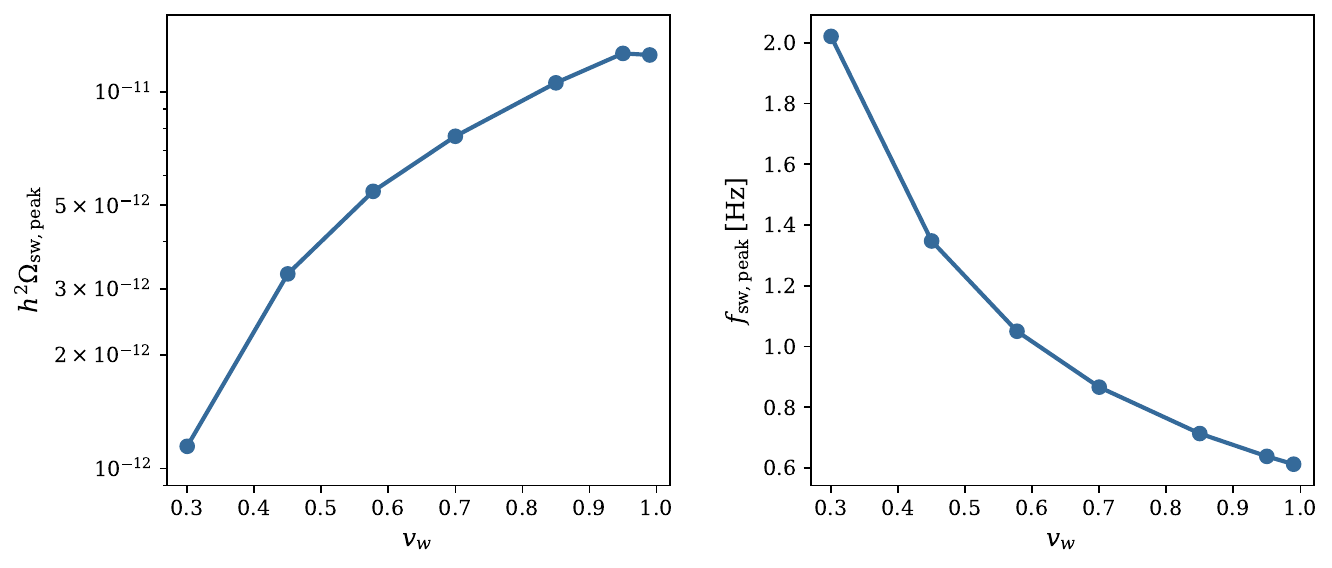}
 \caption{Conditional dependence of the sound wave peak on the wall velocity.
 The efficiency factors in this scan are taken from hydrodynamic fits for the indicated flow regimes.
 Since the microscopic wall friction has not been calculated, the curves show the range of possible signals rather than a fitted value of $v_w$.}
 \label{fig:wall-speed-dependence}
\end{figure}

The wall speed sets the distance a bubble expands before collision and affects the fraction of energy transferred to fluid motion.
To display this uncertainty, we evaluate a set of conditional hydrodynamic wall-speed scenarios.
Across the range $0.3\leq v_w\leq0.99$, the peak of the sound wave varies from
\begin{equation}
 1.14\times10^{-12}\ \lesssim\ h^2\Omega_{\rm sw,peak}
 \lesssim\ 1.27\times10^{-11},
 \qquad
 0.612~\mathrm{Hz}\ \lesssim\ f_{\rm sw,peak}
 \lesssim\ 2.02~\mathrm{Hz}.
 \label{eq:wall-speed-envelope}
\end{equation}
The larger amplitude at high $v_w$ is mainly due to the longer acoustic lifetime, while the peak moves to lower frequency as $f_{\rm sw,0}^{\rm peak}\propto v_w^{-1}$.
Figure~\ref{fig:wall-speed-dependence} shows the signal predicted for each assumed wall speed.

\subsection{Detectability and experimental reach}
\label{subsec:gw-detectability}

\begin{figure}[h]
 \centering
\includegraphics[width=0.92\textwidth]{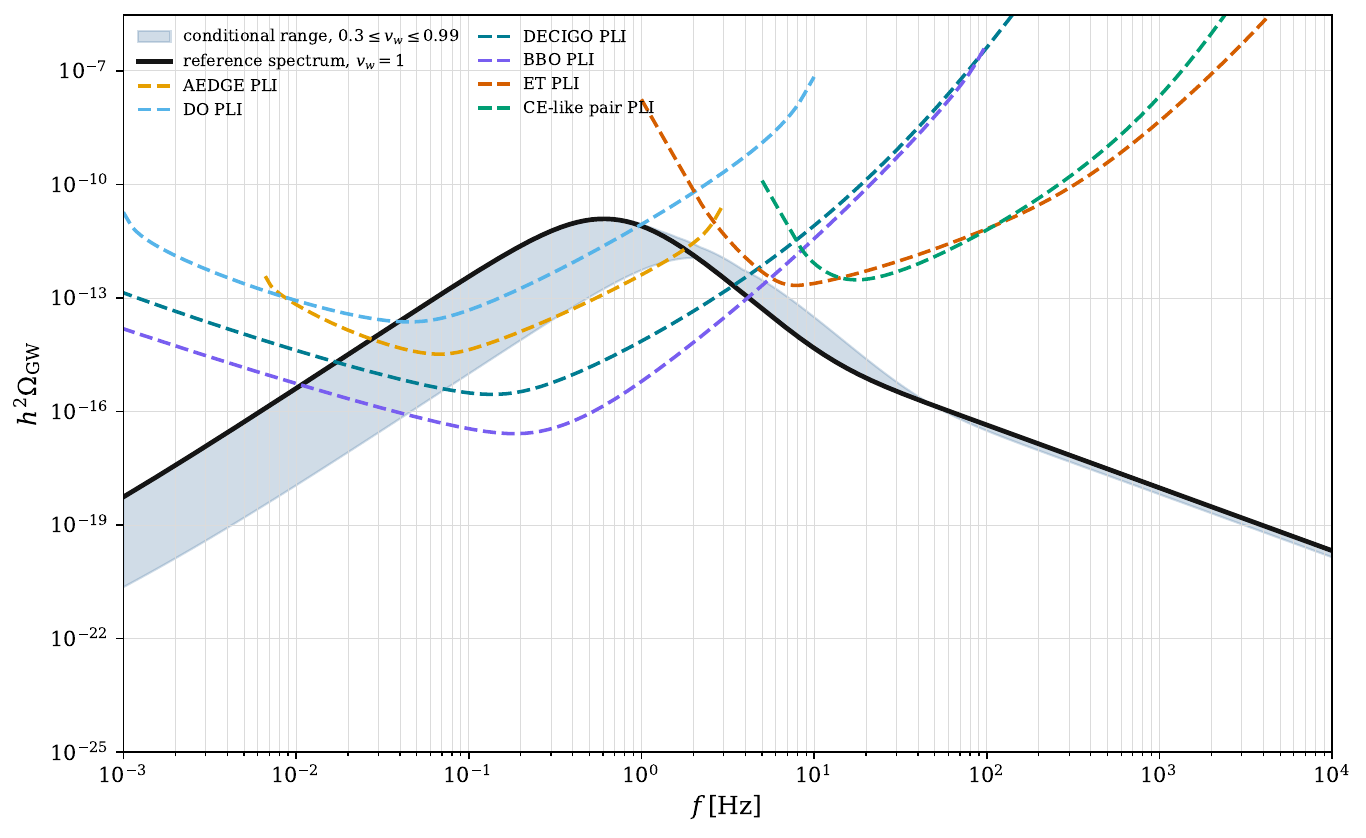}
 \caption{Comparison of the predicted gravitational-wave spectrum with noise-only PLI sensitivities for future detector concepts.
 The black line shows the total spectrum in the $v_w\to1$ limit, and the blue band covers the spectra calculated for $0.3\leq v_w\leq0.99$.
 The dashed detector curves follow the common four-year, $\rho_{\rm thr}=10$ convention described in the text.
 They are not point-by-point exclusion limits and do not include foreground residuals or ground-network geometry.
 The comparison shows that the predicted peak lies in the deci-Hz band targeted by AEDGE, DO, DECIGO, and BBO, rather than in the primary frequency range of ET or CE.}
 \label{fig:detector-comparison}
\end{figure}

The reference spectrum has a peak in the deci-Hz band rather than in the main frequency band of ground-based interferometers.
This follows directly from the few-TeV transition temperature together with the moderately large value of $\beta/H_*$ in \eqref{eq:gw-reference-inputs}.
DECIGO and BBO are designed to probe this frequency range and are therefore especially relevant to the predicted signal.
Figure~\ref{fig:detector-comparison} overlays the reference spectrum and the conditional wall-speed range with power-law-integrated (PLI) sensitivities for the Atomic Experiment for Dark Matter and Gravity Exploration (AEDGE), the Deci-Hertz Observatory
(DO), DECIGO, BBO, the Einstein Telescope (ET), and a CE-like pair.
The underlying strain-noise curves are taken from the public SGWBProbe data release associated with~\cite{Campeti:2020xwn};
the mission concepts and their design sensitivities are described in~\cite{AEDGE:2019nxb,Sedda:2019uro,Yagi:2011wg,Crowder:2005nr,Punturo:2010zz,Reitze:2019iox};
see~\cite{Thrane:2013oya,Schmitz:2020syl} for the use of PLI curves in stochastic-background comparisons.

A PLI curve is a convenient way of displaying a stochastic background sensitivity for specified observing assumptions;
it is not a detector noise curve itself or a point-by-point exclusion boundary.
We use a common observation time $T_{\rm obs}=4$ yr and define the PLI threshold by $\rho_{\rm thr}=10$.
The construction begins with a \emph{trial} power-law
spectrum,
\begin{equation}
 h^2\Omega_n(f)=A_n
 \left(\frac{f}{f_{\rm ref}}\right)^n,
 \qquad f_{\rm ref}=1~\mathrm{Hz}.
 \label{eq:pli-trial-power-law}
\end{equation}
Here $n$ is the trial spectral index, and $A_n$ is the amplitude at the reference frequency.
For each $n$, we determine the amplitude required to reach the chosen SNR threshold.

Using the strain convention of the SGWBProbe detector inputs, Eq.~(4.11) of~\cite{Campeti:2020xwn} gives
\begin{equation}
 S_{h,n}^{\rm sig}(f)=
 \frac{3H_0^2}{4\pi^2h^2f^3}
 A_n\left(\frac{f}{f_{\rm ref}}\right)^n.
 \label{eq:pli-trial-strain}
\end{equation}
Substituting \eqref{eq:pli-trial-strain} into the noise-only SNR yields
\begin{equation}
 \rho_D^2=n_D T_{\rm obs}\int {\rm d}f\,
\left[\frac{S_{h,n}^{\rm sig}(f)}
 {S_{h,D}^{\rm noise}(f)}\right]^2
 =A_n^2\,\mathcal{I}_D(n),
 \label{eq:pli-snr}
\end{equation}
where \(n_D\) is the effective number of independent detector channels defined in Appendix~\ref{app:detector-forecast}.
In this notation,
\begin{equation}
 \mathcal{I}_D(n)=n_D T_{\rm obs}\int {\rm d}f\,
 \left[\frac{3H_0^2}{4\pi^2h^2f^3}
 \frac{(f/f_{\rm ref})^n}{S_{h,D}^{\rm noise}(f)}\right]^2.
 \label{eq:pli-integral}
\end{equation}
The important point is that the SNR is proportional to $A_n$: once the slope and detector are fixed, the integral $\mathcal{I}_D(n)$ is just a number.
Requiring $\rho_D=\rho_{\rm thr}$ therefore fixes the threshold amplitude directly,
\begin{equation}
 A_n^{\rm thr}=
 \frac{\rho_{\rm thr}}{\sqrt{\mathcal{I}_D(n)}}.
 \label{eq:pli-amplitude}
\end{equation}
We repeat the calculation for a range of slopes and take the largest threshold power law at each frequency,
\begin{equation}
 h^2\Omega_D^{\rm PLI}(f)=
 \max_n\left[A_n^{\rm thr}
 \left(\frac{f}{f_{\rm ref}}\right)^n\right],
 \label{eq:pli-envelope}
\end{equation}
following the power-law-integrated construction of~\cite{Thrane:2013oya}.
This procedure gives smooth curves directly from the tabulated $S_h(f)$ data rather than from graphical anchor points.
For the actual phase-transition spectrum, we do not replace it by a power law:
we insert the calculated $h^2\Omega_{\rm GW}(f)$ directly into \eqref{eq:pli-snr} and quote the resulting SNR in Table~\ref{tab:pli-snr}.
The factor $n_D$ specifies the effective number of independent detector correlations used in this comparison.

The detector inputs, the meaning of $n_D$, and the limitations of this noise-only comparison are collected in Appendix~\ref{app:detector-forecast}.
In particular, the CE entry is an idealized CE-like pair rather than a site-specific detector network, and astrophysical foreground residuals and ground-network overlap-reduction functions are not included.

Within this deliberately limited comparison, the signal is well placed for deci-Hz observatories, but the reach is not identical for every concept.
The direct integration in \eqref{eq:pli-snr} gives the noise-only SNRs listed in Table~\ref{tab:pli-snr}.
DECIGO and BBO remain far above the reference threshold throughout the displayed wall-speed range.
AEDGE reaches the threshold for the fast-wall part of the range but is marginal for the slowest wall case, while the conservative DO configuration covers only part of the wall-speed envelope.
By contrast, ET and the CE-like pair become useful only at frequencies of several Hz or above, so their displayed curves do not probe the peak of the present reference signal.
This does not mean that ground-based experiments are unimportant in general:
a transition at a higher temperature or with a much larger $\beta/H_*$ would move its peak toward their frequency range.

\begin{table}[h]
 \centering
 \caption{Noise-only SNRs from Eq.~\eqref{eq:pli-snr}.
 The reference column uses $v_w=1$, while the final column spans the discrete conditional wall speed scan $0.3\leq v_w\leq0.99$.
 These values inherit the approximations stated in the text and are not mission-specific discovery forecasts.
 The value at \(v_w=1\) is not part of the scanned range; slower walls shift the peak to higher frequencies, closer to the ET band, hence the larger SNRs.
 The large SNRs for DECIGO and BBO reflect the idealized noise-only configuration and should not be interpreted as mission-level discovery significance.}
 \label{tab:pli-snr}
 \small
 \begin{tabular}{lcc}
  \hline
  Detector configuration & $\rho(v_w=1)$ & conditional $\rho$ range \\
  \hline
  AEDGE & $1.28\times10^{3}$ & $9.20$--$1.27\times10^{3}$ \\
  DO (conservative) & $9.72\times10^{1}$ & $5.03\times10^{-1}$--$9.66\times10^{1}$ \\
  DECIGO & $8.42\times10^{4}$ & $6.01\times10^{2}$--$8.41\times10^{4}$ \\
  BBO & $1.50\times10^{6}$ & $1.11\times10^{4}$--$1.50\times10^{6}$ \\
  ET & $1.07$ & $1.13$--$5.89$ \\
  CE-like pair & $6.30\times10^{-2}$ & $6.50\times10^{-2}$--$3.78\times10^{-1}$ \\
  \hline
 \end{tabular}
\end{table}

The leading numerical variations are summarized in Table~\ref{tab:index678-uncertainty}.
Varying the renormalization scale over $30{~\rm TeV}\leq Q\leq150{~\rm TeV}$ changes $T_n$, $\alpha_n$, and $\beta/H_n$ by no more than about $7\%$, $12\%$, and $7\%$, respectively.
The coherent $\pm30\%$ thermal screening stress test has a larger effect on the released energy, changing $\alpha_n$ by roughly $-23\%$ to $+27\%$, while retaining a completed transition and a sound wave peak near $10^{-11}$. 
These ranges show how the one-loop results change when the renormalization scale and thermal screening inputs are varied.
The wall speed is a separate macroscopic uncertainty.
It does not change the potential, bounce action, or transition parameters in the present treatment, but it changes the fluid efficiency, acoustic lifetime, and hence the amplitude of gravitational wave by about one order of magnitude.

\begin{table}[h]
 \centering
 \caption{Sensitivity of the transition at the reference point and gravitational-wave prediction to the leading numerical assumptions.
 The first row gives the result at $Q=65$ TeV with the thermal masses calculated from the retained spectrum and $v_w=1$.
 The $Q$ range is $30$--$150$ TeV.
 The screening tests multiply $(c_\phi^{\rm base},c_\Phi,c_H,c_A)$ by $0.7$ or $1.3$, one at a time (individual) or all together (joint).
 The portal contribution to $c_\phi^{\rm ring}$ is fixed.
 The wall-speed scan changes only the hydrodynamic gravitational-wave source, so the phase transition quantities in that row are fixed by construction.
 The quoted peak is the sound wave peak;
 the turbulence correction to the total peak is sub-percent for this point.}
 \label{tab:index678-uncertainty}
 \small
 \renewcommand{\arraystretch}{1.2}
 \begin{tabular}{lcccc}
  \hline
  Variation & $T_n$ [TeV] & $\alpha_n$ & $\beta/H_n$ & $h^2\Omega_{\rm peak}$ \\
  \hline
  Reference & 6.728 & 1.901 & 410 & $1.232\times10^{-11}$ \\
  $Q=30$--$150$ TeV & 6.486--6.978 & 1.703--2.131 & 381--436 &
     $(0.977$--$1.583)\times10^{-11}$ \\
  Individual $c_i$ & 6.406--7.083 & 1.562--2.295 & 404--414 &
  $(1.041$--$1.436)\times10^{-11}$ \\
  Joint $c_i$ & 6.328--7.188 & 1.473--2.412 & 398--417 &
     $(1.003$--$1.475)\times10^{-11}$ \\
  $0.3\leq v_w\leq0.99$ & fixed & fixed & fixed &
     $1.144\times10^{-12}$--$1.265\times10^{-11}$ \\
  \hline
 \end{tabular}
\end{table}

With the finite sound wave lifetime included, the predicted signal remains within the noise-only reach of DECIGO and BBO across the wall speeds studied here.
This makes the transition a promising target for deci-Hz observations.
A forecast for a specific mission would also account for astrophysical foregrounds, observing efficiency, and detector configuration, as discussed in Appendix~\ref{app:detector-forecast}.

\section{Discussion and conclusion}
\label{sec:conclusion}

We have studied the first-order phase transition in a non-supersymmetric EFT
motivated by a four-stack intersecting D6-brane Pati-Salam construction.
The construction fixes the gauge group, scalar representations, and
multiplicities of the Higgs-like and transition-sector states.
Below $M_{\rm SUSY}=300$~TeV, the scalar masses and quartic couplings are
treated as renormalized EFT parameters.
This gives a thermal theory that retains the spectrum and gauge-coupling
matching of the ultraviolet construction while remaining entirely
non-supersymmetric during the transition.

For the representative spectrum, the thermal potential develops a broken
$SU(2)_{R_1}\times SU(2)_{R_2}\to SU(2)_R$ phase separated from the symmetric
phase by a barrier.  The diagonal breaking produces three heavy gauge bosons
with $m_A(\phi_c)\simeq42.76$~TeV, which sets the transition scale; the
low-mass rescaled points are constrained by dilepton searches, with the
benchmark points below about 5.2~TeV excluded in the aligned benchmark.  The critical temperature is
$T_c=13.879$~TeV with $\phi_c/T_c=6.190$.  A complete transition history
requires following the evolution beyond $T_c$: we find thermal nucleation at
$T_n=6.728$~TeV, percolation at $T_p=6.397$~TeV, and completion at
$T_{\rm completion}=6.375$~TeV.  The false-vacuum physical volume is
decreasing at percolation, and the reheating estimate,
$T_{\rm reh}=8.573$~TeV, remains below $T_c$.  With these checks, a complete
transition history is obtained.

This transition is strongly supercooled.  At nucleation we obtain
\begin{equation}
 \alpha_n=1.9013,\qquad \frac{\beta}{H_n}=410.4.
\end{equation}
The diagnostic of the leading-order friction gives
$P_{\rm LO}/\Delta\mathcal{F}=1.721$, so the adopted leading-order estimate does not support a runaway wall for this point.
We consequently treat sound waves as the leading source of the gravitational wave, include a
sub-leading turbulence contribution, and do not use a scalar-field collision
signal as a central source.  The finite sound wave lifetime is important:
for the fast-wall reference case it gives
$\Upsilon_{\rm sw}=1.22\times10^{-2}$, a suppression of about two orders of
magnitude relative to a source lasting a full Hubble time, which prevents an
overestimate of the acoustic signal.  The equation-of-state integrals use
bare spectator masses, a prescription retained in the screening scan.
At $T_*=T_n$ we have $g_{\rho,*}=235.67$ and $g_{s,*}=234.31$.

Using the nucleation-temperature convention for the source parameters, the
reference $v_w=1$ spectrum peaks at
\begin{equation}
 f_{\rm peak}=0.611~\mathrm{Hz},\qquad
 h^2\Omega_{\rm GW,peak}=1.232\times10^{-11}.
\end{equation}
The peak therefore lies in the deci-Hz band.  The scaling relations of
Section~\ref{subsec:global-rescaling} connect this benchmark to lower-mass
points.  Using $T_p$ instead of $T_n$ changes $\alpha$ and $f_{\rm peak}$ by
amounts comparable to the screening variations, without affecting the
deci-Hz peak.

The microscopic wall friction is not computed here, so $v_w$ is treated as a
free input.  Over the scanned range $0.3\leq v_w\leq0.99$, the sound wave
peak moves between $0.612$ and $2.02$~Hz and its amplitude ranges from
$1.14\times10^{-12}$ to $1.27\times10^{-11}$.  The scan uses the
corresponding hydrodynamic efficiency fits; the fast-wall formula is used
only for $v_w\to1$.

The detector comparison points to the same frequency range.  In the common
noise-only PLI convention adopted here, DECIGO and BBO have strong frequency
overlap with the whole wall-speed envelope.  AEDGE is sensitive to the
fast-wall portion of the range, with the lowest scanned value just below the
threshold $\rho_{\rm thr}=10$, while the conservative DO configuration probes
only part of it.  The peak lies at deci-Hz frequencies, below the main band
of ET and an idealized pair of Cosmic Explorer (CE)-like ground-based
instruments, so their noise-only SNRs are below threshold.  These are
noise-only comparisons, not mission-level discovery forecasts; the large
SNRs for DECIGO and BBO reflect the idealized configuration and should not be
interpreted as mission-level discovery significance.

We have tested two important numerical systematics of the one-loop
calculation.  Varying the renormalization scale over
$30~\mathrm{TeV}\leq Q\leq150~\mathrm{TeV}$ and varying the
spectrum-informed screening coefficients by $\pm30\%$ both preserve
nucleation, percolation, and completion.  The resulting sound wave peak
remains at the $10^{-11}$ level for $v_w=1$.  Together with the positive
transverse spectator curvatures along the tunneling path, these checks show
that the completed transition, its strong supercooling, and its deci-Hz
frequency scale do not arise from one isolated numerical prescription; the
scale and screening studies are robustness tests that leave the main
physical picture unchanged over the variations considered here.  The
transition history is obtained in a one-field reduction with a diagonal
spectator ansatz, and a full multi-field bounce with nonzero cross-quartics
remains to be done.  A residual gauge dependence remains, as discussed in
Section~\ref{sec:finite-temperature-potential}.

The calculation also points to the next steps for a more precise prediction.
A transport calculation of plasma friction and a full multi-field bounce
would determine the wall speed and test the radial tunnelling path.  For the
detector comparison, foreground subtraction, duty cycles, response
functions, and overlap reduction functions should be included for specified
networks; higher-order thermal corrections would improve the effective
potential and its gauge dependence.

The main result is that the EFT admits a completed, strongly supercooled
first-order transition with a stochastic gravitational-wave signal in the
deci-Hz band.  The particle content and gauge structure connect this class of
extended Pati-Salam theories to a cosmological probe that complements
collider and low-energy searches.  These steps will determine how accurately
the deci-Hz prediction can be tested.

\section*{Acknowledgements}
HH acknowledges support from the Doctoral Student Program of the Young Elite Scientists Sponsorship Program by CAST.
TL is supported in part by the National Key Research and Development Program of China Grant No. 2020YFC2201504, by the Projects No. 11875062, No. 11947302, No. 12047503, and No. 12275333 supported by the National Natural Science Foundation of China, by the Key Research Program of the Chinese Academy of Sciences, Grant No. XDPB15, by the Scientific Instrument Developing Project of the Chinese Academy of Sciences, Grant No. YJKYYQ20190049, by the International Partnership Program of Chinese Academy of Sciences for Grand Challenges, Grant No. 112311KYSB20210012, and by the Henan Province Outstanding Foreign Scientist Studio Project, No.GZS2025008. This work was supported by the High Performance Computing Platform of Henan Normal University. RS is supported by the Fundamental Research Funds for the Central Universities Grant No.~E4EQ0102X2.
 WZ is supported in part by the National Natural Science Foundation of China under Grant No. 12405120, Start-up Funds for Young Talents of Hebei University (No. 521100224226), and Funded by Science Research Project of Hebei Education Department No. QN2026608. 
\appendix

\section{Thermal Tadpole in the $O(N)$ Scalar Sector}
\label{app:thermal-tadpole}

For $N$ real scalars with
\begin{equation}
 V_{\rm self}=\frac{\lambda}{4}(\varphi_i\varphi_i)^2,
\end{equation}
the thermal part of the massless bosonic tadpole is
\begin{equation}
 I_B(T)=\int\frac{{\rm d}^3p}{(2\pi)^3}\frac{n_B(|\mathbf p|)}{|\mathbf p|}
 =\frac{T^2}{2\pi^2}\int_0^\infty {\rm d}x\,\frac{x}{e^x-1}
 =\frac{T^2}{12},
 \label{eq:bosonic-tadpole}
\end{equation}
where $n_B(E)=(e^{E/T}-1)^{-1}$ and $\int_0^\infty dx\,x/(e^x-1)=\pi^2/6$.
For finite mass, the integrand is
$n_B(\sqrt{p^2+m^2})/\sqrt{p^2+m^2}$; Eq.~\eqref{eq:bosonic-tadpole} is its high-temperature limit~\cite{Dolan:1973qd,Carrington:1991hz}.

The quartic vertex is
\begin{equation}
 \lambda_{abcd}=2\lambda\left(
 \delta_{ab}\delta_{cd}+\delta_{ac}\delta_{bd}
 +\delta_{ad}\delta_{bc}\right).
\end{equation}
The tadpole self-energy is
\begin{equation}
 \Pi^{\rm self}_{ab}=\frac12\lambda_{abcc}I_B(T).
\end{equation}
Note that only the repeated internal index $c$ is summed.
Performing this sum gives
\begin{equation}
 \lambda_{abcc}=2\lambda(N+2)\delta_{ab},
 \qquad
 \Pi^{\rm self}_{ab}=\delta_{ab}\frac{N+2}{12}\lambda T^2.
\end{equation}
The indices $a,b$ label the external scalar components and remain free; the factor $\delta_{ab}$ shows that the thermal mass matrix is diagonal and degenerate in an $O(N)$-symmetric scalar sector.
For the $O(4)$ sector used in the main text, $\lambda=\lambda_\phi$ and $N=4$, yielding $c_\phi^{\rm self}=\lambda_\phi/2$.

\section{Detector forecast conventions}
\label{app:detector-forecast}

This appendix records the conventions used to produce the detector comparison in Fig.~\ref{fig:detector-comparison} and the SNR values in Table \ref{tab:pli-snr}.
Its purpose is reproducibility rather than a mission-specific sensitivity study.
In particular, the calculation uses instrumental strain-noise curves and a common observing convention, but does not model astrophysical foreground residuals or the geometry of a particular
ground-based detector network.

\subsection{Signal and noise conventions}

The phase-transition calculation gives the spectrum $h^2\Omega_{\rm GW}(f)$ today.
To compare it with a strain-noise power spectral density, we use
\begin{equation}
 S_h^{\rm sig}(f)=\frac{3H_0^2}{4\pi^2h^2f^3}
 h^2\Omega_{\rm GW}(f),
 \label{eq:app-omega-to-strain}
\end{equation}
where $h=0.6736$ and $H_0=100h~\mathrm{km\,s^{-1}\,Mpc^{-1}}$.
We use the polarization convention of Eqs.~(4.5) and (4.11) of~\cite{Campeti:2020xwn}, which is also used in the SGWBProbe data.
Here $S_h^{\rm sig}$ is the strain spectrum per polarization for an unpolarized background.
Summing the two polarizations gives $S_{h,\rm sum}=2S_h^{\rm sig}=3H_0^2\Omega_{\rm GW}/(2\pi^2f^3)$.
The signal and effective strain-noise curves in Eq.~\eqref{eq:app-snr} use the same normalization.
Changing that normalization rescales both curves together and leaves their ratio, and hence the SNR and PLI curves, unchanged.
The factor $h^2$ in the reported spectrum is the conventional cosmological normalization; it is not an additional source
parameter.

For a detector configuration $D$, we use the noise-only estimator
\begin{equation}
 \rho_D^2=n_D T_{\rm obs}\int_{f_{\rm min}}^{f_{\rm max}}{\rm d}f\,
 \left[\frac{S_h^{\rm sig}(f)}{S_{h,D}^{\rm noise}(f)}\right]^2.
 \label{eq:app-snr}
\end{equation}
The integration limits are the frequency range of each tabulated detector curve.
The quantity $n_D$ is an effective count of independent correlations or channels in the simplified configuration.
It increases the available information, and hence multiplies $\rho_D^2$, but it should not be confused with a number of particle degrees of freedom.
Equation~\eqref{eq:app-snr} is the independent-channel, noise-only limit of the standard stochastic cross correlation statistic~\cite{Allen:1997ad}.
A complete network forecast would replace this expression by the appropriate sum over detector pairs, including the overlap-reduction functions, detector responses, duty cycles, and foreground residuals.

\subsection{Power-law-integrated curves}

For the PLI curves, we introduce a trial power law with spectral index $n$,
\begin{equation}
 h^2\Omega_n(f)=A_n
 \left(\frac{f}{f_{\rm ref}}\right)^n,
 \qquad f_{\rm ref}=1~\mathrm{Hz}.
 \label{eq:app-trial-power-law}
\end{equation}
The reference frequency only fixes the definition of the normalization $A_n$;
it has no physical significance.
Substituting the corresponding strain spectrum into \eqref{eq:app-snr} gives
\begin{equation}
 \rho_D^2=A_n^2\mathcal{I}_D(n),
\end{equation}
with
\begin{equation}
 \mathcal{I}_D(n)=n_D T_{\rm obs}\int {\rm d}f\,
 \left[\frac{3H_0^2}{4\pi^2h^2f^3}
 \frac{(f/f_{\rm ref})^n}{S_{h,D}^{\rm noise}(f)}\right]^2.
 \label{eq:app-pli-integral}
\end{equation}
For the common threshold $\rho_{\rm thr}=10$, the minimum normalization of this trial spectrum is therefore
\begin{equation}
 A_n^{\rm thr}=\frac{\rho_{\rm thr}}
 {\sqrt{\mathcal{I}_D(n)}}.
 \label{eq:app-pli-amplitude}
\end{equation}
Scanning slopes $-8\leq n\leq8$ and taking the upper envelope gives
\begin{equation}
 h^2\Omega_D^{\rm PLI}(f)=\max_n\left[
 A_n^{\rm thr}\left(\frac{f}{f_{\rm ref}}\right)^n\right],
 \label{eq:app-pli-envelope}
\end{equation}
following the construction of~\cite{Thrane:2013oya}.
The actual phase-transition spectrum is not replaced by \eqref{eq:app-trial-power-law}:
we insert the calculated spectrum directly into \eqref{eq:app-snr} to obtain the SNRs quoted in the main text.

\subsection{Detector inputs and simplifying assumptions}

The strain-noise data are taken from the public SGWBProbe release associated with~\cite{Campeti:2020xwn}.
We use the same exposure $T_{\rm obs}=4$ yr and the same PLI threshold $\rho_{\rm thr}=10$ for all entries, so that the figure provides a uniform comparison.
The numerical implementation evaluates \eqref{eq:app-snr} and \eqref{eq:app-pli-integral} directly on the tabulated detector frequency grids, using a trapezoidal frequency integral.
The displayed PLI envelope uses 321 trial slopes uniformly spaced over $-8\leq n\leq8$.

\begin{table}[h]
 \centering
 \caption{Detector inputs used in the noise-only PLI comparison.
 The strain-noise curves are taken from the public SGWBProbe release.
 All entries use \(T_{\rm obs}=4\)~yr, and CE denotes Cosmic Explorer.
 The CE-like entry is an idealized pair of identical CE-sensitivity instruments;
 it is not a forecast for a specified Cosmic Explorer site network.}
 \label{tab:detector-forecast-configurations}
 \small
 \begin{tabular}{lccc}
  \hline
  Detector configuration & adopted strain-noise curve & $n_D$ & $T_{\rm obs}$ \\
  \hline
  AEDGE & AEDGE design $S_h(f)$ & 1 & 4 yr \\
  DO (conservative) & DO conservative $S_h(f)$ & 1 & 4 yr \\
  DECIGO & DECIGO design $S_h(f)$ & 2 & 4 yr \\
  BBO & BBO design $S_h(f)$ & 2 & 4 yr \\
  ET & ET design $S_h(f)$ & 3 & 4 yr \\
  CE-like pair & CE design $S_h(f)$ & 2 & 4 yr \\
  \hline
 \end{tabular}
\end{table}

The AEDGE, DO, DECIGO, and BBO entries should be read as noise-only mission concept comparisons under the common conventions above.
For ET, the adopted effective channel count follows the triangular-detector treatment used in the public input framework.
For CE, only a single-instrument noise curve is available in the adopted data set.
We therefore display an idealized pair of identical CE-like instruments with $n_D=2$, without choosing detector sites, arm orientations, or an overlap-reduction function.
This is sufficient to show the frequency mismatch between the present deci-Hz signal and the main ground-based band, but it is not an official CE network forecast.

\FloatBarrier
\section{Numerical implementation}
\label{app:numerical-implementation}

This appendix specifies the numerical prescription used for the effective potential, the bounce, the percolation integral, and the tables quoted in the main text.
Masses, fields, and temperatures are evaluated in GeV, and potentials in $\mathrm{GeV}^4$.
The reference calculation was reproduced with Python 3.13.9, NumPy 2.3.5, SciPy 1.16.3, and the \texttt{CosmoTransitions} 2.0.7 distribution.

\subsection{Potential and thermal functions}

In $V_{\rm CW}$, $V_T$, and $V_{\rm ring}$, every tree-level squared mass is replaced by
\begin{equation}
 \widehat m_i^2=\max(m_i^2,10^{-30}~\mathrm{GeV}^2).
 \label{eq:numerical-mass-regulator}
\end{equation}
In particular, each ring difference is evaluated as
$(\widehat m_i^2+c_iT^2)^{3/2}-(\widehat m_i^2)^{3/2}$, with the cutoff applied before adding the thermal mass.
The screening coefficients are
\begin{align}
 c_\phi^{\rm base}&=0.337561358555,&
 c_\phi^{\rm ring}&=2.344975435567,\notag\\
 c_\Phi&=0.306471463974,&
 c_H&=0.295949104752,&
 c_A&=2.146955466037.
 \label{eq:numerical-screening-inputs}
\end{align}

For $y=\widehat m^2/T^2\geq0.02$, the thermal-function table is built from
\begin{align}
 J_B(y)&=-y\sum_{k=1}^{N}\frac{K_2(k\sqrt y)}{k^2},\qquad
 J_F(y)=y\sum_{k=1}^{N}\frac{(-1)^{k+1}K_2(k\sqrt y)}{k^2},\notag\\
 N&=\max\!\left(30,\left\lceil\frac{35}{\sqrt y}\right\rceil\right),
 \label{eq:numerical-bessel-series}
\end{align}
where $K_2$ is the modified Bessel function.
For $y<0.02$ we use
\begin{align}
 J_B(y)&=-\frac{\pi^4}{45}+\frac{\pi^2y}{12}
 -\frac{\pi y^{3/2}}6-\frac{y^2}{32}\ln\frac{y}{a_B},\notag\\
 J_F(y)&=\frac{7\pi^4}{360}-\frac{\pi^2y}{24}
 -\frac{y^2}{32}\ln\frac{y}{a_F},\qquad
 a_B=16\pi^2e^{3/2-2\gamma_E},\quad a_F=a_B/16,
 \label{eq:numerical-small-mass-expansion}
\end{align}
with $J_B(0)=-\pi^4/45$ and $J_F(0)=7\pi^4/360$.
These expansions follow the standard Dolan--Jackiw form~\cite{Dolan:1973qd}.
The grid contains 1001 uniformly spaced points on $0\leq y\leq1$ and 1000 logarithmically spaced points on $1.002\leq y\leq500$.
We use piecewise cubic Hermite interpolation (PCHIP) and set both functions to zero for $y\geq500$.
For the thermal-bath integrals in Eq.~\eqref{eq:thermal-bath-integrals}, adaptive quadrature uses absolute and relative tolerances $10^{-8}$ and $10^{-7}$, with at most 150 subintervals; $m/T$ is rounded to six decimal places for caching.

Writing $F(\phi)=V_{\rm CW}(\phi)$, the counterterm conditions give
\begin{equation}
 \delta\lambda_\phi=\frac{F'(v_\phi)/v_\phi-F''(v_\phi)}{2v_\phi^2},
 \qquad
 \delta\mu_\phi^2=\delta\lambda_\phi v_\phi^2+\frac{F'(v_\phi)}{v_\phi}.
 \label{eq:numerical-counterterms}
\end{equation}
Both derivatives use central differences with $h_{\rm CT}=\max(10^{-3}v_\phi,1~\mathrm{GeV})$.
For the reference inputs this yields
$\delta\mu_\phi^2=-223.7211802~\mathrm{TeV}^2$ and $\delta\lambda_\phi=0.02607910275$.
These counterterms enforce the renormalization conditions of Eq.~\eqref{eq:counterterm-conditions}.

\subsection{Minima and bounce solutions}

At each temperature, we sample 241 field values uniformly on
$0\leq\phi\leq\max(250~\mathrm{TeV},2.5v_\phi)$ and refine each nonzero local minimum by bounded minimization within its neighboring grid interval.
The field tolerance is $10^{-5}~\mathrm{GeV}$, with at most 500 iterations; the lowest nonzero minimum defines the broken branch.
To locate $T_c$, we use 140 uniformly spaced temperatures from $1~\mathrm{TeV}$ to $\max(250~\mathrm{TeV},2v_\phi)$, followed by Brent root finding on $\Delta\mathcal F$, with absolute tolerance $10^{-4}~\mathrm{GeV}$.
The origin curvature is evaluated as
\begin{equation}
 V_{\rm eff}''(0,T)\simeq\frac{2[V_{\rm eff}(h_0,T)-V_{\rm eff}(0,T)]}{h_0^2},
 \qquad h_0=\max(10~\mathrm{GeV},10^{-3}T).
 \label{eq:numerical-origin-curvature}
\end{equation}
Its zero defines the spinodal temperature, $T_{\rm sp}=4.89091836~\mathrm{TeV}$ for the reference point. This temperature is used only to set the lower bound of the nucleation search.

For the bounce, we introduce $\Lambda=1~\mathrm{TeV}$, $x=\phi/\Lambda$, $\widehat r=\Lambda r$, and
\begin{equation}
 U(x,T)=\frac{V_{\rm eff}(\Lambda x,T)-V_{\rm eff}(0,T)}{\Lambda^4},
 \qquad \frac{S_3(T)}{T}=\frac{\Lambda}{T}\widehat S_3[U].
 \label{eq:numerical-bounce-rescaling}
\end{equation}
The gradient of $U$ uses a central difference with
$h_x=\max(10^{-5},10^{-4}\max(|x|,1))$, switching to one-sided differences within $h_x$ of either minimum.
We use \texttt{SingleFieldInstanton} with its radial-friction parameter set to 2 and \texttt{phi\_eps}$=10^{-3}$.
The profile settings are \texttt{xtol}$=10^{-4}$, \texttt{phitol}$=10^{-4}$, \texttt{thinCutoff}$=0.01$, and \texttt{npoints}$=500$.
The radial bounds are \texttt{rmin}$=10^{-4}$ and \texttt{rmax}$=10^4$, in units of the solver's automatically determined radial scale; the remaining profile settings use the package defaults.

\subsection{Nucleation and temperature derivatives}

We first evaluate $\ln(\Gamma/H^4)$ at 48 temperatures uniformly spaced between $1.001T_{\rm sp}$ and $0.995T_c$.
The sign change brackets $T_n$, which is refined with Brent's method using absolute tolerance $0.2~\mathrm{GeV}$ and relative tolerance $2\times10^{-5}$.
The Hubble rate is Eq.~\eqref{eq:percolation-hubble-rate}, with $M_{\rm Pl}=1.2209\times10^{19}~\mathrm{GeV}$.
The solution is $T_n=6.72766952~\mathrm{TeV}$, where
$S_3/T=129.99974$, $H_n=1.60917\times10^{-10}~\mathrm{GeV}$, and $\ln(\Gamma/H^4)=1.20\times10^{-3}$ within the root tolerance.

For the temperature derivatives, let $A(T)=S_3(T)/T$. We use
\begin{align}
 \frac{{\rm d}\Delta\mathcal F}{{\rm d}T}
 &\simeq\frac{\Delta\mathcal F(T+h_\alpha)-\Delta\mathcal F(T-h_\alpha)}{2h_\alpha},
 &h_\alpha&=\max(2~\mathrm{GeV},10^{-3}T),\notag\\
 \frac{\beta}{H}
 &\simeq T\frac{A(T+h_\beta)-A(T-h_\beta)}{2h_\beta},
 &h_\beta&=\max(5~\mathrm{GeV},2\times10^{-3}T).
 \label{eq:numerical-temperature-derivatives}
\end{align}
At $T_n$, the steps are $6.72767~\mathrm{GeV}$ and $13.45534~\mathrm{GeV}$.
Table~\ref{tab:nucleation-derivative-samples} gives direct evaluations around this point.
Together with Table~\ref{tab:false-phase-bath}, they give $\alpha_n=1.901318$ and $\beta/H_n=410.4237$.

\begin{table}[htbp]
 \centering
 \caption{Direct evaluations around $T_n$. The central row is \(T_n\); the inner pair is used for the $\Delta\mathcal F$ derivative and the outer pair for the $S_3/T$ derivative.}
 \label{tab:nucleation-derivative-samples}
 \small
 \begin{tabular}{rrr}
  \toprule
  $T$ [TeV] & $S_3/T$ & $\Delta\mathcal F$ [$\mathrm{TeV}^4$] \\
  \midrule
  $6.714214185$ & $129.1740543$ & $242862.375051$ \\
  $6.720941854$ & $129.5863412$ & $242803.456021$ \\
  $6.727669524$ & $129.9997433$ & $242744.321814$ \\
  $6.734397193$ & $130.4006343$ & $242684.972133$ \\
  $6.741124863$ & $130.8157489$ & $242625.406497$ \\
  \bottomrule
 \end{tabular}
\end{table}

\subsection{Percolation and reheating}

The central percolation calculation uses 96 uniformly spaced temperatures from
$T_{\min}=\max(1.002T_{\rm sp},0.60T_n)$ to $0.995T_c$.
These bounds are $4.90070020$ and $13.80973796~\mathrm{TeV}$ for the reference point.
We evaluate the action and Hubble rate at each point, accumulate $\int {\rm d}T/H$ with the trapezoidal rule, and then evaluate the outer integral in Eq.~\eqref{eq:percolation-integral} on the same grid, with $v_w=1$.
The upper numerical bound is $0.995T_c$, where nucleation is already strongly suppressed.
PCHIP interpolation of the resulting $I(T)$ locates $I=0.34$ and $I=4.6$; its derivative gives $3+T_pI'(T_p)=-373.20$.

The reheating temperature is found by solving Eq.~\eqref{eq:reheating-condition} with the same thermal-bath prescription as in Section~\ref{subsec:total-potential}.
At $T_p$, $(g_\rho,g_s)=(235.238097,233.806312)$; at completion they are $(235.207535,233.770164)$.
For the scale and screening comparisons we use 72 percolation temperatures with the same integration method.
In the scale study, the gauge and Yukawa couplings are interpolated linearly in $\ln Q$ and the counterterms are recalculated, while the scalar inputs and baseline screening coefficients are held at their reference values.
The screening study varies those coefficients individually and together, keeping the portal contribution to $c_\phi^{\rm ring}$ fixed.

\FloatBarrier
\bibliography{reference}

@article{Croon:2024mde,
    author = "Croon, Djuna Lize and Weir, David James",
    title = "{Gravitational waves from phase transitions}",
    eprint = "2410.21509",
    archivePrefix = "arXiv",
    primaryClass = "hep-ph",
    reportNumber = "HIP-2024-23/TH",
    doi = "10.1080/00107514.2024.2423496",
    journal = "Contemp. Phys.",
    volume = "65",
    number = "2",
    pages = "75--96",
    year = "2024"
}

@article{Athron:2023xlk,
    author = "Athron, Peter and Bal{\'a}zs, Csaba and Fowlie, Andrew and Morris, Lachlan and Wu, Lei",
    title = "{Cosmological phase transitions: From perturbative particle physics to gravitational waves}",
    eprint = "2305.02357",
    archivePrefix = "arXiv",
    primaryClass = "hep-ph",
    doi = "10.1016/j.ppnp.2023.104094",
    journal = "Prog. Part. Nucl. Phys.",
    volume = "135",
    pages = "104094",
    year = "2024"
}

@article{Bernardo:2024bdc,
    author = "Bernardo, Reginald Christian and Ng, Kin-Wang",
    title = "{Charting the nanohertz gravitational wave sky with pulsar timing arrays}",
    eprint = "2409.07955",
    archivePrefix = "arXiv",
    primaryClass = "astro-ph.CO",
    doi = "10.1142/S0218271825400139",
    journal = "Int. J. Mod. Phys. D",
    volume = "34",
    number = "04",
    pages = "2540013",
    year = "2025"
}

@article{Caprini:2024ofd,
    author = "Caprini, Chiara and Pujol{\`a}s, Oriol and Quelquejay-Leclere, Hippolyte and Rompineve, Fabrizio and Steer, Dani{\`e}le A.",
    title = "{Primordial gravitational wave backgrounds from phase transitions with next generation ground based detectors}",
    eprint = "2406.02359",
    archivePrefix = "arXiv",
    primaryClass = "astro-ph.CO",
    reportNumber = "CERN-TH-2024-065",
    doi = "10.1088/1361-6382/ad9a48",
    journal = "Class. Quant. Grav.",
    volume = "42",
    number = "4",
    pages = "045015",
    year = "2025"
}

@article{Caprini:2025trt,
    author = "Caprini, Chiara",
    title = "{Probing Beyond the Standard Model with Gravitational Waves from Phase Transitions}",
    eprint = "2602.02861",
    archivePrefix = "arXiv",
    primaryClass = "astro-ph.CO",
    reportNumber = "CERN-TH-2025-269",
    doi = "10.53941/hihep.2025.100022",
    journal = "HiHEP",
    volume = "1",
    number = "3",
    pages = "22",
    year = "2025"
}

@article{An:2020fff,
    author = "An, Haipeng and Lyu, Kun-Feng and Wang, Lian-Tao and Zhou, Siyi",
    title = "{A unique gravitational wave signal from phase transition during inflation*}",
    eprint = "2009.12381",
    archivePrefix = "arXiv",
    primaryClass = "astro-ph.CO",
    doi = "10.1088/1674-1137/ac76a7",
    journal = "Chin. Phys. C",
    volume = "46",
    number = "10",
    pages = "101001",
    year = "2022"
}

@article{An:2023jxf,
    author = "An, Haipeng and Su, Boye and Tai, Hanwen and Wang, Lian-Tao and Yang, Chen",
    title = "{Phase transition during inflation and the gravitational wave signal at pulsar timing arrays}",
    eprint = "2308.00070",
    archivePrefix = "arXiv",
    primaryClass = "astro-ph.CO",
    doi = "10.1103/PhysRevD.109.L121304",
    journal = "Phys. Rev. D",
    volume = "109",
    number = "12",
    pages = "L121304",
    year = "2024"
}

@article{Hindmarsh:2015qta,
    author = "Hindmarsh, Mark and Huber, Stephan J. and Rummukainen, Kari and Weir, David J.",
    title = "{Numerical simulations of acoustically generated gravitational waves at a first order phase transition}",
    eprint = "1504.03291",
    archivePrefix = "arXiv",
    primaryClass = "astro-ph.CO",
    doi = "10.1103/PhysRevD.92.123009",
    journal = "Phys. Rev. D",
    volume = "92",
    number = "12",
    pages = "123009",
    year = "2015"
}

@article{Hindmarsh:2017gnf,
    author = "Hindmarsh, Mark and Huber, Stephan J. and Rummukainen, Kari and Weir, David J.",
    title = "{Shape of the acoustic gravitational wave power spectrum from a first order phase transition}",
    eprint = "1704.05871",
    archivePrefix = "arXiv",
    primaryClass = "astro-ph.CO",
    doi = "10.1103/PhysRevD.96.103520",
    journal = "Phys. Rev. D",
    volume = "96",
    number = "10",
    pages = "103520",
    year = "2017"
}

@article{Berkooz:1996km,
    author = "Berkooz, Micha and Douglas, Michael R. and Leigh, Robert G.",
    title = "{Branes intersecting at angles}",
    eprint = "hep-th/9606139",
    archivePrefix = "arXiv",
    reportNumber = "RU-96-44",
    doi = "10.1016/S0550-3213(96)00452-X",
    journal = "Nucl. Phys. B",
    volume = "480",
    pages = "265--278",
    year = "1996"
}

@article{Blumenhagen:2000wh,
    author = "Blumenhagen, Ralph and Goerlich, Lars and Kors, Boris and Lust, Dieter",
    title = "{Noncommutative compactifications of type I strings on tori with magnetic background flux}",
    eprint = "hep-th/0007024",
    archivePrefix = "arXiv",
    reportNumber = "HUB-EP-00-27",
    doi = "10.1088/1126-6708/2000/10/006",
    journal = "JHEP",
    volume = "10",
    pages = "006",
    year = "2000"
}

@article{Klein:2000qw,
    author = "Klein, Matthias and Rabadan, Raul",
    title = "{Z(N) $\times$ Z(M) orientifolds with and without discrete torsion}",
    eprint = "hep-th/0008173",
    archivePrefix = "arXiv",
    reportNumber = "FTUAM-00-19, IFT-UAM-CSIC-00-24",
    doi = "10.1088/1126-6708/2000/10/049",
    journal = "JHEP",
    volume = "10",
    pages = "049",
    year = "2000"
}

@article{Aldazabal:2000cn,
    author = "Aldazabal, G. and Franco, S. and Ibanez, Luis E. and Rabadan, R. and Uranga, A. M.",
    title = "{Intersecting brane worlds}",
    eprint = "hep-ph/0011132",
    archivePrefix = "arXiv",
    reportNumber = "CAB-IB-2919500, CERN-TH-2000-320, MIT-CTP-3042, FTUAM-00-23, IFT-UAM-CSIC-00-37",
    doi = "10.1088/1126-6708/2001/02/047",
    journal = "JHEP",
    volume = "02",
    pages = "047",
    year = "2001"
}

@article{Aldazabal:2000dg,
    author = "Aldazabal, G. and Franco, S. and Ibanez, Luis E. and Rabadan, R. and Uranga, A. M.",
    title = "{D = 4 chiral string compactifications from intersecting branes}",
    eprint = "hep-th/0011073",
    archivePrefix = "arXiv",
    reportNumber = "CAB-IB-2918200, CERN-TH-2000-321, CTP-MIT-3041, FTUAM-00-22, IFT-UAM-CSIC-00-36, MIT-CTP-3041",
    doi = "10.1063/1.1376157",
    journal = "J. Math. Phys.",
    volume = "42",
    pages = "3103--3126",
    year = "2001"
}

@article{Ibanez:2001nd,
    author = "Ibanez, Luis E. and Marchesano, F. and Rabadan, R.",
    title = "{Getting just the standard model at intersecting branes}",
    eprint = "hep-th/0105155",
    archivePrefix = "arXiv",
    reportNumber = "FTUAM-01-09, IFT-UAM-CSIC-01-15",
    doi = "10.1088/1126-6708/2001/11/002",
    journal = "JHEP",
    volume = "11",
    pages = "002",
    year = "2001"
}

@article{Cvetic:2001tj,
    author = "Cvetic, Mirjam and Shiu, Gary and Uranga, Angel M.",
    title = "{Three family supersymmetric standard - like models from intersecting brane worlds}",
    eprint = "hep-th/0107143",
    archivePrefix = "arXiv",
    reportNumber = "UPR-948-T, CERN-TH-2001-190",
    doi = "10.1103/PhysRevLett.87.201801",
    journal = "Phys. Rev. Lett.",
    volume = "87",
    pages = "201801",
    year = "2001"
}

@article{Blumenhagen:2002gw,
    author = "Blumenhagen, Ralph and Gorlich, Lars and Ott, Tassilo",
    title = "{Supersymmetric intersecting branes on the type 2A T6 / Z(4) orientifold}",
    eprint = "hep-th/0211059",
    archivePrefix = "arXiv",
    reportNumber = "HUB-EP-02-46",
    doi = "10.1088/1126-6708/2003/01/021",
    journal = "JHEP",
    volume = "01",
    pages = "021",
    year = "2003"
}

@article{Lykken:1998ec,
    author = "Lykken, Joseph D. and Poppitz, Erich and Trivedi, Sandip P.",
    title = "{Branes with GUTs and supersymmetry breaking}",
    eprint = "hep-th/9806080",
    archivePrefix = "arXiv",
    reportNumber = "UCSD-PTH-98-16, FERMILAB-PUB-98-177-T",
    doi = "10.1016/S0550-3213(98)00794-9",
    journal = "Nucl. Phys. B",
    volume = "543",
    pages = "105--121",
    year = "1999"
}

@article{Cvetic:1999hb,
    author = "Cvetic, Mirjam and Plumacher, Michael and Wang, Jing",
    title = "{Three family type IIB orientifold string vacua with nonAbelian Wilson lines}",
    eprint = "hep-th/9911021",
    archivePrefix = "arXiv",
    reportNumber = "UPR-865-T, NSF-ITP-99-132, FERMILAB-PUB-99-328-T",
    doi = "10.1088/1126-6708/2000/04/004",
    journal = "JHEP",
    volume = "04",
    pages = "004",
    year = "2000"
}

@article{Aldazabal:2000sa,
    author = "Aldazabal, G. and Ibanez, Luis E. and Quevedo, F. and Uranga, A. M.",
    title = "{D-branes at singularities: A Bottom up approach to the string embedding of the standard model}",
    eprint = "hep-th/0005067",
    archivePrefix = "arXiv",
    reportNumber = "FTUAM-00-10, IFT-UAM-CSIC-00-18, DAMTP-2000-42, CAB-IB-2405300, CERN-TH-2000-127",
    doi = "10.1088/1126-6708/2000/08/002",
    journal = "JHEP",
    volume = "08",
    pages = "002",
    year = "2000"
}

@article{Cvetic:2004ui,
    author = "Cvetic, Mirjam and Li, Tianjun and Liu, Tao",
    title = "{Supersymmetric Pati-Salam models from intersecting D6-branes: A Road to the standard model}",
    eprint = "hep-th/0403061",
    archivePrefix = "arXiv",
    reportNumber = "UPR-1068-T",
    doi = "10.1016/j.nuclphysb.2004.07.036",
    journal = "Nucl. Phys. B",
    volume = "698",
    pages = "163--201",
    year = "2004"
}

@article{Li:2019nvi,
    author = "Li, Tianjun and Mansha, Adeel and Sun, Rui",
    title = "{Revisiting the supersymmetric Pati{\textendash}Salam models from intersecting D6-branes}",
    eprint = "1910.04530",
    archivePrefix = "arXiv",
    primaryClass = "hep-th",
    doi = "10.1140/epjc/s10052-021-08839-w",
    journal = "Eur. Phys. J. C",
    volume = "81",
    number = "1",
    pages = "82",
    year = "2021"
}

@article{Li:2021pxo,
    author = "Li, Tianjun and Mansha, Adeel and Sun, Rui and Wu, Lina and He, Weikun",
    title = "{N=1 supersymmetric $SU(12)_C \times SU (2)_L \times SU(2)_R$ models,~$SU(4)_C \times SU(6)_L \times SU(2)_R$~models, and $SU(4)_C \times SU(2)_L \times SU(6)_R$~ models from intersecting D6-branes}",
    doi = "10.1103/PhysRevD.104.046018",
    journal = "Phys. Rev. D",
    volume = "104",
    number = "4",
    pages = "046018",
    year = "2021"
}

@article{He:2021gug,
    author = "He, Weikun and Li, Tianjun and Sun, Rui",
    title = "{The complete search for the supersymmetric Pati-Salam models from intersecting D6-branes}",
    eprint = "2112.09632",
    archivePrefix = "arXiv",
    primaryClass = "hep-th",
    doi = "10.1007/JHEP08(2022)044",
    journal = "JHEP",
    volume = "08",
    pages = "044",
    year = "2022"
}

@article{He:2021kbj,
    author = "He, Weikun and Li, Tianjun and Sun, Rui and Wu, Lina",
    title = "{The final model building for the supersymmetric Pati{\textendash}Salam models from intersecting D6-branes}",
    eprint = "2112.09630",
    archivePrefix = "arXiv",
    primaryClass = "hep-th",
    doi = "10.1140/epjc/s10052-022-10663-9",
    journal = "Eur. Phys. J. C",
    volume = "82",
    number = "8",
    pages = "710",
    year = "2022"
}

@article{Li:2022cqk,
    author = "Li, Tianjun and Sun, Rui and Wu, Lina",
    title = "{String-scale gauge coupling relations in the supersymmetric Pati-Salam models from intersecting D6-branes}",
    eprint = "2212.05875",
    archivePrefix = "arXiv",
    primaryClass = "hep-th",
    doi = "10.1007/JHEP03(2023)210",
    journal = "JHEP",
    volume = "03",
    pages = "210",
    year = "2023"
}

@article{Li:2025xpn,
    author = "Li, Tianjun and Sun, Qi and Sun, Rui and Wu, Lina",
    title = "{Generalized Three-Family Supersymmetric Pati-Salam Models from Type IIA Intersecting D6-Branes}",
    eprint = "2511.16565",
    archivePrefix = "arXiv",
    primaryClass = "hep-th",
    month = "11",
    year = "2025"
}

@article{Mansha:2023kwq,
    author = "Mansha, Adeel and Li, Tianjun and Wu, Lina",
    title = "{The hidden sector variations in the $\mathcal{N}=1$ supersymmetric three-family Pati{\textendash}Salam models from intersecting D6-branes}",
    eprint = "2303.02864",
    archivePrefix = "arXiv",
    primaryClass = "hep-th",
    doi = "10.1140/epjc/s10052-023-12167-6",
    journal = "Eur. Phys. J. C",
    volume = "83",
    number = "11",
    pages = "1067",
    year = "2023"
}

@article{Machacek:1983tz,
    author = "Machacek, Marie E. and Vaughn, Michael T.",
    title = "{Two Loop Renormalization Group Equations in a General Quantum Field Theory. 1. Wave Function Renormalization}",
    reportNumber = "NUB-2590, HUTP-83/A003",
    doi = "10.1016/0550-3213(83)90610-7",
    journal = "Nucl. Phys. B",
    volume = "222",
    pages = "83--103",
    year = "1983"
}

@article{Machacek:1983fi,
    author = "Machacek, Marie E. and Vaughn, Michael T.",
    title = "{Two Loop Renormalization Group Equations in a General Quantum Field Theory. 2. Yukawa Couplings}",
    reportNumber = "NUB 2611",
    doi = "10.1016/0550-3213(84)90533-9",
    journal = "Nucl. Phys. B",
    volume = "236",
    pages = "221--232",
    year = "1984"
}

@article{Machacek:1984zw,
    author = "Machacek, Marie E. and Vaughn, Michael T.",
    title = "{Two Loop Renormalization Group Equations in a General Quantum Field Theory. 3. Scalar Quartic Couplings}",
    reportNumber = "NUB-2653-REV, NUB-2653",
    doi = "10.1016/0550-3213(85)90040-9",
    journal = "Nucl. Phys. B",
    volume = "249",
    pages = "70--92",
    year = "1985"
}

@article{Cvetic:1998uw,
    author = "Cvetic, G. and Kim, C. S. and Hwang, S. S.",
    title = "{Higgs mediated flavor changing neutral currents in the general framework with two Higgs doublets: An RGE analysis}",
    eprint = "hep-ph/9806282",
    archivePrefix = "arXiv",
    doi = "10.1103/PhysRevD.58.116003",
    journal = "Phys. Rev. D",
    volume = "58",
    pages = "116003",
    year = "1998"
}

@article{Gogoladze:2010in,
    author = "Gogoladze, Ilia and He, Bin and Shafi, Qaisar",
    title = "{New Fermions at the LHC and Mass of the Higgs Boson}",
    eprint = "1004.4217",
    archivePrefix = "arXiv",
    primaryClass = "hep-ph",
    doi = "10.1016/j.physletb.2010.05.076",
    journal = "Phys. Lett. B",
    volume = "690",
    pages = "495--500",
    year = "2010"
}

@article{Barger:1992ac,
    author = "Barger, Vernon D. and Berger, M. S. and Ohmann, P.",
    title = "{Supersymmetric grand unified theories: Two loop evolution of gauge and Yukawa couplings}",
    eprint = "hep-ph/9209232",
    archivePrefix = "arXiv",
    reportNumber = "MAD-PH-711",
    doi = "10.1103/PhysRevD.47.1093",
    journal = "Phys. Rev. D",
    volume = "47",
    pages = "1093--1113",
    year = "1993"
}

@article{Barger:1993gh,
    author = "Barger, Vernon D. and Berger, M. S. and Ohmann, P.",
    title = "{The Supersymmetric particle spectrum}",
    eprint = "hep-ph/9311269",
    archivePrefix = "arXiv",
    reportNumber = "MAD-PH-801",
    doi = "10.1103/PhysRevD.49.4908",
    journal = "Phys. Rev. D",
    volume = "49",
    pages = "4908--4930",
    year = "1994"
}

@article{Martin:1993zk,
    author = "Martin, Stephen P. and Vaughn, Michael T.",
    title = "{Two loop renormalization group equations for soft supersymmetry breaking couplings}",
    eprint = "hep-ph/9311340",
    archivePrefix = "arXiv",
    reportNumber = "NUB-3081-93-TH",
    doi = "10.1103/PhysRevD.50.2282",
    journal = "Phys. Rev. D",
    volume = "50",
    pages = "2282",
    year = "1994",
    note = "[Erratum: Phys.Rev.D 78, 039903 (2008)]"
}

@article{Huang:2020bbe,
    author = "Huang, Wei-Chih and Sannino, Francesco and Wang, Zhi-Wei",
    title = "{Gravitational Waves from Pati-Salam Dynamics}",
    eprint = "2004.02332",
    archivePrefix = "arXiv",
    primaryClass = "hep-ph",
    reportNumber = "CP3-Origins-2020-05 DNRF90",
    doi = "10.1103/PhysRevD.102.095025",
    journal = "Phys. Rev. D",
    volume = "102",
    number = "9",
    pages = "095025",
    year = "2020"
}

@article{Huangfu:2025uhu,
    author = "Huangfu, Haotian and Li, Tianjun and Sun, Qi and Sun, Rui",
    title = "{The ${\cal N}=1$ supersymmetric Pati-Salam models with extra $SU(2)_{L_2/R_2}$ gauge symmetry from intersecting D6-branes}",
    eprint = "2512.09057",
    archivePrefix = "arXiv",
    primaryClass = "hep-th",
    month = "12",
    year = "2025"
}

@article{Arnold:1992rz,
    author = "Arnold, Peter Brockway and Espinosa, Olivier",
    title = "{The Effective potential and first order phase transitions: Beyond leading-order}",
    eprint = "hep-ph/9212235",
    archivePrefix = "arXiv",
    reportNumber = "UW-PT-92-18, USM-TH-60",
    doi = "10.1103/PhysRevD.47.3546",
    journal = "Phys. Rev. D",
    volume = "47",
    pages = "3546",
    year = "1993",
    note = "[Erratum: Phys.Rev.D 50, 6662 (1994)]"
}

@article{Carrington:1991hz,
    author = "Carrington, M. E.",
    title = "{The Effective potential at finite temperature in the Standard Model}",
    reportNumber = "TPI-MINN-91-48-T-REV, TPI-MINN-91-48-T",
    doi = "10.1103/PhysRevD.45.2933",
    journal = "Phys. Rev. D",
    volume = "45",
    pages = "2933--2944",
    year = "1992"
}

@article{Dolan:1973qd,
    author = "Dolan, L. and Jackiw, R.",
    title = "{Symmetry Behavior at Finite Temperature}",
    reportNumber = "MIT-CTP-406",
    doi = "10.1103/PhysRevD.9.3320",
    journal = "Phys. Rev. D",
    volume = "9",
    pages = "3320--3341",
    year = "1974"
}

@inproceedings{Quiros:1999jp,
    author = "Quiros, Mariano",
    title = "{Finite temperature field theory and phase transitions}",
    booktitle = "{ICTP Summer School in High-Energy Physics and Cosmology}",
    eprint = "hep-ph/9901312",
    archivePrefix = "arXiv",
    reportNumber = "IEM-FT-187-99",
    pages = "187--259",
    month = "1",
    year = "1999"
}

@book{Laine:2016hma,
    author = "Laine, Mikko and Vuorinen, Aleksi",
    title = "{Basics of Thermal Field Theory}",
    eprint = "1701.01554",
    archivePrefix = "arXiv",
    primaryClass = "hep-ph",
    doi = "10.1007/978-3-319-31933-9",
    publisher = "Springer",
    volume = "925",
    year = "2016"
}

@book{KolbTurner1990,
  author    = {Kolb, Edward W. and Turner, Michael S.},
  title     = {The Early Universe},
  publisher = {Addison-Wesley},
  address   = {Redwood City, CA},
  year      = {1990}
}

@article{Linde:1981zj,
    author = "Linde, Andrei D.",
    title = "{Decay of the False Vacuum at Finite Temperature}",
    reportNumber = "LEBEDEV-81-265",
    doi = "10.1016/0550-3213(83)90072-X",
    journal = "Nucl. Phys. B",
    volume = "216",
    pages = "421",
    year = "1983",
    note = "[Erratum: Nucl.Phys.B 223, 544 (1983)]"
}

@article{Guth:1982pn,
    author = "Guth, Alan H. and Weinberg, Erick J.",
    title = "{Could the Universe Have Recovered from a Slow First Order Phase Transition?}",
    reportNumber = "MIT-CTP-950",
    doi = "10.1016/0550-3213(83)90307-3",
    journal = "Nucl. Phys. B",
    volume = "212",
    pages = "321--364",
    year = "1983"
}

@article{Ellis:2018mja,
    author = "Ellis, John and Lewicki, Marek and No, Jos{\'e} Miguel",
    title = "{On the Maximal Strength of a First-Order Electroweak Phase Transition and its Gravitational Wave Signal}",
    eprint = "1809.08242",
    archivePrefix = "arXiv",
    primaryClass = "hep-ph",
    reportNumber = "KCL-PH-TH/2018-46, CERN-TH/2018-197, IFT-UAM/CSIC-18-94, CERN-TH-2018-197",
    doi = "10.1088/1475-7516/2019/04/003",
    journal = "JCAP",
    volume = "04",
    pages = "003",
    year = "2019"
}

@article{Megevand:2016lpr,
    author = "Megevand, Ariel and Ramirez, Santiago",
    title = "{Bubble nucleation and growth in very strong cosmological phase transitions}",
    eprint = "1611.05853",
    archivePrefix = "arXiv",
    primaryClass = "astro-ph.CO",
    doi = "10.1016/j.nuclphysb.2017.03.009",
    journal = "Nucl. Phys. B",
    volume = "919",
    pages = "74--109",
    year = "2017"
}

@article{Rintoul:1997tze,
    author = "Rintoul, M. D. and Torquato, S.",
    title = "{Precise determination of the critical threshold and exponents in a three-dimensional continuum percolation model}",
    doi = "10.1088/0305-4470/30/16/005",
    journal = "J. Phys. A",
    volume = "30",
    number = "16",
    pages = "L585",
    year = "1997"
}

@article{Bodeker:2009qy,
    author = "Bodeker, Dietrich and Moore, Guy D.",
    title = "{Can electroweak bubble walls run away?}",
    eprint = "0903.4099",
    archivePrefix = "arXiv",
    primaryClass = "hep-ph",
    doi = "10.1088/1475-7516/2009/05/009",
    journal = "JCAP",
    volume = "05",
    pages = "009",
    year = "2009"
}

@article{Espinosa:2010hh,
    author = "Espinosa, Jose R. and Konstandin, Thomas and No, Jose M. and Servant, Geraldine",
    title = "{Energy Budget of Cosmological First-order Phase Transitions}",
    eprint = "1004.4187",
    archivePrefix = "arXiv",
    primaryClass = "hep-ph",
    reportNumber = "CERN-PH-TH-2010-027",
    doi = "10.1088/1475-7516/2010/06/028",
    journal = "JCAP",
    volume = "06",
    pages = "028",
    year = "2010"
}

@article{Caprini:2015zlo,
    author = "Caprini, Chiara and others",
    title = "{Science with the space-based interferometer eLISA. II: Gravitational waves from cosmological phase transitions}",
    eprint = "1512.06239",
    archivePrefix = "arXiv",
    primaryClass = "astro-ph.CO",
    reportNumber = "DESY-15-246",
    doi = "10.1088/1475-7516/2016/04/001",
    journal = "JCAP",
    volume = "04",
    pages = "001",
    year = "2016"
}

@article{Caprini:2019egz,
    author = "Caprini, Chiara and others",
    title = "{Detecting gravitational waves from cosmological phase transitions with LISA: an update}",
    eprint = "1910.13125",
    archivePrefix = "arXiv",
    primaryClass = "astro-ph.CO",
    reportNumber = "DESY-19-159, IPPP/19/27, HIP-2019-14/TH, MITP/19-066, IFT-UAM/CSIC-19-139",
    doi = "10.1088/1475-7516/2020/03/024",
    journal = "JCAP",
    volume = "03",
    pages = "024",
    year = "2020"
}

@article{AEDGE:2019nxb,
    author = "El-Neaj, Yousef Abou and others",
    collaboration = "AEDGE",
    title = "{AEDGE: Atomic Experiment for Dark Matter and Gravity Exploration in Space}",
    eprint = "1908.00802",
    archivePrefix = "arXiv",
    primaryClass = "gr-qc",
    reportNumber = "KCL-PH-TH/2019-65, CERN-TH-2019-126",
    doi = "10.1140/epjqt/s40507-020-0080-0",
    journal = "EPJ Quant. Technol.",
    volume = "7",
    pages = "6",
    year = "2020"
}

@article{Sedda:2019uro,
    author = "Sedda, Manuel Arca and others",
    title = "{The missing link in gravitational-wave astronomy: discoveries waiting in the decihertz range}",
    eprint = "1908.11375",
    archivePrefix = "arXiv",
    primaryClass = "gr-qc",
    doi = "10.1088/1361-6382/abb5c1",
    journal = "Class. Quant. Grav.",
    volume = "37",
    number = "21",
    pages = "215011",
    year = "2020"
}

@article{Crowder:2005nr,
    author = "Crowder, Jeff and Cornish, Neil J.",
    title = "{Beyond LISA: Exploring future gravitational wave missions}",
    eprint = "gr-qc/0506015",
    archivePrefix = "arXiv",
    doi = "10.1103/PhysRevD.72.083005",
    journal = "Phys. Rev. D",
    volume = "72",
    pages = "083005",
    year = "2005"
}

@article{Punturo:2010zz,
    author = "Punturo, M. and others",
    editor = "Ricci, Fulvio",
    title = "{The Einstein Telescope: A third-generation gravitational wave observatory}",
    doi = "10.1088/0264-9381/27/19/194002",
    journal = "Class. Quant. Grav.",
    volume = "27",
    pages = "194002",
    year = "2010"
}

@article{Reitze:2019iox,
    author = "Reitze, David and others",
    title = "{Cosmic Explorer: The U.S. Contribution to Gravitational-Wave Astronomy beyond LIGO}",
    eprint = "1907.04833",
    archivePrefix = "arXiv",
    primaryClass = "astro-ph.IM",
    reportNumber = "LIGO-P1900316",
    journal = "Bull. Am. Astron. Soc.",
    volume = "51",
    number = "7",
    pages = "035",
    year = "2019"
}

@article{Schmitz:2020syl,
    author = "Schmitz, Kai",
    title = "{New Sensitivity Curves for Gravitational-Wave Signals from Cosmological Phase Transitions}",
    eprint = "2002.04615",
    archivePrefix = "arXiv",
    primaryClass = "hep-ph",
    reportNumber = "CERN-TH-2020-018",
    doi = "10.1007/JHEP01(2021)097",
    journal = "JHEP",
    volume = "01",
    pages = "097",
    year = "2021"
}

@article{Thrane:2013oya,
    author = "Thrane, Eric and Romano, Joseph D.",
    title = "{Sensitivity curves for searches for gravitational-wave backgrounds}",
    eprint = "1310.5300",
    archivePrefix = "arXiv",
    primaryClass = "astro-ph.IM",
    doi = "10.1103/PhysRevD.88.124032",
    journal = "Phys. Rev. D",
    volume = "88",
    number = "12",
    pages = "124032",
    year = "2013"
}

@article{Campeti:2020xwn,
    author = "Campeti, Paolo and Komatsu, Eiichiro and Poletti, Davide and Baccigalupi, Carlo",
    title = "{Measuring the spectrum of primordial gravitational waves with CMB, PTA and Laser Interferometers}",
    eprint = "2007.04241",
    archivePrefix = "arXiv",
    primaryClass = "astro-ph.CO",
    doi = "10.1088/1475-7516/2021/01/012",
    journal = "JCAP",
    volume = "01",
    pages = "012",
    year = "2021"
}

@article{Yagi:2011wg,
    author = "Yagi, Kent and Seto, Naoki",
    title = "{Detector configuration of DECIGO/BBO and identification of cosmological neutron-star binaries}",
    eprint = "1101.3940",
    archivePrefix = "arXiv",
    primaryClass = "astro-ph.CO",
    doi = "10.1103/PhysRevD.83.044011",
    journal = "Phys. Rev. D",
    volume = "83",
    pages = "044011",
    year = "2011",
    note = "[Erratum: Phys.Rev.D 95, 109901 (2017)]"
}

@article{Allen:1997ad,
    author = "Allen, Bruce and Romano, Joseph D.",
    title = "{Detecting a stochastic background of gravitational radiation: Signal processing strategies and sensitivities}",
    eprint = "gr-qc/9710117",
    archivePrefix = "arXiv",
    reportNumber = "WISC-MILW-97-TH-14",
    doi = "10.1103/PhysRevD.59.102001",
    journal = "Phys. Rev. D",
    volume = "59",
    pages = "102001",
    year = "1999"
}

@article{Patt:2006fw,
    author = "Patt, Brian and Wilczek, Frank",
    title = "{Higgs-field portal into hidden sectors}",
    eprint = "hep-ph/0605188",
    archivePrefix = "arXiv",
    reportNumber = "MIT-CTP-3745",
    month = "5",
    year = "2006"
}

@article{Lebedev:2021xey,
    author = "Lebedev, Oleg",
    title = "{The Higgs portal to cosmology}",
    eprint = "2104.03342",
    archivePrefix = "arXiv",
    primaryClass = "hep-ph",
    doi = "10.1016/j.ppnp.2021.103881",
    journal = "Prog. Part. Nucl. Phys.",
    volume = "120",
    pages = "103881",
    year = "2021"
}

@article{ATLAS:2026Dilepton,
 author = "{ATLAS Collaboration}",
 title = "Search for high-mass dilepton resonances in $pp$ collisions at $\sqrt{s}=13.6$ TeV combined with 13 TeV results using the ATLAS detector",
 eprint = "2607.28334",
 archivePrefix = "arXiv",
 primaryClass = "hep-ex",
 year = "2026"
}

@article{NNPDF:2013QED,
 author = "Ball, Richard D. and others",
 collaboration = "NNPDF",
 title = "Parton distributions with QED corrections",
 eprint = "1308.0598",
 archivePrefix = "arXiv",
 primaryClass = "hep-ph",
 doi = "10.1016/j.nuclphysb.2013.10.010",
 year = "2013"
}

\end{document}